\documentclass[12pt]{iopart}

\usepackage{iopams}  
\usepackage{graphicx}
\usepackage[ruled,linesnumbered]{algorithm2e}
\usepackage{lscape}

\begin{document}

\title[]{Surface Code Error Correction on a Defective Lattice}
\author{Shota Nagayama$^1$, Austin G. Fowler$^2$, Dominic Horsman$^3$, Simon J. Devitt$^4$ and Rodney Van Meter$^1$}
\address{
  $^1$Keio University, 5322 Endo, Fujisawa-shi, Kanagawa 252-0882 Japan\\
  $^2$Google Inc., Santa Barbara, CA 93117, USA\\
  $^3$Department of Physics, Durham University, South Road, Durham DH1 3LE, UK\\
  $^4$Center for Emergent Matter Science, RIKEN, Wakoshi, Saitama 315-0198, Japan\\
}
\ead{kurosagi@sfc.wide.ad.jp}
 \begin{abstract}
  The yield of physical qubits fabricated in the laboratory is much lower than that of classical
  transistors in production semiconductor fabrication.
  Actual implementations of quantum computers will be susceptible
  to loss in the form of physically faulty qubits.
  Though these physical faults must negatively affect the computation,
  we can deal with them by adapting error correction schemes.
  In this paper 
  We have simulated 
  statically placed single-fault lattices
  and lattices with randomly placed faults at functional qubit yields of
  80\%, 90\% and 95\%,
  showing practical performance of a defective surface code by employing actual circuit constructions
  and realistic
  errors on every gate, including identity gates.
  We extend Stace et al.'s superplaquettes solution 
  against dynamic losses for the surface code
  to handle static losses such as physically faulty qubits~\cite{stace:200501}.
  The single-fault analysis shows that a static loss at the periphery
  of the lattice has less negative effect than a static loss at the center.
  The randomly-faulty analysis shows that
  95\% yield is good enough to build a large scale quantum computer.
  The local gate error rate threshold is $\sim 0.3\%$, and a code distance of seven suppresses the residual error rate below the original error rate at $p=0.1\%$.
  90\% yield is also good enough when we discard badly fabricated quantum computation chips, while
  80\% yield does not show enough error suppression even when discarding 90\% of the chips.
  We evaluated several metrics for predicting chip performance, and found that
  the average of the product of the number of data qubits and the cycle time of a stabilizer measurement of stabilizers
  gave the strongest correlation
  with post-correction residual error rates.
  Our analysis will help with selecting usable
  quantum computation chips from among the pool of all fabricated chips.
 \end{abstract}

\clearpage
\section{Introduction}
Fully scalable quantum computers are required in order to solve
meaningful problems~\cite{nielsen-chuang:qci,VanMeter:2013:BBQ:2507771.2494568,cirac2000scalable,yao10:_scal_room_temp_arch,chiaverini05:qft-impl,PhysRevB.76.174507}.
For example, processing Shor's algorithm to
factor a number described with $N$ bits requires a quantum register with at least $2N+2$ high-quality qubits 
~\cite{
shor:1994factor,
takahashi06:small-shor}.
Many architectures have been proposed for a scalable quantum computer and
their feasibility depends on the physical systems in which they are implemented
and the physical operations they use
~\cite{van-meter10:dist_arch_ijqi,
Jones:2012Layered_Architecture_for_Quantum_Computing,
devitt:2009Architectural_design_for_a_topological_cluster_state_quantum_computer,
Whitney:2009:FTA:1555815.1555802}.
The coherence time of a quantum state is limited as the
quantum state is easily changed by noise;
fault tolerant quantum computation (FTQC) is therefore
required~\cite{Shor:1996:FQC:874062.875509,
preskill98:_reliab_quant_comput,
1402-4896-2009-T137-014020,
1367-2630-11-1-013061,
svore2007noise}.
The surface code is one of the most feasible current proposals
for FTQC, requiring a 2-D square-lattice of qubits and
interaction only between nearest neighbors, maintaining good scalability and
having a higher threshold
than other codes on equivalently constrained architectures~\cite{
Fowler:2009High-threshold_universal_quantum_computation_on_the_surface_code,
kitaev2003ftq,
bravyi1998qcl,
raussendorf07:_2D_topo,
raussendorf07:_topol_fault_toler_in_clust}.
The surface code qubits are grouped in ``plaquettes'' which,
in the absence of faulty components, consist of four neighboring
qubits in the lattice. Each plaquette is associated with a stabilizer measurement.
There are two types of stabilizers -- Z stabilizers and X stabilizers -- enabling the correction of arbitrary errors.
Error syndromes are associated with pairs of sequential stabilizer measurements that differ.

In reality, the problems we face include not only state errors but also losses of quanta.
Some examples of loss mechanisms are:
static loss such as devices incapable of trapping single electrons for use as qubits,
dynamic loss such as photon generation failure or dynamic loss of other qubit carriers.
There are many proposals for 2-D nearest neighbor architectures
on which the surface code runs.
However, each of them suffers from the problems we mentioned above;
if a qubit is missing, there will be a hole in the code.
DiVincenzo offered an architecture of superconducting hardware for the surface code~\cite{1402-4896-2009-T137-014020},
in which a superconducting loop which does not show the appropriate quantum effect will be a hole.
Jones et al. proposed an architecture for scalable quantum computation
with self-assembled quantum dots used 
to trap electrons, which are used as qubits~\cite{Jones:2012Layered_Architecture_for_Quantum_Computing}.
There very likely will be defective quantum dots which cannot trap a single electron,
leaving holes in the code.

The surface code is robust against unintended changes of quantum state,
provided these changes are local in space and time,
but it does not address loss.
To resolve this problem, we have two choices:
design a microarchitecture to work around missing qubits,
or adapt the syndrome collection and processing to tolerate loss.
Van Meter et al. proposed a system in which the microarchitecture can create
the regular 2-D lattice even when some qubits are faulty~\cite{van-meter10:dist_arch_ijqi}.
However, this requires the ability to couple qubits across a distance spanning several qubit sites.
Stace et al. showed that qubit loss is acceptable
when performing the surface code
and that there is a tradeoff between the loss rate and the state error rate~\cite{stace:200501}.
They introduced the concept of a ``superplaquette'' which 
consists of several plaquettes that surround defective qubits.
They showed that, under the assumption that the superplaquette operators can be measured perfectly,
a threshold error rate exists for qubit loss rates below 50\%.
Barrett et al. showed that dynamic loss in the 3-D topological quantum computation
is acceptable up to $p_{loss}=24.9\%$~\cite{Barrett:PhysRevLett.105.200502}.
This latter approach, however, cannot be used if a device used to bond together qubits in the 3-D topological lattice
in the quantum computer is permanently faulty, leading to a column in time of lost qubits.

In this paper,
we examine these theoretical limits in the context of errors in the state
and stabilizer measurement process.
We give the realistic relationship of the 2-D surface code
between static loss tolerance and state error tolerance
by employing explicit stabilizer circuit constructions.
We generalize the concept of a ``superplaquette'' to a ``superunit''
and measure error syndromes working around faulty devices.
Additionally, we introduce the concept of a defective stabilizer whose syndrome qubit -- the ancilla qubit
in the center of the stabilizer used to measure the error syndrome -- is defective.
Finally we show the acceptability of such special stabilizer units with examples of some stabilizer circuits and
graphs of the relationships between qubit yield, code distance and effective code distance.
The effective code distance is the value characterizing how many gate errors the code can truly handle given the presence of defective qubits.
We analyze the correlation between the logical error rates we find on each defective lattice and the characteristics of those defective lattices
to find indicators to judge whether a defective lattice is good enough to use.
We simulate yields of 0.95, 0.90 and 0.80.
Our results show that once the fabrication yield reaches 90\%,
it becomes possible to build large-scale systems, by culling the poorer 50\% of chips during post-fabrication testing.
Yield 0.80 is not usable even when discarding 90\% of generated lattices.
We find that the average of the product of the number of data qubits and the cycle time of a stabilizer measurement of stabilizers
has the largest correlation with the logical error rate of the lattice
and the biggest number of data qubits owned by a stabilizer has the next largest correlation.
Large scale quantum computation requires distributed quantum computation and an ensemble of sufficiently-fault-tolerant quantum computation chips
~\cite{RevModPhys.82.665,buhrman03:_dist_qc,buhrman1998quantum,PhysRevA.89.022317,chien15:_ft-blind,broadbent2010measurement,crepeau:_secur_multi_party_qc}.
Our results will contribute to guiding the construction of this ensemble.

\section{Surface code quantum computation}
\label{surface_code}
The surface code is a means for encoding logical qubits on a form of entangled 2-D lattice, consisting of many qubits,
made by interaction only between nearest neighbor qubits.
This fact makes it potentially possible to fabricate devices using planar photolithography, including quantum dot, superconducting, and planar ion trap structures.
It gives the quantum processor extensibility by adding one more row of qubit and control devices along the outside edge of the lattice,
making it one of the most feasible current proposals for building a scalable quantum computer.

The lattice is divided into many plaquettes and 
the state of the lattice is maintained by repeatedly
measuring sets of stabilizers.
By definition, a stabilizer generator is a set of Pauli operators $U$ that do not change a state $|\Psi\rangle$,
such as $ \vert \Psi \rangle = U \vert \Psi \rangle$.
We check a stabilizer by extracting its eigenvalue.
$X$ ($Z$) errors are checked using the $Z$ ($X$) stabilizers
We refer to these normal stabilizers involving four data qubits and one syndrome qubit as ``unit stabilizers''.

\begin{figure}[t]
 \begin{center}
  (a)
  \includegraphics[width=6cm]{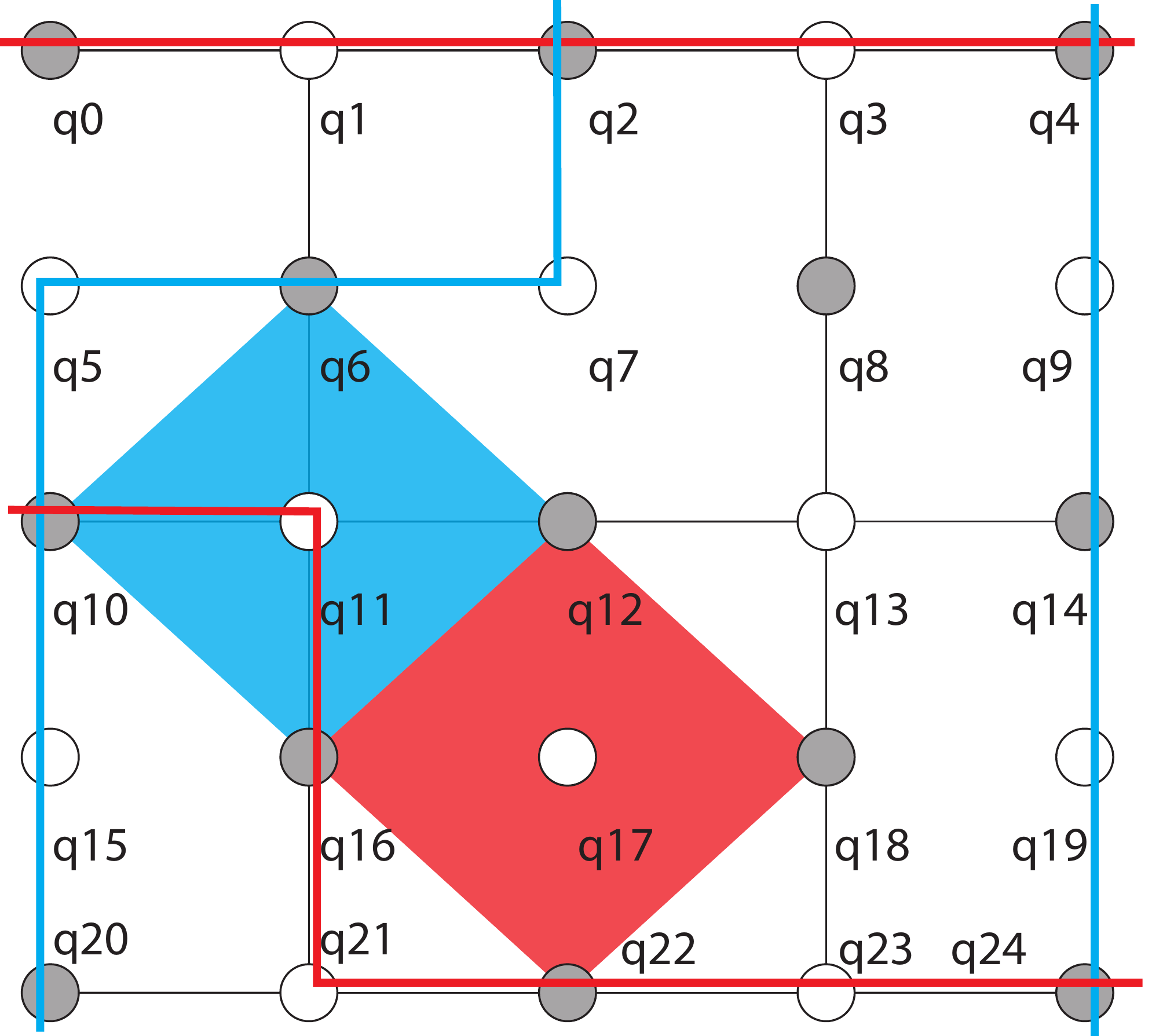}
  (b)
  \includegraphics[width=6cm]{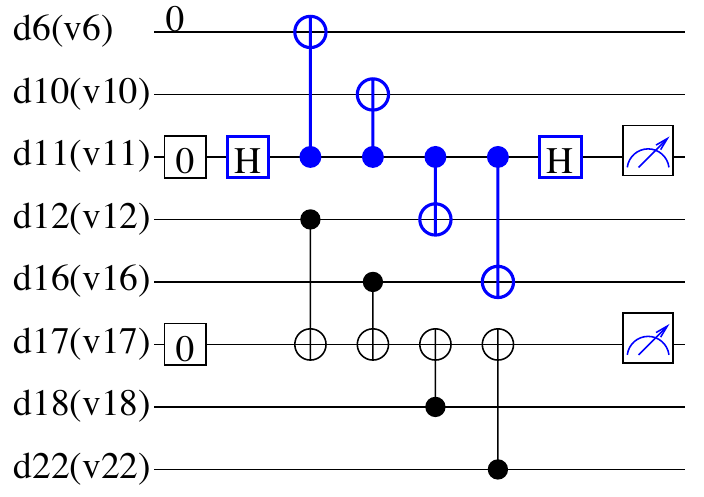}
  \caption{
  (a)
  Example of a surface code encoding a single logical qubit,
  describing a Z stabilizer (red diamond), an X stabilizer (blue diamond),
  two instances of the Z operator (red lines) and two instances of the X operator (blue lines).
  The gray (white) circles are data (syndrome) qubits.
  Qubits q12, q16, q18 and q22 (q6, q10, q12 and q16) are included in the $Z$ ($X$) stabilizer.
  Other qubits are also involved in other corresponding stabilizers.
  The west and the east boundaries of the lattice are for the $Z$ operator.
  Other possible lines between the west and the east boundaries also have the same $Z$ operator.
  The $X$ operator runs between the north and the south boundaries.
  Other equivalent logical operators are formed by multiplying a line by associated $Z$ ($X$) stabilizers.
  (b)
  Circuits for each stabilizer colored in Fig \ref{fig:sc:stabilizers} (a).
  The ``0'' on the top of the figure is step number.
  The top (bottom) half is an $X$ ($Z$) stabilizer.
  Each face and star in Figure \ref{fig:sc:stabilizers} (a) has the same stabilizer circuit.
  The only gates required for stabilizer operation are
  INIT in Z basis, CNOT, SWAP and H gates and MEASUREMENT in Z basis.
  Boxes containing 0 are the INIT gates.}
  \label{fig:sc:stabilizers}
 \end{center}
\end{figure}

The surface code corrects errors in each unit and the code space is protected as a whole.
Figure \ref{fig:sc:stabilizers} (a) shows the layout of normal unit stabilizers of a planar code, which is the form of the surface code
we employ for our simulation~\cite{bravyi1998qcl}.
The black lines in the figure are just a visual guide demarking plaquettes; each syndrome qubit is actually physically coupled to four neighbors.
Figure \ref{fig:sc:stabilizers} (b) and Table \ref{tab:sc:stabilizers} show the stabilizer representation and the circuits of the stabilizers marked
in Figure \ref{fig:sc:stabilizers} (a), respectively.
\begin{table}[t]
  \begin{center}
   \caption{Stabilizer representation of the stabilizers in Figure \ref{fig:sc:stabilizers} (b). The upper line is a Z stabilizer and the lower is an X stabilizer.}
   \label{tab:sc:stabilizers}
   \begin{tabular}[t]{cccccc}
    q6 &q10 &q12 &q16 &q18 &q22\\
    \hline
       &    &Z   &Z   &Z   &Z\\
    X  &X   &X   &X   &    & \\
   \end{tabular}
  \end{center}
\end{table}
The stabilizers measure the parity of the data qubits involved.
Normally, the parity is even (+1 eigenstate). When the states of
an odd number of qubits that belong to the stabilizer are flipped,
the parity becomes odd (-1 eigenstate).

The planar code defines a logical operator, using the degree of freedom introduced at a set of lattice boundaries.
A lattice boundary is a terminal of a logical operator; hence a pair of boundaries introduces a logical operator and two pairs of different boundaries can generate a set of a logical X operator and a logical Z operator so that a single logical qubit is introduced.
Any path between a pair of boundaries defines the same logical operator.
The planar code performs logical two-qubit operations by transversal operations
or lattice surgery~\cite{Horsman:2012lattice_surgery}.
To measure a logical qubit, take the parity of the measurement results on the physical qubits composing a logical operator.
Parities of measurements on any path should have the same value, so that the logical measurement has redundancy against measurement failure
as shown in Figure \ref{fig:sc:stabilizers} (a).

A change in the error syndrome of a stabilizer indicates 
that the stabilizer is the termination of an error chain.
Therefore we execute minimum weight perfect matching to find the most likely error pattern that results in the observed error syndromes.
See \ref{sec:app:ec} for details of the error correction of the surface code.
The distance between the two boundaries for an operator is the code distance of the surface code,
as shown with the blue line on the right and with the red line on the top in Figure \ref{fig:sc:stabilizers} (a).
The longer the code distance, the higher tolerance against errors.
If the two boundaries were farther apart,
a longer chain of errors would be required to cause the error correction to fail.

A \textit{nest} is created and used as a network for minimum weight perfect matching.
Each vertex of the nest corresponds to a stabilizer value
and each edge corresponds to a possible error chain.
Details are in \ref{sec:appendix:matching}.

\section{The structure of the surface code on a defective lattice}
\label{sec:solution}
The difficulty in working around faulty devices arises from the nearest neighbor architecture and the two separate roles for qubits.
Distant qubits have to interact around faulty devices but the nearest neighbor architecture does not provide the capability for such qubits to interact directly.
SWAP gates are brought in to solve this problem.
The solutions for faulty syndrome qubits and for faulty data qubits differ.
To tolerate faulty data qubits, we introduce the ``superunit'' that Stace et al. called the ``superplaquette'' \cite{stace:200501}.
The idea is to maintain error correction by modifying the shape of stabilizers around lost data qubits (faulty devices).
On the other hand, we do not have to modify the unit of stabilizers when syndrome qubits are faulty.
We can gather error syndromes onto another syndrome qubit instead of the faulty syndrome qubit, by using SWAP gates.

\subsection{Stabilizer reconfiguration}
\begin{figure}[t]
 \begin{center}
(a)
  \includegraphics[width=5cm]{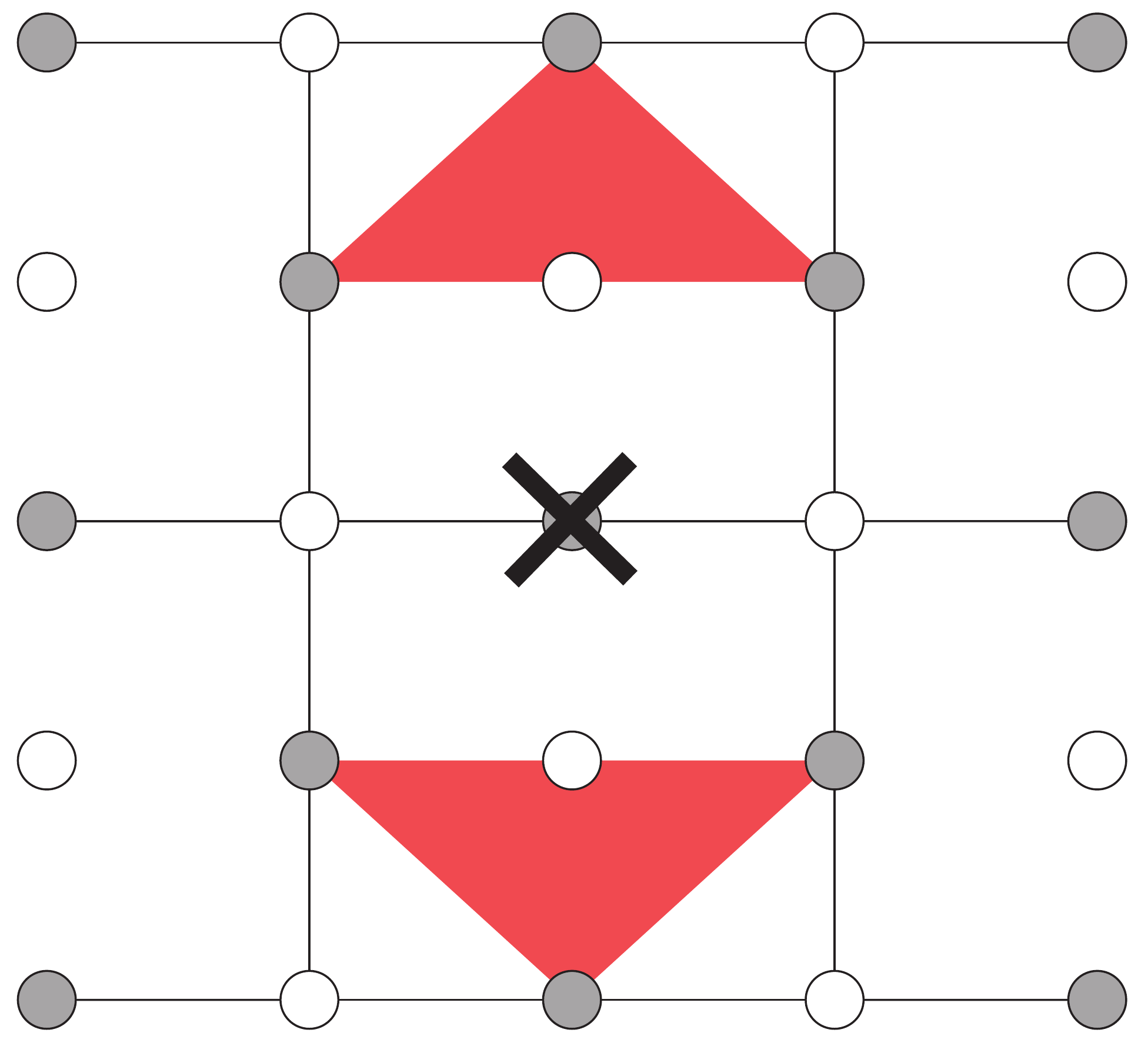}
(b)
  \includegraphics[width=5cm]{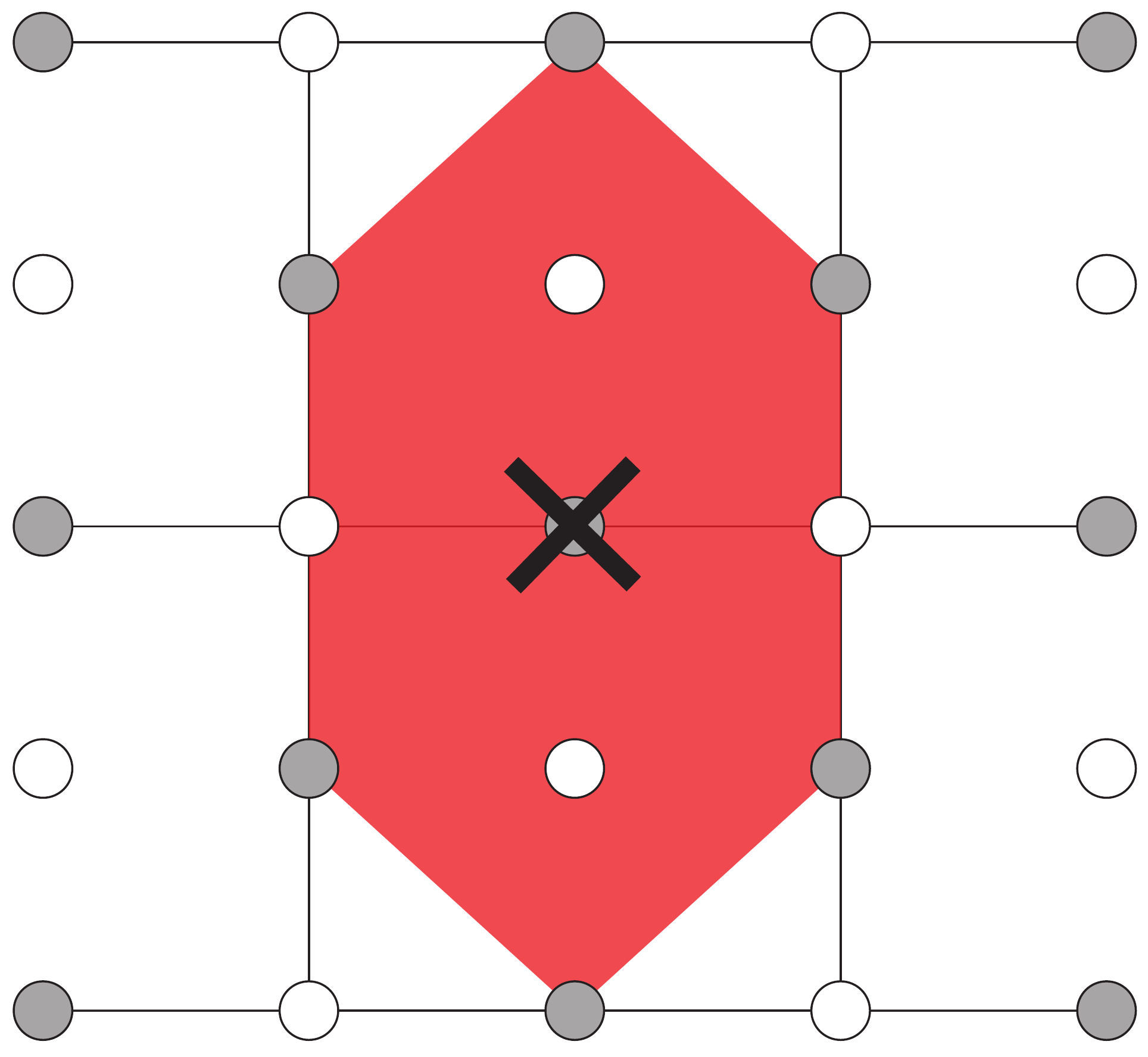}
  \caption{Modified stabilizers around a faulty device marked with the black cross. (a) A pair of Z triangular stabilizers. (b) A superunit Z stabilizer.}
  \label{fig:sol:modified_unit}
 \end{center}
\end{figure}

There are two ways to reconfigure around a faulty device.
The first is to introduce two triangular stabilizers by purging the broken qubit from stabilizers that involve it, leaving two stabilizers composed of three data qubits and one syndrome qubit, as depicted in Figure \ref{fig:sol:modified_unit}(a). 
\begin{figure}[t]
 \begin{center}
  (a)
  \includegraphics[width=5cm]{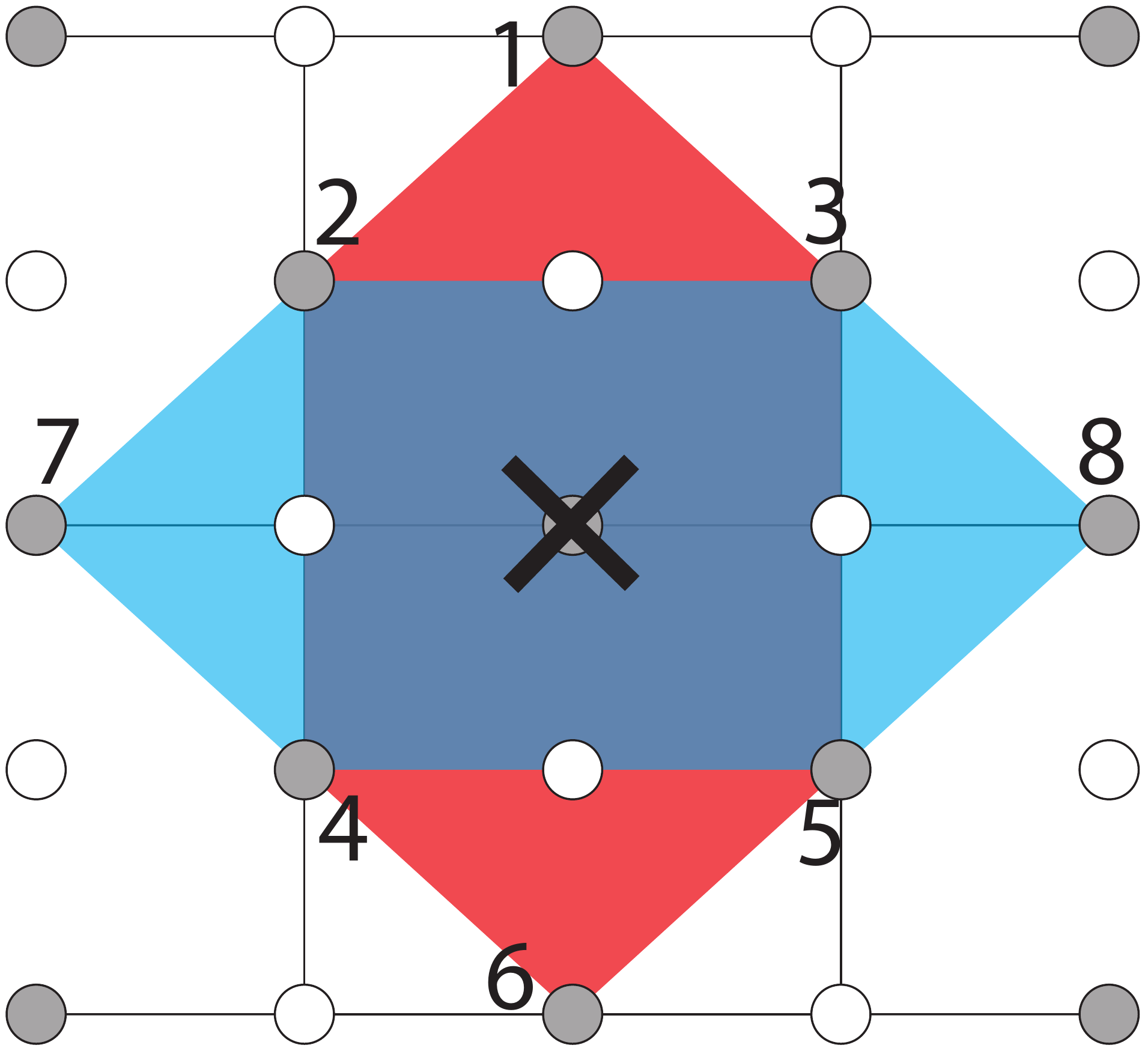}
  (b)
  \includegraphics[width=5cm]{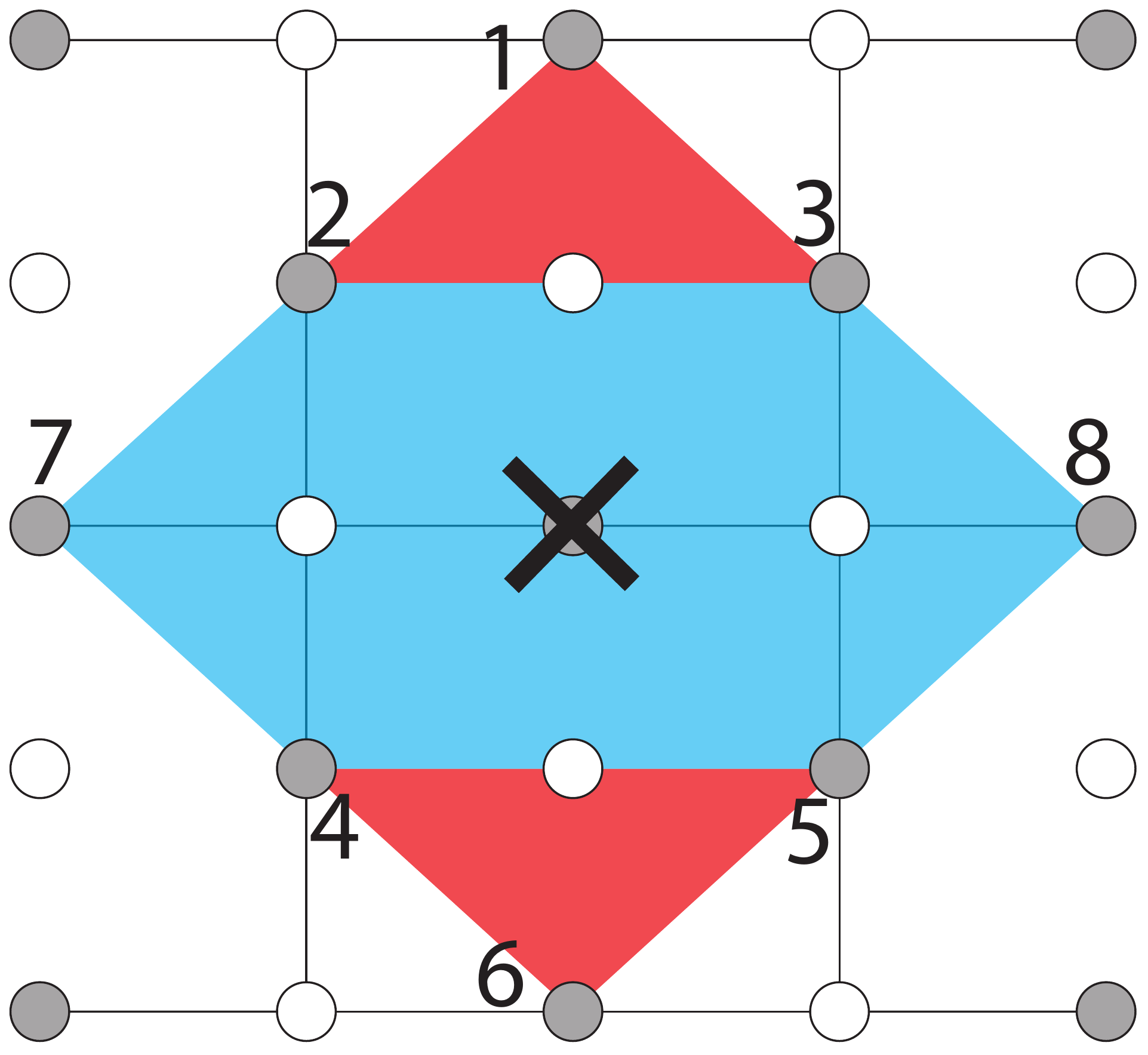}
  \caption{Two sets of modified stabilizers that commute. The corresponding stabilizers are shown in Table \ref{tab:sol:commute-stabilizers}. (a) Superunit stabilizer is adopted both for the Z stabilizer and for the X stabilizer. (b) Triangular stabilizer is adopted for Z stabilizer and superunit stabilizer is adopted for the X stabilizer. }
  \label{fig:sol:commute}
 \end{center}
\end{figure}
\begin{figure}[t]
 \begin{center}
  \includegraphics[width=5cm]{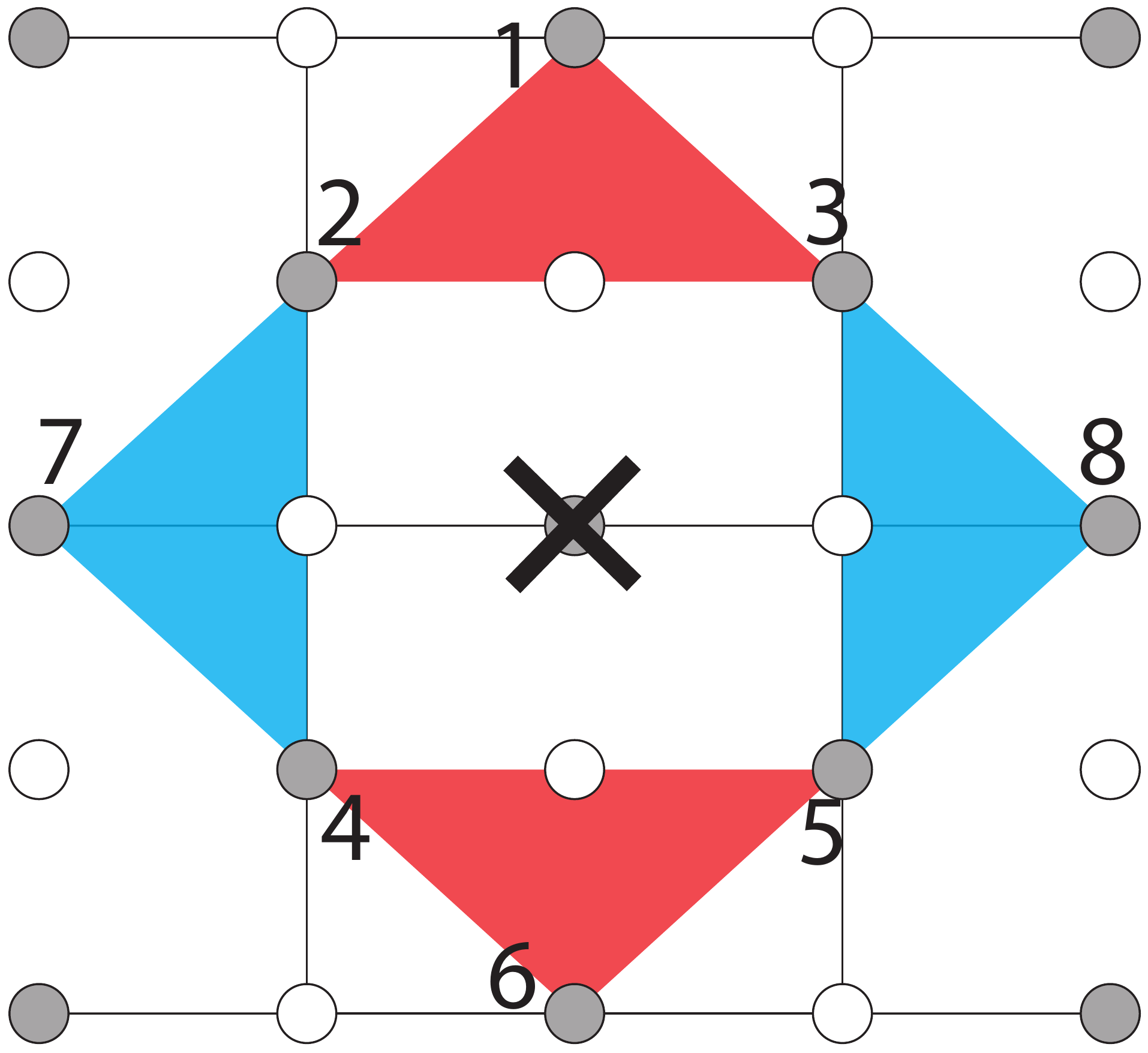}
  \caption{A set of modified stabilizers which anti-commute. The corresponding stabilizers are shown in Table \ref{tab:sol:anticommute-stabilizers}.}
  \label{fig:sol:anticommute}
 \end{center}
\end{figure}
\begin{table}[t]
  \begin{center}
   \caption{Stabilizers of two sets of modified unit stabilizers.
   (a) Superunit stabilizers are adopted both to the Z stabilizer and to the X stabilizer (Figure~\ref{fig:sol:commute}(a)).
   (b) A triangular stabilizers are adopted to the Z stabilizer and a superunit stabilizer is adopted to the X stabilizer (Figure~\ref{fig:sol:commute}(b)). }
   \label{tab:sol:commute-stabilizers}
   (a)
   \begin{tabular}[t]{cccccccc}
    1 &2 &3 &4 &5 &6 &7 &8 \\
    \hline
    Z &Z &Z &Z &Z &Z & & \\
    &X &X &X &X & &X &X \\
   \end{tabular}
   (b)
   \begin{tabular}[t]{cccccccc}
    1 &2 &3 &4 &5 &6 &7 &8 \\
    \hline
    Z &Z &Z & & & & & \\
    & & &Z &Z &Z & & \\
    &X &X &X &X & &X &X \\
   \end{tabular}
  \end{center}
\end{table}
\begin{table}[t]
  \begin{center}
   \caption{Stabilizers of four triangular unit stabilizers as in Figure \ref{fig:sol:anticommute}. Some pairs anti-commute.}
   \label{tab:sol:anticommute-stabilizers}
   \begin{tabular}[t]{cccccccc}
    1 &2 &3 &4 &5 &6 &7 &8 \\
    \hline
    Z &Z &Z & & & & & \\
    & & &Z &Z &Z & & \\
    &X & &X & & &X & \\
    & &X & &X & & &X \\
   \end{tabular}
  \end{center}
\end{table}
It is impossible to adopt triangular stabilizers for both stabilizers around a faulty device since neighboring Z triangular stabilizers and X triangular stabilizers do not commute when they have only one qubit in common,
as shown in Figure~\ref{fig:sol:anticommute}.
Note that those four triangles cannot be stabilizers, but can be gauge operators
for the subsystem code~\cite{bacon05:_operator-self-qec,PhysRevLett.95.230504}.

The second approach is to generate a superunit stabilizer by merging the two broken unit stabilizers, depicted in Figure \ref{fig:sol:modified_unit}(b).
At least one lattice unit must adopt a superunit stabilizer.
In this paper, we form superunit stabilizers for both stabilizers after Stace et al. and
Barrett et al.~\cite{stace:200501,Barrett:PhysRevLett.105.200502}.
Forming superunit stabilizers for both stabilizers around a faulty device produces a degree of freedom which results in a logical qubit by code deformation~\cite{1751-8121-42-9-095302}.
However, this logical qubit does not affect planar code qubits and defect-based qubits
on the same lattice without dedicated operations,
so that we can ignore its presence for our purposes in this paper.

\subsection{Stabilizer circuits around faulty devices}
\label{subsec:stab}
\begin{figure}[t]
 \begin{center}
  (a)
  \includegraphics[width=6cm]{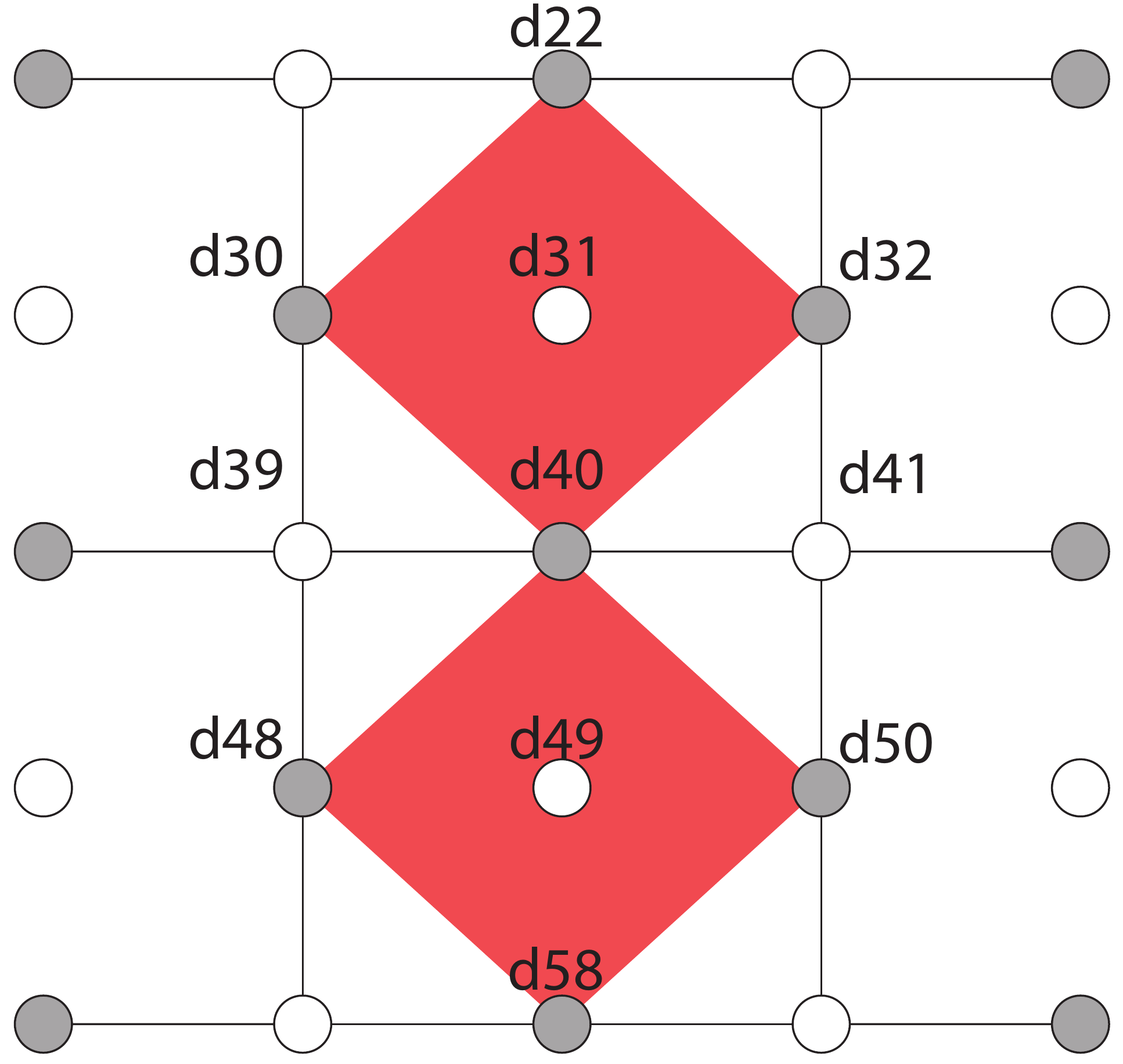}
  \includegraphics[height=5cm]{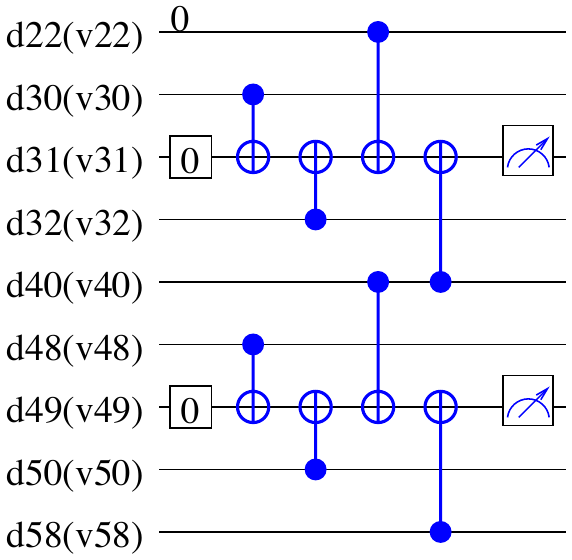}\\
  (b)
  \includegraphics[width=6cm]{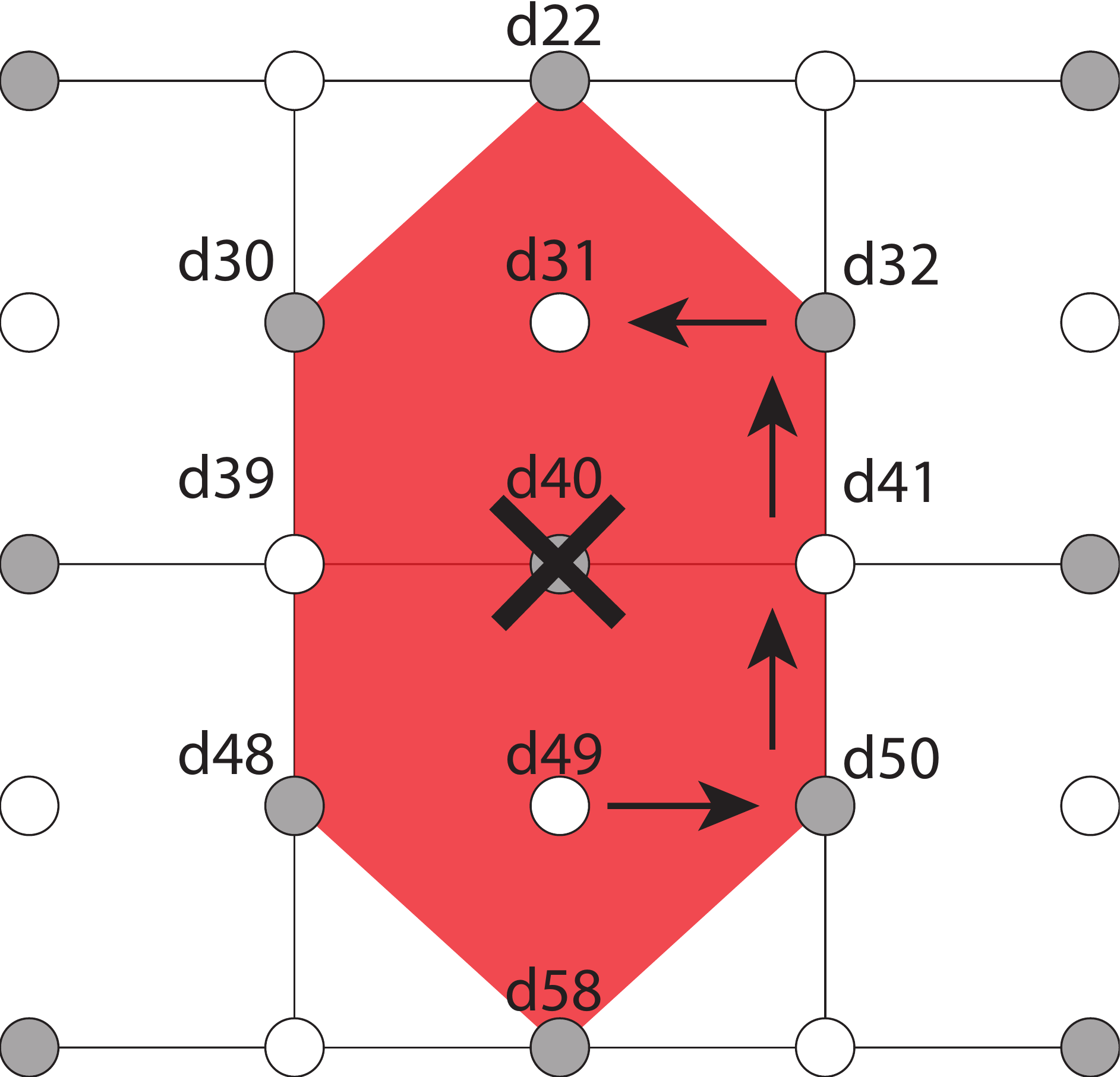}
  \includegraphics[height=5cm]{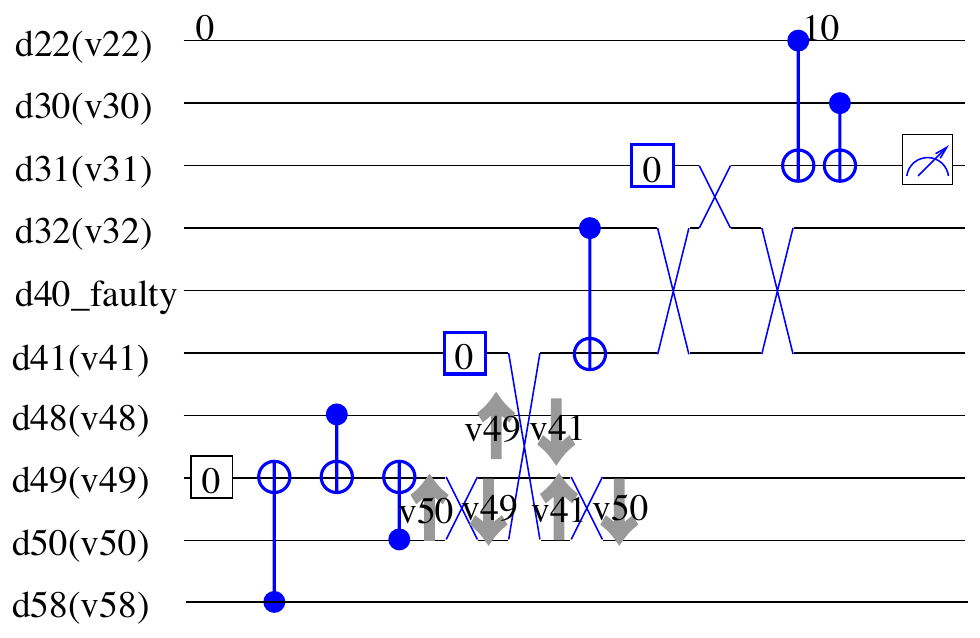}
  \caption{
  Stabilizers and their circuits.
  (a) A set of normal $Z$ stabilizer circuits around d40 for the case where d40 is properly functional. 
  The stabilizers are $Z_{v58}Z_{v48}Z_{v50}Z_{v40}$ and $Z_{v40}Z_{v32}Z_{v22}Z_{v30}$.
  (b) An example of a superunit stabilizer circuit.
  The qubit device d40 is faulty and two units are connected.
  The new stabilizer is $Z_{v58}Z_{v48}Z_{v50}Z_{v32}Z_{v22}Z_{v30}$.
  This stabilizer circuit is isolated from the whole circuit shown in Figure \ref{fig:sol:whole_circuit_single_faulty_center}.
  }
  \label{fig:sol:a_2-units_superunit}
 \end{center}
\end{figure}

Stabilizer-measurement circuits working around faulty devices have different shape and depth from the circuits of normal stabilizers.
Figure \ref{fig:sol:a_2-units_superunit}(b) shows the shape of a superunit in which two units are connected by a faulty device and its circuit.
We call a circuit for an individual stabilizer a ``stabilizer circuit'' and the circuit for a complete lattice the ``whole circuit''.
We define two terms, ``qubit device'' and ``qubit variable''. A qubit device is the physical structure that holds the qubit variable, such as the semiconductor quantum dot or loop of superconducting wire. A qubit variable is the information encoded on a qubit device.
This distinction corresponds to the difference between a register or memory location in a classical computer, and the program variable held in that location.
In Figure \ref{fig:sol:a_2-units_superunit}(a), the horizontal lines correspond to qubit devices, distinguished with the ``d'' labels (numbers).
The ``v'' labels of qubit variables
share the same number as
the label of the qubit device of the qubit variable's original position.

In Figure~\ref{fig:sol:a_2-units_superunit}(b), the qubit device labeled d40 is faulty,
hence the variable v40 does not exist.
The data qubit variables v58, v48, v50, v32, v22 and v30, initially held respectively in
the qubit devices d58, d48, d50, d32, d22 and d30, are stabilized by the red stabilizer.
The syndrome qubit variable v49 is initialized while residing in d49,
then moves around using SWAP gates to gather error syndromes of those data qubits.
After gathering three error syndromes from v58, v48 and v50, v49 moves into d41 via d50.
The data qubit variable v50 is moved onto d49 by the first SWAP gate between d49 and d50.
After moving v49 from d50 to d41, the data qubit v50 on d49 is moved back to d50 by the second SWAP gate.
v41, now in d49, is disentangled from other qubits, hence we can initialize d49 any time.
v49 eventually moves to d31, finishes gathering all error syndromes and gets measured.
Figure \ref{fig:sol:a_2-units_superunit} (b) summarizes the move of v49 from d49 to d31.

In Figure \ref{fig:sol:a_2-units_superunit}, we see that the superunit circuit is deeper than the normal unit stabilizer circuit.
In general, superunit stabilizers require more steps to gather error syndromes than normal stabilizers.
Obviously, the deeper stabilizer will have more
opportunities to accumulate physical errors.
Thus, an important engineering goal is to create stabilizer circuits that are as shallow as possible.

We present a basic algorithm for composing a stabilizer circuit, shown in Algorithm~\ref{alg:sol:stabilizer_circuit} in \ref{subsec:app:alg}.
A syndrome qubit variable travels one way to gather error syndromes.
In this algorithm, we search for the shortest traversable path in which error syndromes can be gathered from all data qubits.

\subsection{Building a whole circuit from stabilizer circuits}
\label{subsec:whole}
On a perfect lattice, the stabilizer circuits are highly synchronous and easily scheduled efficiently.
The circuits for a defective lattice must be asynchronous on account of the different depth of stabilizers.
Such asynchronicity introduces a problem when several stabilizers try to access a qubit at the same time. We have to assign priorities to stabilizers.
Stabilizers with lower priority have to wait, so that they have more opportunities to accumulate physical
errors on data qubits and
ancilla qubits.
Therefore we give higher priority to stabilizers which have deeper stabilizer circuits to deter error opportunities from concentrating there,
since a shorter error chain is obviously preferred for error correction.
The scheduling algorithm, shown in Algorithm~\ref{alg:sol:scheduling_algorithm} in \ref{subsec:app:alg}, is
\begin{enumerate}
 \item Sort stabilizers in order of depth, longest first. If they tie, stabilizers in the upper left of the lattice have priority. (lines 1-2)
 \item The deepest stabilizer is scheduled. The step when the deepest stabilizer finishes is saved (currentCeil). (line 9)
 \item Each non-deepest stabilizer is scheduled once, in order of decreasing depth. (lines 10-13)
 \item Each non-deepest stabilizer which does not exceed the currentCeil is scheduled once again, in order of depth. 
	 Short ones may be scheduled twice or more before the loop terminates. (lines 14-21)
 \item If all of the non-deepest stabilizers exceed the currentCeil, return to step (ii). Otherwise, return to step (iv). (lines 21-22)
\end{enumerate}
Our algorithm must enforce an important restriction,
different types of stabilizers which share an even number of data qubits must
access those qubits in the same order.
For example, if we have two stabilizers $X_1X_2$ and $Z_1Z_2$ on qubits $1$ and $2$,
we have to execute them as $X_1X_2$ then $Z_1Z_2$ (or reverse order).
$X_1Z_2$ then $Z_1X_2$ is not allowed because it will entangle syndrome qubits.
We postpone stabilizers of low priority to resolve conflicts by simply adding identity gates.
See ~\ref{sec:appendix:conflict} for details.
Optimization around this strategy is left for later research.

\subsection{Adapting matching to asynchronous operation}
Irregular stabilizer circuits degrade the parallelism of stabilizer measurements of
the whole circuit, so that the surface code on a defective lattice has irregular error matching nests.
A superunit stabilizer is measured in a longer cycle than normal stabilizers and
a vertex corresponding to a superunit stabilizer has many edges.
The network which Blossom V is fed is generated on this adapted nest to achieve a more correct solution of the matching problem.

\section{Numerical Results}
In our simulations,
\begin{itemize}
 \item static loss placement is accurately determined by scanning before quantum computation;
 \item static loss does not occur after fabrication; and
 \item stabilizer circuits are created before quantum computation.
\end{itemize}

We assume a circuit-based error occurrence model,
including imperfect syndrome extraction, summarized by Landahl et al. \cite{landahl:arXiv:1108.5738}.
This circuit-based error model assumes that each gate acts ideally, followed by a noisy channel, and that each measurement acts ideally, after a noisy channel.
Errors may occur at every gate in the circuit.
Our error channel for a single-qubit gate has error probability $p$, meaning that each error (X, Z or Y) occurs with probability $p$/3.
In a similar fashion, for two-qubit gates, our error model has probability $p$/15 for each two-qubit error (IX, IZ, IY, XI, XX, XZ, XY, ZI, ZX, ZZ, ZY, YI, YX, YZ, YY).
We assume that the set of physical gates available includes CNOT, SWAP and Hadamard gates.
We assume that INIT and measurement in Z basis have X error probability $p$.
All operations require one time step.


Our circuit is asynchronous in the sense that stabilizers are measured at different frequencies.
Stabilizers whose circuits have shallower depth may be measured more times than those whose circuits have deeper depth.
To achieve proper syndrome matching,
the surface code requires that the lattice be covered by stabilizers.
Otherwise, an unstabilized area works as a defect-based qubit which may serve as an end of error chains, leading to undetectable logical errors.
Hence, after all stabilizers covering the lattice
have been measured at least once since the last execution of the matching algorithm, the matching algorithm is re-executed.
Typically, this timing is dependent on the deepest stabilizer circuit.
From the result of matching, we make a map of Pauli frames which describes where Pauli frame corrections should be applied for error correction.
Because our circuit is asynchronous and there might be SWAP gates, we must keep track of the location of qubit variables to combine the error information about data qubits and the map to check the result of error correction.

We have conducted extensive simulations, beginning with a perfect lattice, then extending to imperfect ones.
First, we show the numerical results of several basic test cases:
only a single faulty device exists, in the center of the lattice;
only a single faulty device exists, in the west of the lattice; and
only a single faulty device exists, in the northwest of the lattice,
all for the distances 5, 7, 9 and 13.
Our simulation holds $d$ temporal rounds of measured stabilizer values for error correction.
Hence $d$ measurements are saved for the stabilizer with the deepest circuit and
more measurements are saved for normal stabilizers, because of the scheduling algorithm shown in Subsection~\ref{subsec:whole}.
After finishing an error correction cycle, the oldest round is discarded, a new round is created by new measurements and error correction is re-executed.
Next, we show the numerical results for randomly generated lattices for three different yields, 80\%, 90\% and 95\%.
We generated 30 lattices for each pair of yield and code distance of 5, 7, 9, 13, 17 and 21.
Some defective lattices cannot encode a logical qubit for the code distance becomes 0 as a result of merging stabilizers, so that ultimately
we simulated 474 randomly generated lattices (details described in subsection~\ref{subsec:num_res_random})
for physical error rates of $0.1\%$, $0.2\%$, $0.3\%$, $0.4\%$, $0.5\%$, $0.6\%$, $0.7\%$, $0.8\%$, $0.9\%$, $1\%$ and $2\%$.
It is hard to collect enough logical errors in Monte Carlo simulation
as the logical error rate is exponentially suppressed,
therefore we choose $0.1\%$ as the lowest physical error rate
for our simulation.
Therefore we simulated 5214 parameter combinations.

The computational resource devoted to circuit simulation, excluding chip generations and circuit constructions, was more than 100,000 CPU days, executed on the 
StarBED project testbed~\cite{Miyachi:2006:SSL:1190095.1190133}.
Each preparation of stabilizer circuits which solves the traveling salesman problem
required up to 1 CPU day.
After construction of the nest, for example, the simulation of $d=5$ of single-faulty-northwest for $p=10^{-3}$
consisted of 370945 rounds of error correction to find 500 logical $X$ errors in 1424.98 seconds.
The simulation of $d=13$ of single-faulty-northwest for $p=10^{-3}$
consisted of 315550 rounds of error correction in 5.8 days but found 0 logical $X$ errors.

Peak memory sizes are estimated to be 30GB for 316 lattices,
63GB for 133 lattices and
more than 100GB for 25 lattices.
The greatest memory consumption is during nest building, shown in Figure B5, B6 and B7.
To give accurate weights to the edges of the ``nests'', Autotune virtually creates errors on every qubit at every physical step, and traces their propagation.
Roughly speaking, the size of the error structure is 136 bytes.
A lattice includes 1089 qubits for distance 17.
Let us assume: 200 physical steps per error correction cycle due to the asynchronous stabilizers;
each error propagates to 10 physical qubits on average;
each error remains for 100 physical steps on average.
Then memory consumption is
$136 * 1089 * 200 * 10 * 100 = 29620800000$ bytes, roughly 30 GB.
Several factors affect this rough estimate.
Faulty devices reduce the number of qubits
and
other structures generated to create the nests,
but the memory consumption remains on the order of tens giga bytes.

Those peak memory sizes are big, however,
they do not affect the quantum computation in practice.
This is because the heavy operations that
Autotune virtually creates errors on every qubit at every physical step,
and traces their propagation by the circuits
to give accurate weights to the edges of the nests
can be executed preliminarily.
To avoid redundant execution of heavy creation of nests,
all 11 physical gate error rates for a single lattice are simulated in parallel on a single simulation node, allowing us to share a single in-memory copy of the nest.
%
We attempted to simulate distance 21, but failed because
we cannot accumulate enough logical errors
to have valid data points, for one of several reasons:
good lattices have strong tolerance against errors;
even bad lattices have strong error tolerance at lower physical error rate;
at higher physical error rates, simulating an error correction cycle takes too much computation time because
many physical errors occur in our extended asynchronous error correction cycle,
taxing the scalability of the matching algorithm;
or because the simulation requires more than 128GB memory, the maximum available in our system.

\subsection{perfect lattice}
\begin{figure}[t]
 \begin{center}
  \includegraphics[width=13cm]{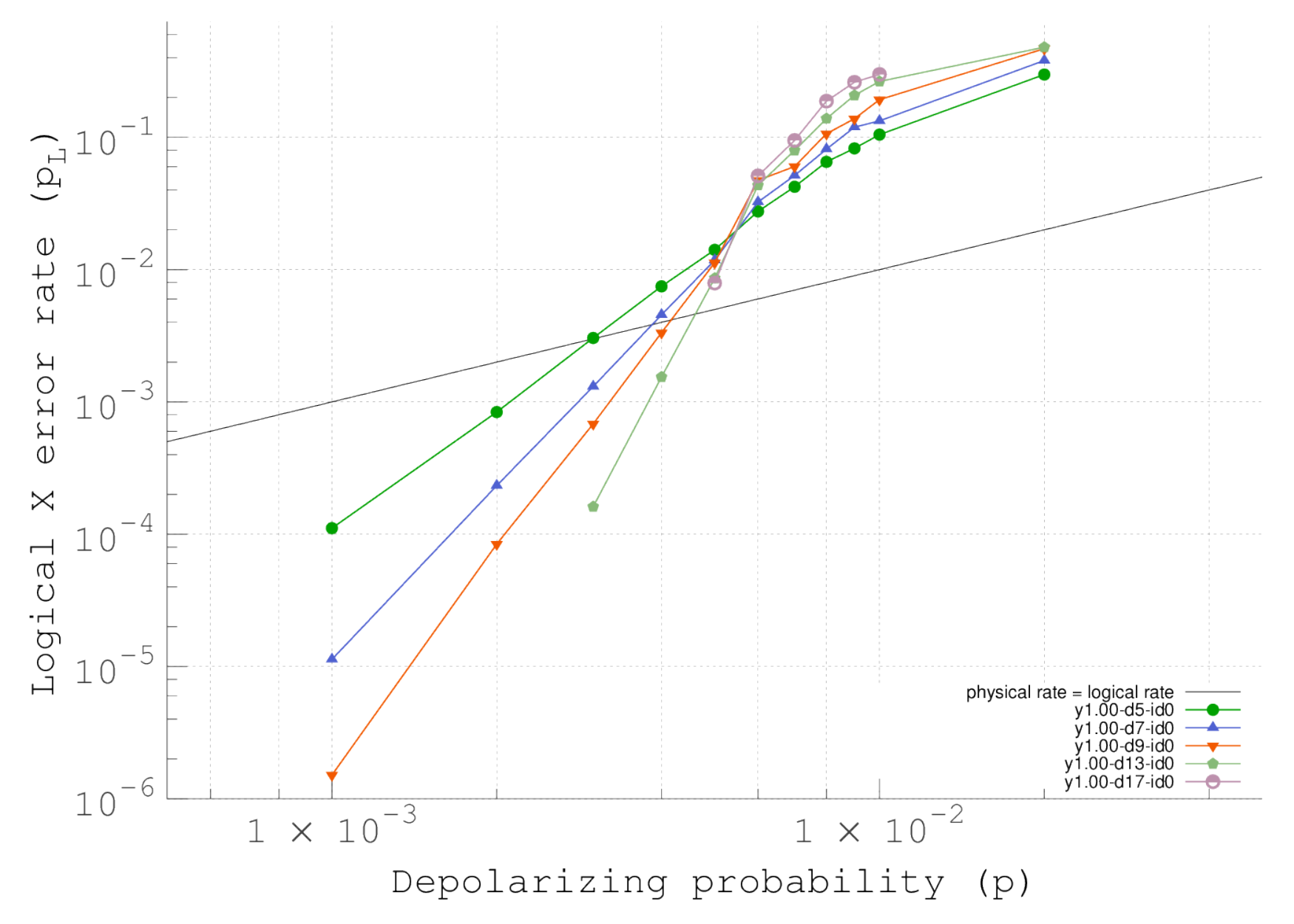}
  \caption{Results of baseline simulations of perfect lattices of code distance 5, 7, 9, 13 and 17.
  The average number of steps per error correction cycle for every code distance is
  $8.1$, $8.0$, $8.0$, $8.0$ and $8.0$ respectively.
  The black line is the break-even line.
  The threshold seems to be around $0.58\%$.
  Each data point has $50\sim 1500$ logical errors.
  The irregularity of the point at $p=0.6\%$ of distance 9 may come from statistical variance.
  }
  \label{fig:graph_perfect_lattice}
 \end{center}
\end{figure}

Figure \ref{fig:graph_perfect_lattice} depicts the results of simulation of perfect lattices, used as our baseline for comparison.
Each curve represents a set of simulations for a lattice of a particular code distance, for varying physical gate error rates.
Points below the break-even line are conditions in which the logical error rate
in the logical state is below that of a bare, unencoded physical qubit for a single physical gate time. 
Distance 9 achieves break-even at $p=0.3\%$.
The crossing point of the curves, each of which describes a code distance, is called the \emph{threshold}, the physical error rate below which
the larger code distance has the lower logical error rate.
Above the threshold, the error correction process introduces more errors than it corrects,
and the higher code distance has the higher logical error rate.

The threshold indicated by this simulation is around $0.58\%$, similar to the $0.60\%$ reported in
~\cite{Fowler:2009High-threshold_universal_quantum_computation_on_the_surface_code}.
This related work employs the assumptions most similar to our perfect lattice simulation,
other than the asynchronous scheduling of stabilizers.
Our error correction circuits are designed to omit identity gates to shrink the asynchronous circuit
depth, whereas circuits of related work achieve perfect synchronization and parallelism through careful insertion of identity gates.
For example, identity gates on the qubit d17 in Fig \ref{fig:sc:stabilizers} (b) between the initialization and the CNOT gates or between the CNOT gates and the measurement are omitted in our simulation.
We infer that our baseline simulation follows the related work, our baseline simulation is valid and the effect of asynchronicity to the perfect lattice is small.

\subsection{lattice with a single faulty device}
\begin{figure}[t]
 \begin{center}
  (a)
  \includegraphics[width=10cm]{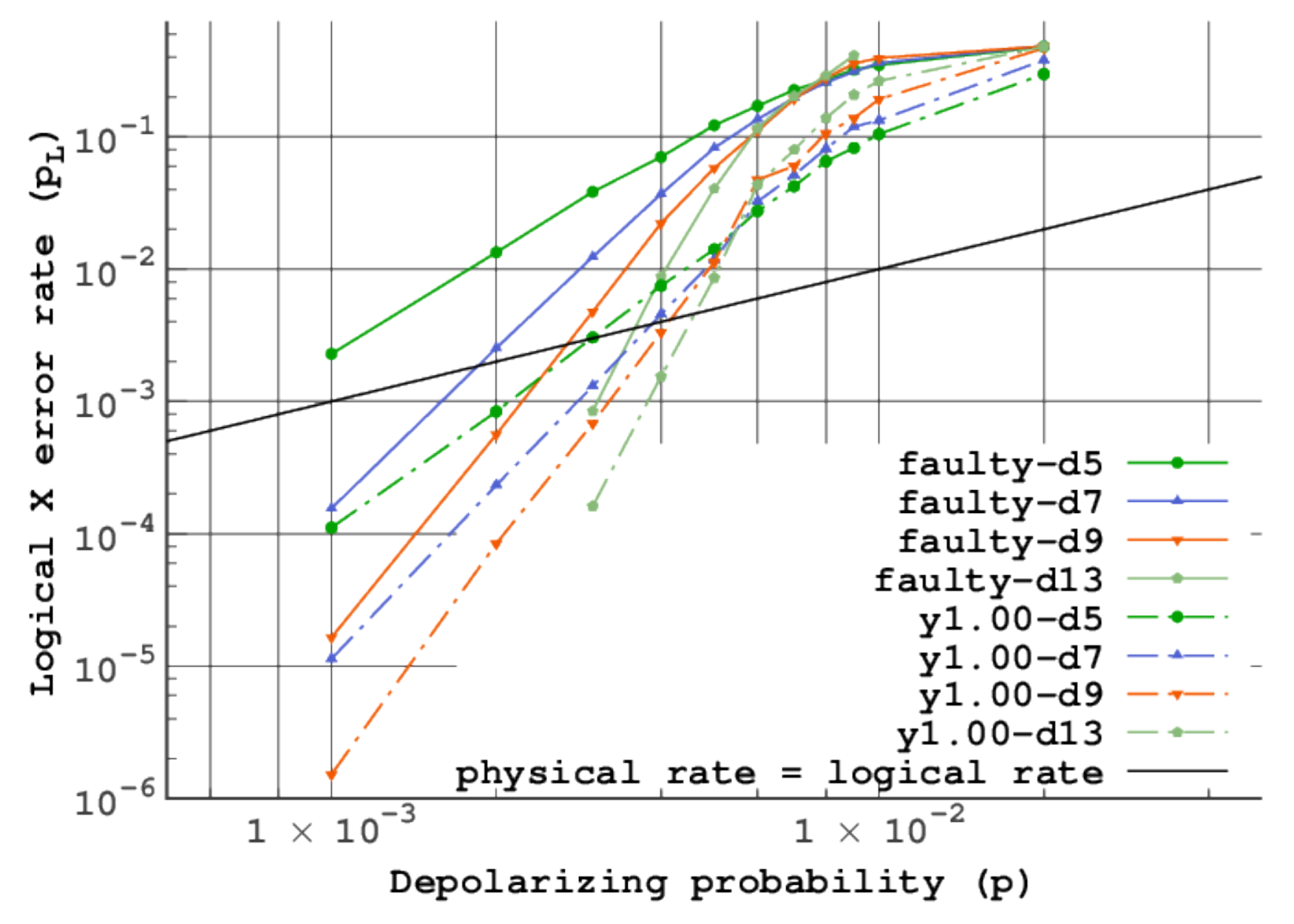}\\
  (b)
  \includegraphics[width=10cm]{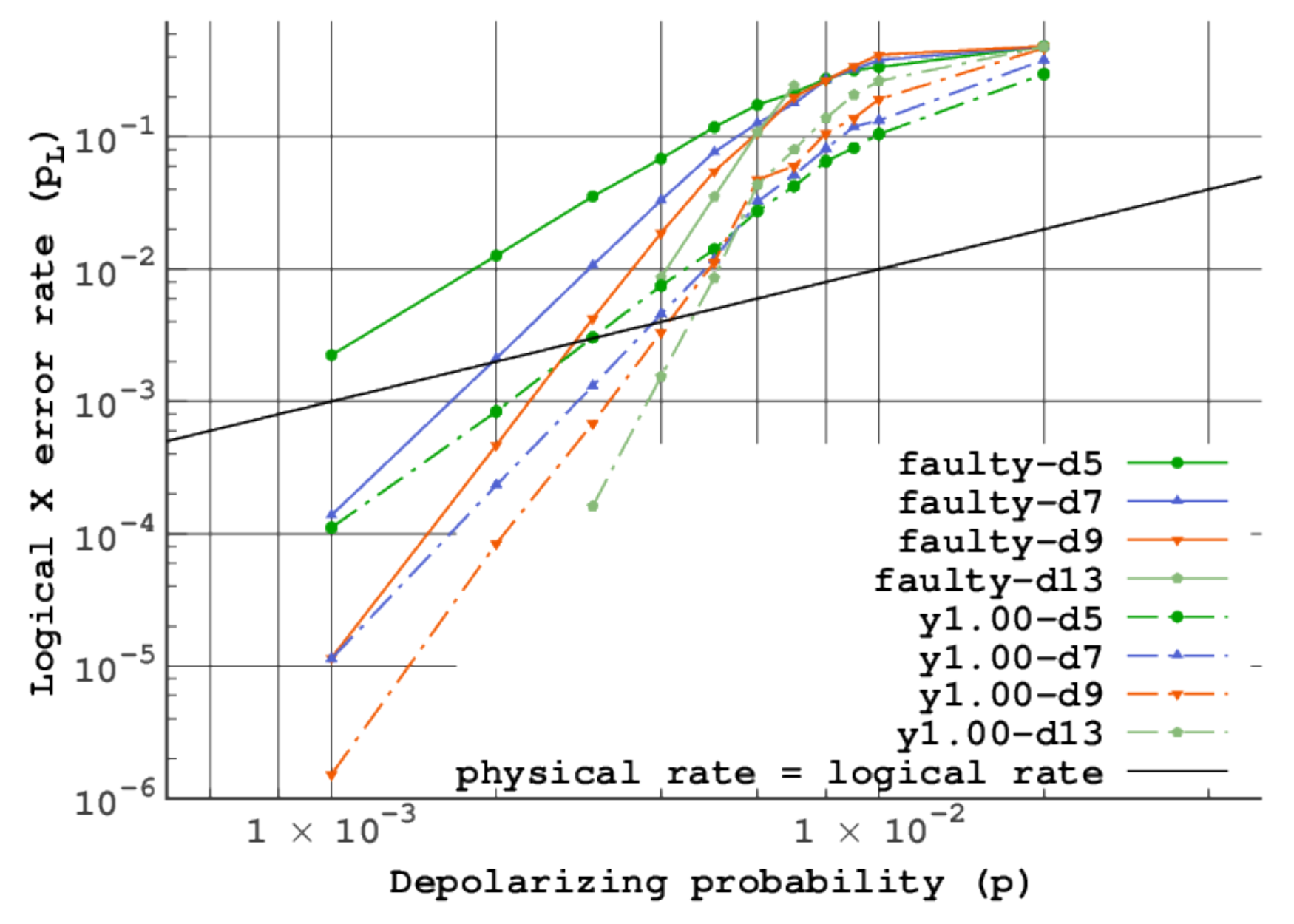}\\
  (c)
  \includegraphics[width=10cm]{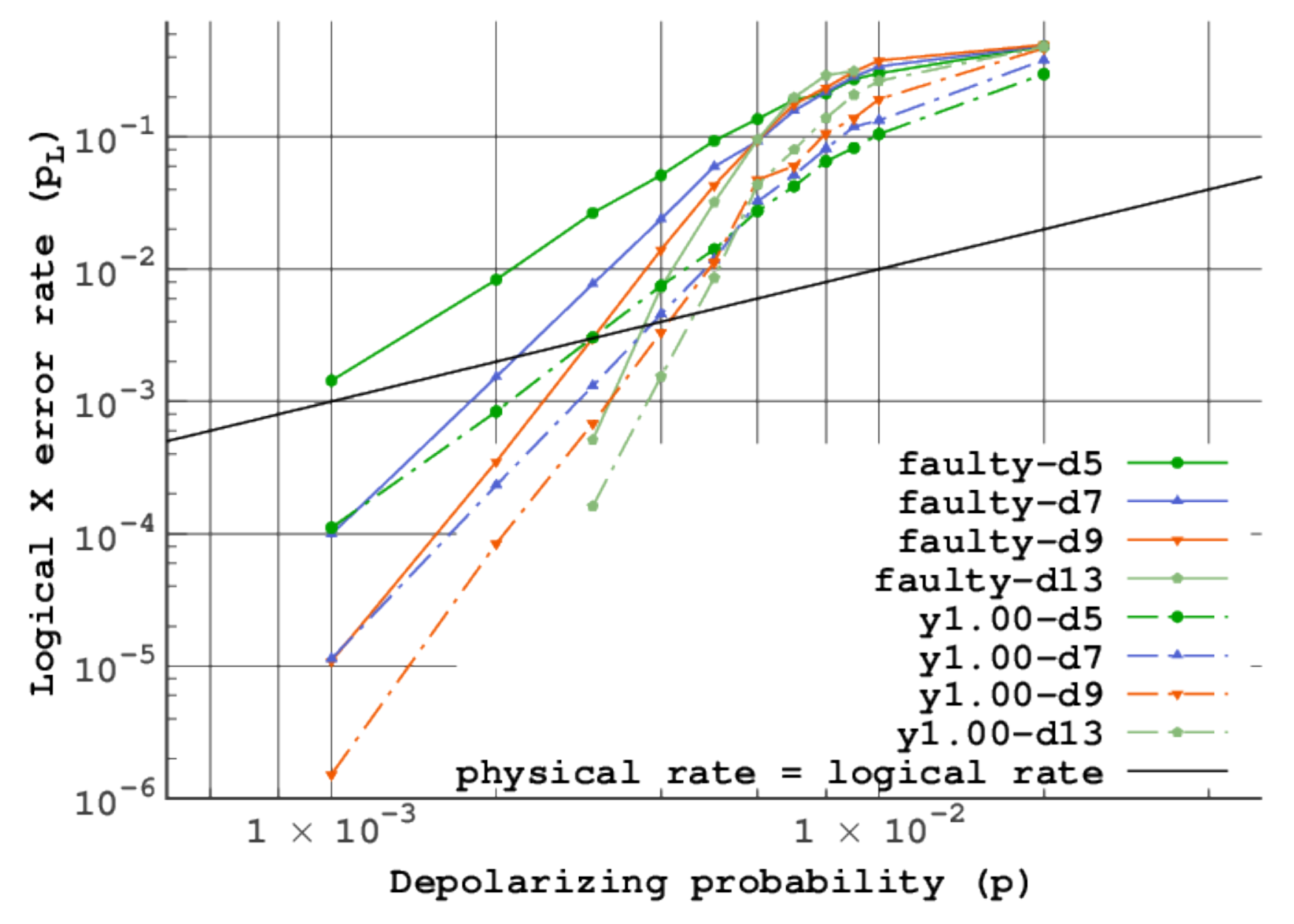}\\
  \caption{
  Results of simulations of defective lattices that have a single faulty device
  (a) in the center of the lattice,
  (b) in the west of the lattice
  and
  (c) in the northwest of the lattice respectively.
  Dashed lines are of the perfect lattices for reference.
  The code distances are 5, 7, 9 and 13.
  The average number of steps per error correction cycle is $32.5$ for every code distance and fault location.
  }
  \label{fig:graph_single_faulty}
 \end{center}
\end{figure}

Figure \ref{fig:graph_single_faulty} (a), (b) and (c)
depict the results of simulations
to investigate the effect of a single faulty device in the center, on the west edge
and on the northwest corner of the lattice, respectively.
The plots show that our approach works properly
because the larger code distance has the lower logical error rate at lower physical error rates.

Each single-fault residual error rate is worse than that of the corresponding perfect lattice.
The slope of each code distance of single-fault chips is lower than that of the corresponding perfect lattice.
The gap grows slightly as the physical error rate is reduced, visible as the less-steep curve for the defective lattice.

There are differences depending on the single-fault location.
Comparing the points $d=9$ of the perfect lattice with those of single-faulty-center, single-faulty-west and single-faulty-northwest at $p=0.1\%$,
faulty lattices are $10.9\times$, $7.60\times$ and $7.20\times$ worse than the perfect lattice, respectively.
Single-faulty-northwest has a lower residual error rate than the others.
This may be because the big stabilizer that causes asynchronous scheduling of stabilizers is
on the periphery, so that the number of stabilizers that are close to the big stabilizer and hence
which have stronger scheduling restrictions than more remote stabilizers is smaller than other single-faulty chips.
Across the range of our simulations, the negative impact is $6\times\sim 11\times$ depending on location, distance and error rate.

From the point of view of absolute logical error rate,
the penalty for having a defect is greater at lower physical error rates.
An ``effective'' code distance is the code distance at the same physical error rate of the perfect lattice which has the closest logical error rate to the defective lattices. 
For single-faulty-center,
at $p=0.3\%$, faulty $d=9$ is $1.5\times$ worse than perfect $d=5$, hence the effective code distance of faulty $d=9$ at $p=0.3\%$ is $\approx 5$.
The effective code distance is useful when considering the resource overhead of modifications.
In the example above, to achieve a logical error rate 
equivalent to that of $d=5$ on the perfect lattice at $p=0.3\%$,
we at least need $d=9$ for the defective lattice. This indicates that $3.5 \times$ the number of physical qubits are required.

From the point of view of the effective code distance,
the penalty for having a defect is smaller at lower physical error rates.
At $p=0.3\%$, faulty $d=9$ is $1.5\times$ worse than perfect $d=5$, and
at $p=0.1\%$, faulty $d=9$ is $14.9\times$ better than perfect $d=5$ while
faulty $d=7$ is $1.4\times$ worse than perfect $d=5$.
Hence, to exceed the effective code distance 5, $p=0.3\%$ requires us to use $d=11$
while $p=0.1\%$ only requires us to use $d=9$.
The trend of the penalty of the effective code distance and that of the absolute logical error rate differ.
This difference is caused by the difference of slopes of each code distance of each lattice.
We have to be mindful of those trends when designing a quantum computer
to achieve an adequate logical error rate.
%

Because the proportional impact of a single fault should lessen as the code distance increases,
the crossing point of the curves is not a good measure of performance here.
Figure \ref{fig:graph_single_faulty} shows that the crossing points of two distances would differ.
The crossing point of distance 9 and 13 appears to be around $0.6\%$ which is
the threshold for the perfect lattice as shown in Figure \ref{fig:graph_perfect_lattice},
whereas the crossing point of distance 5 and 7 is around $0.8\%$.

Table \ref{tab:num_static_result} shows the data of the single-faulty lattice simulations.
The reduced code distance is the minimum distance between corresponding boundaries shortened by merging stabilizers.
The naive hypothesis would be that reduced code distance is a good metric to predict the logical error rate of the lattice,
since the number of physical errors required to cause a logical error is a minimum on the shortest logical operator,
which is the minimum distance between corresponding boundaries.
However, the effect is more complex. We will explore this further in Section~\ref{subsec:metrics} and ~\ref{sec:discussion}.
 \begin{landscape}
 \begin{table}[b]
  \caption{The $X$ error rate of single-faulty lattices and corresponding $Z$ stabilizer data.
  Averages here are arithmetic means.
  ``Faulty location'' is the location of the faulty device (static loss).
  ``\#X and Z stabs'' stands for the total number of X stabilizers and Z stabilizers.
  ``Reduced distance'' is the the minimum distance between corresponding boundaries shortened by merging stabilizers.
  ``\#Z stabs'' is the number of Z stabilizers.
  ``Biggest \#dataq of Z stabs'' is the largest number of data qubits in a Z stabilizer.
  ``Ave. $CDQ$ of Z stabs'' is the average of $CDQ$s (metric is the space-time product of an error correction circuit: the number of data qubits $DQ$ involved, multiplied by the ``cycle'', the sum of the circuit depth and the waiting time for next stabilization $C$, after~\cite{steane02:ft-qec-overhead}) of Z stabilizers.
  ``Biggest Z $CDQ$'' is the largest $CDQ$ for any stabilizer circuit of the chip.
  ``Ave. \#dataq of Z stabs'' is the average of the number of data qubits in Z stabilizers.
  }
  \label{tab:num_static_result}
 \begin{tabular}[t]{c|c|c|c|c|c|c|c|c|c|c|c}
  faulty &code    &\#X and  &residual&reduced &\#Z stabs&biggest&steps per &ave. $CDQ$&biggest&ave. steps &ave. \#dataq\\
location &distance&Z        &X error &distance&         &\#dataq&error     &of Z      &Z $CDQ$&per error  &of Z stabs\\
         &        &stabs    &rate    &        &         &of Z   &correction&stabs     &       &correction & \\
         &        &         &        &        &         &stabs  &cycle     &          &       &of Z stabs & \\
  \hline
center&5&38&7.038E-02&4&19&6&32&35.871&195.194&8.844&3.684\\
west&5&38&6.843E-02&4&19&6&32&35.803&195.194&8.822&3.684\\
  northwest&5&38&3.335E-02&4&19&6&32&35.648&195.194&8.791&3.684\\
  \hline
center&7&82&3.704E-02&6&41&6&32&32.249&195.194&8.149&3.756\\
west&7&82&3.319E-02&6&41&6&32&32.270&195.194&8.153&3.756\\
northwest&7&82&1.969E-02&6&41&6&32&32.505&195.194&8.228&3.756\\
  \hline
center&9&142&2.224E-02&8&71&6&32&31.215&195.194&7.943&3.803\\
west&9&142&1.873E-02&8&71&6&32&31.456&195.194&8.016&3.803\\
northwest&9&142&1.046E-02&8&71&6&32&31.539&195.194&8.033&3.803\\
  \hline
center&13&310&8.776E-03&12&155&6&32&30.711&195.194&7.824&3.858\\
west&13&310&8.662E-03&12&155&6&32&31.028&195.194&7.910&3.858\\
northwest&13&310&4.869E-03&12&155&6&32&31.084&195.194&7.925&3.858\\
 \end{tabular}
 \end{table}
 \end{landscape}
 
 \subsection{Random multiple faulty devices}
 \label{subsec:num_res_random}
We generated 30 randomly defective lattices for each combination of three yields, $80\%$, $90\%$ and $95\%$
and of 5 code distances, 5, 7, 9, 13, 17 and 21 so that
we generated 540 lattices.
66 lattices cannot hold a logical qubit and we were unable to simulate distance 21, hence we ultimately simulated 474 lattices.
Table \ref{tab:num_simulated} shows the number of defective lattices generated and simulated.
On some defective lattices, by chance the faulty qubit placement
results in a lattice for which we are unable to build an effective circuit for 
encoding a logical qubit, so they are not simulated.
Our software successfully built circuits for almost all lattices at $y=0.90$ and above,
but only about two-thirds at $y=0.80$.
  \begin{table}[t]
   \caption{The number of defective lattices generated and simulated.}
   \label{tab:num_simulated}
 \begin{tabular}[t]{c||c|c|c|c|c|c|c|c|c|c|c|c|c|c|c}
  yield         & \multicolumn{5}{|c|}{0.80}& \multicolumn{5}{|c|}{0.90}& \multicolumn{5}{|c}{0.95}\\
  code distance& 5& 7& 9&13&17& 5& 7& 9&13&17& 5& 7&9& 13& 17\\ \hline  \hline
  \#encodable        &     20&     24&     22&     19&     19&     29&     29&     30&     30&     30&     28&     30&     30&     30&     30\\
  \#unencodable      &     10&      6&      8&     11&     11&      1&      1&      0&      0&      0&      2&      0&      0&      0&      0\\
 \end{tabular}
  \end{table}

  \begin{table}[t]
   \caption{The average number of faulty qubits in all generated lattices,
   after culling the weakest 50\% and weakest 90\%,
   respectively.
   The numbers of qubits in a perfect lattice of distance 5, 7, 9, 13 and 17 are 81, 169, 289, 625 and 1089, respectively.
   The correlations between this number and the logical error rate is 0.25 to 0.68, as shown in Tables \ref{tab:linear_correlation} and \ref{tab:exponential_correlation},
   a strong correlation but not as strong as the best metrics in
   Tables \ref{tab:linear_correlation} and \ref{tab:exponential_correlation}.}
   \label{tab:num_faulty}
   \begin{center}
    \begin{tabular}[t]{c|c||c|c|c}
     yield&code distance&all &50\%&90\%\\
     \hline
     \hline
     &5&13.9&13.2&12.3\\
     &7&29.7&28.9&29\\
     0.80&9&54&55.1&55\\
     &13&122.6&122.6&121.3\\
     &17&215.6&214.5&221.7\\
     \hline
     &5&8.4&7.3&5.7\\
     &7&16.2&14.2&9.3\\
     0.90&9&29.6&26.7&23.7\\
     &13&61&56.3&48\\
     &17&107.3&101.7&96\\
     \hline
     &5&3.8&2.9&1.7\\
     &7&8&7.1&4.7\\
     0.95&9&14.7&11.9&9\\
     &13&31.3&28.5&25.3\\
     &17&53.7&51.3&51.3
    \end{tabular}
   \end{center}
  \end{table}
  
  This unencodable condition occurs when a defective data qubit chain stretches from a boundary of the lattice
  to the other boundary of the same type (south and north for Z stabilizer boundary, or west and east for X stabilizer boundary).
  For instance, if a faulty qubit is on a boundary, say, the $qubit 1$ which is stabilized by $Z_1$ of the stabilizer $Z_1Z_2Z_3Z_4$ and
  the qubit is not stabilized by another Z stabilizer, then $Z_1Z_2Z_3Z_4$ cannot be merged with another stabilizer to work around $Z_1$.
  Hence we remove $Z_1Z_2Z_3Z_4$ with $qubit 1$ and eventually $qubit 2$, $qubit 3$ and $qubit 4$ become a part of the boundary instead.
  In general, this adaptation reduces the code distance (shown in Tables \ref{tab:linear_correlation} and \ref{tab:exponential_correlation}).
  Therefore, a lattice of lower yield and of lower code distance
  has a higher probability of being unencodable.
  Though only 30 instances for each condition are too few to collect statistics of useful accuracy, Table \ref{tab:num_simulated} shows
  this trend at yields of 90\% and 95\%.
  $Y=80\%$ might be saturated in
  terms of the percentage of encodable lattices
  because the code distances do not show meaningful differences.

\begin{landscape}
\begin{figure}[t]
 \begin{center}
  \includegraphics[width=7cm]{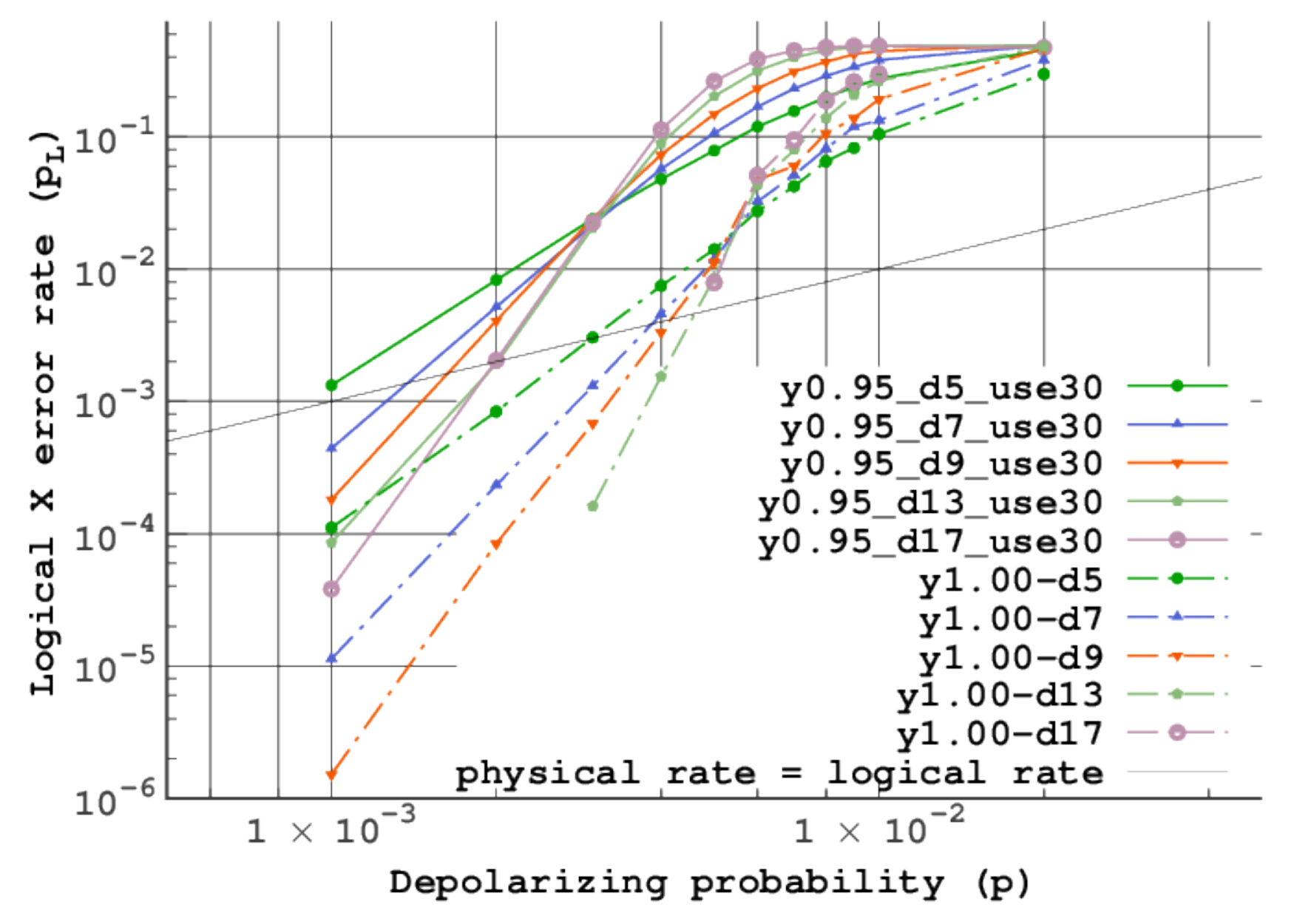}
  \includegraphics[width=7cm]{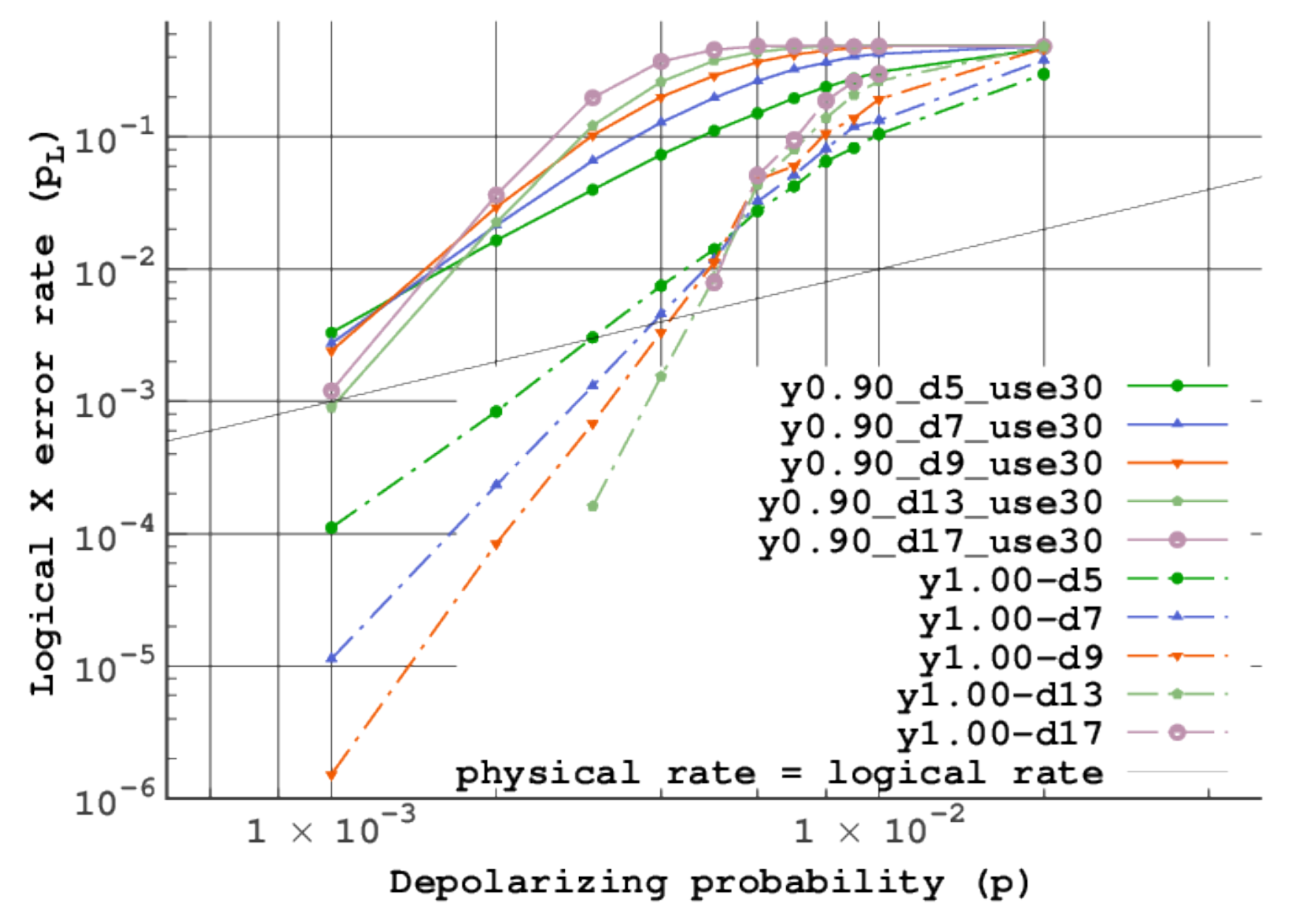}
  \includegraphics[width=7cm]{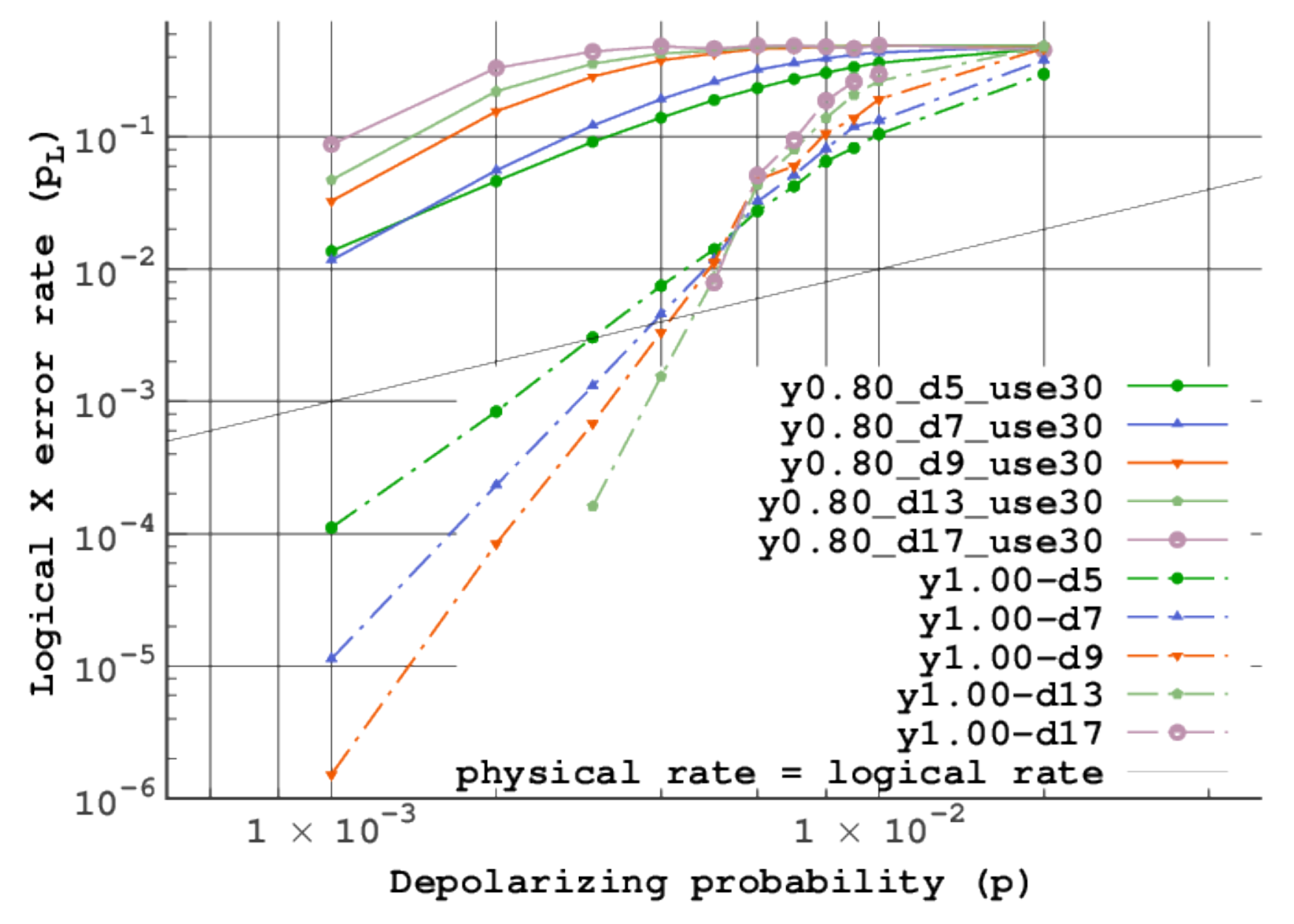}\\
  \includegraphics[width=7cm]{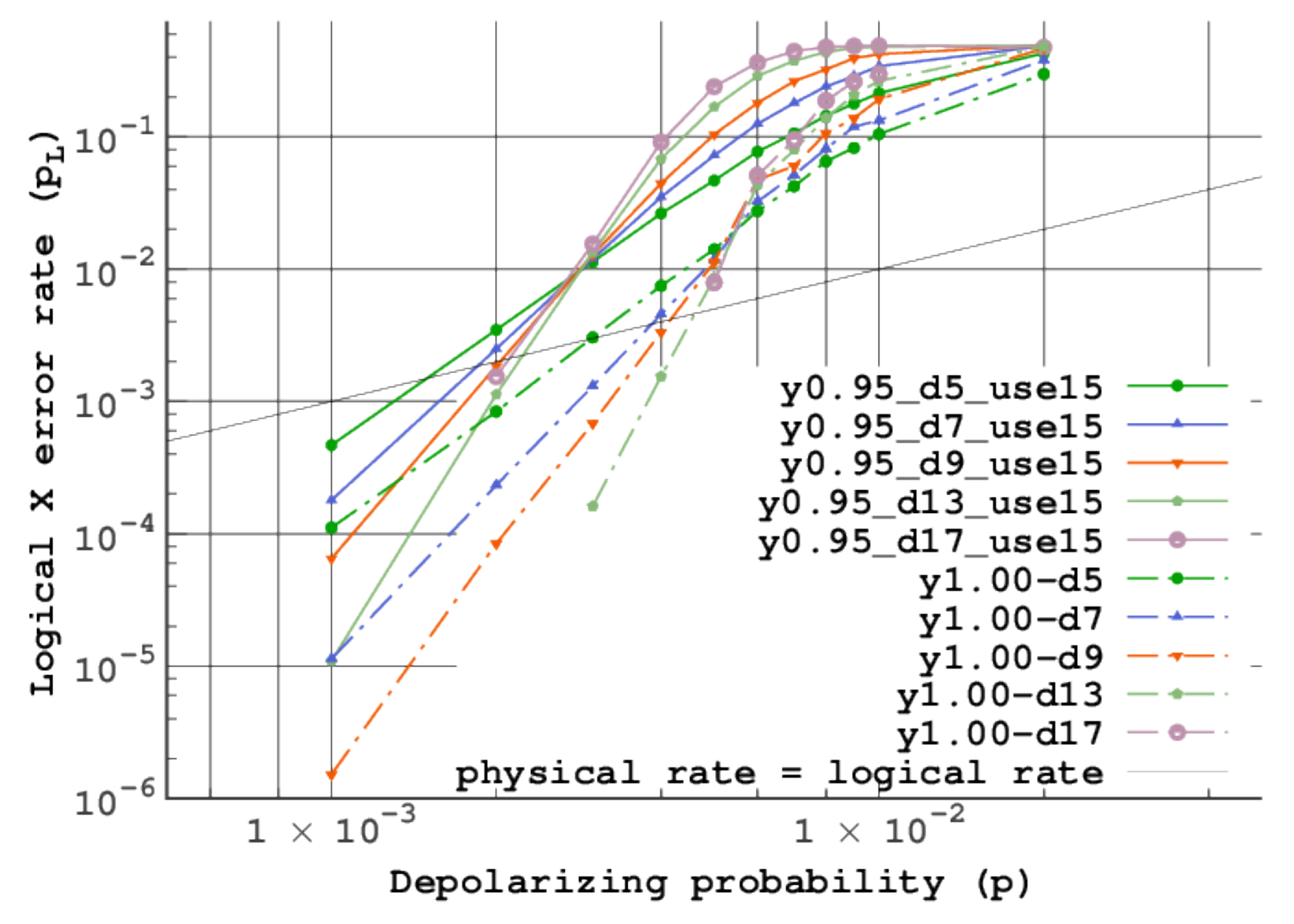}
  \includegraphics[width=7cm]{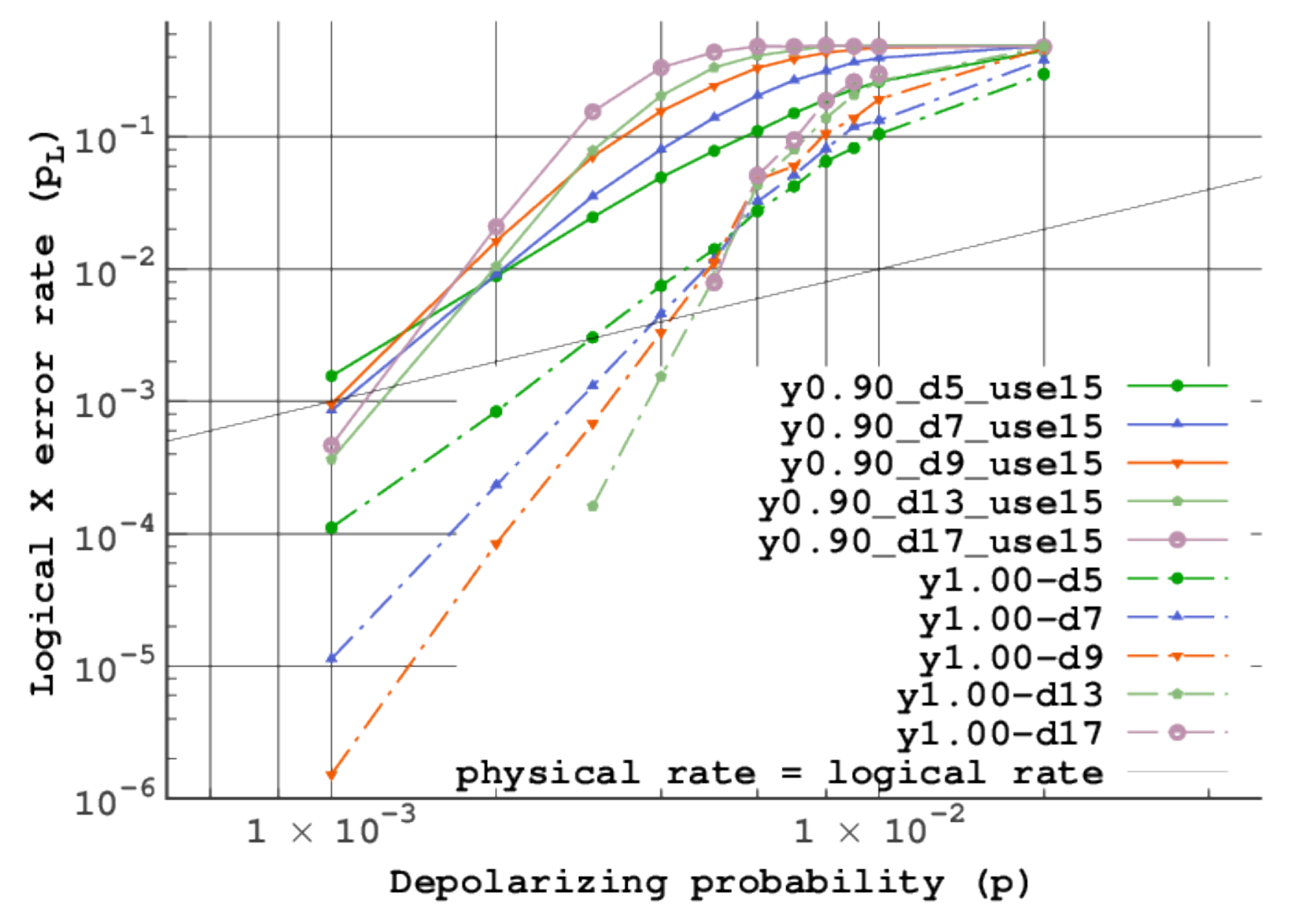}
  \includegraphics[width=7cm]{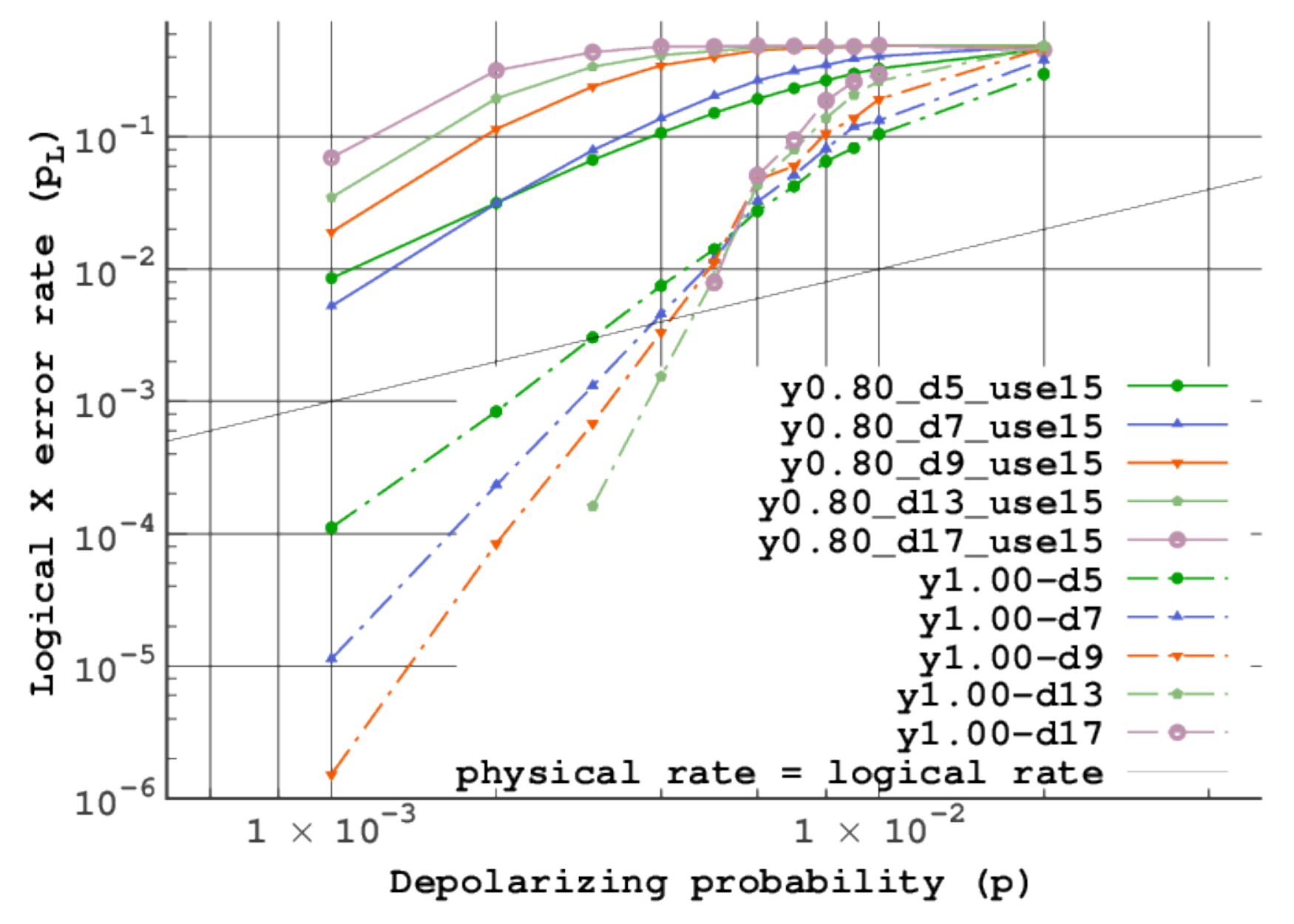}\\
  \includegraphics[width=7cm]{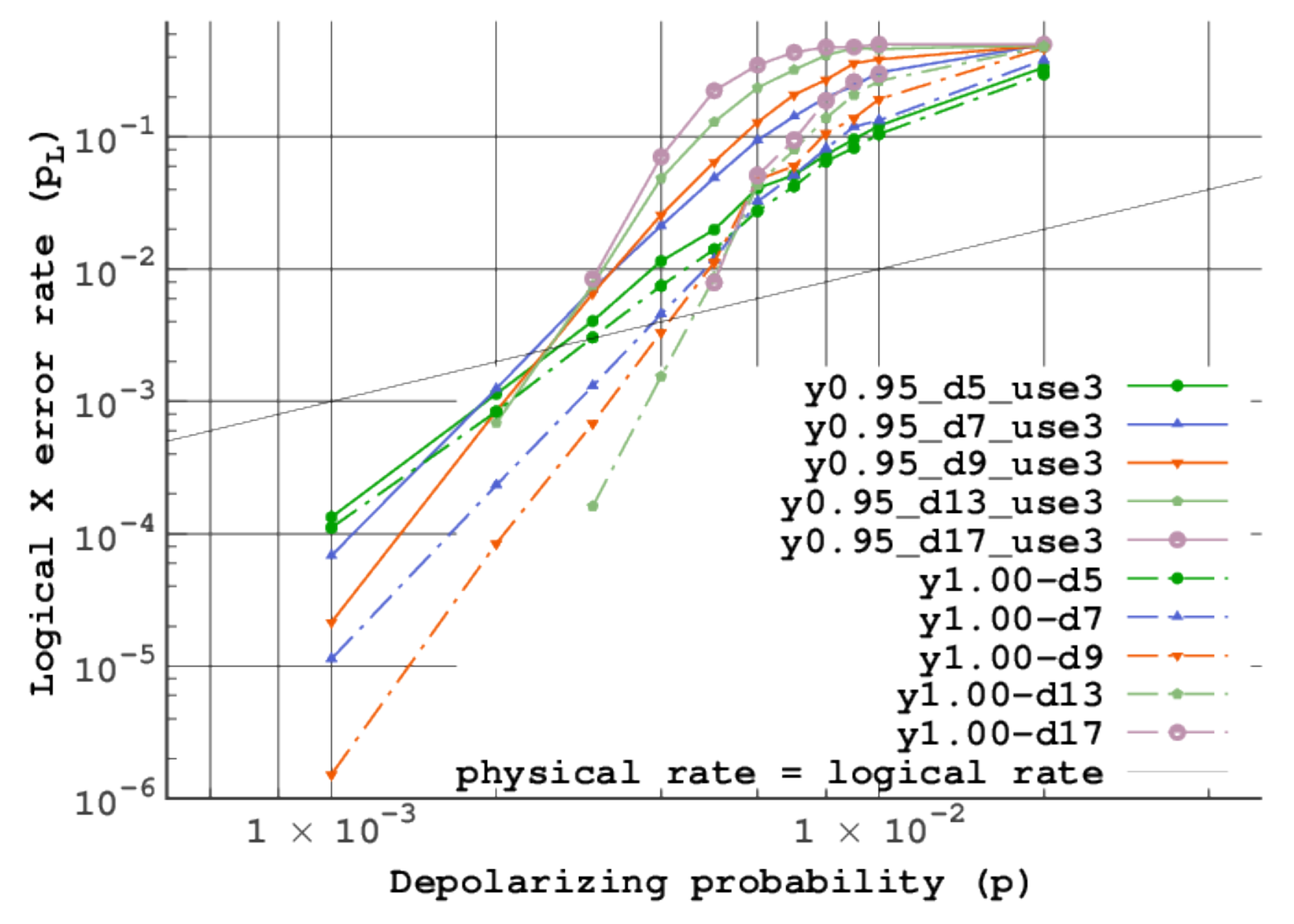}
  \includegraphics[width=7cm]{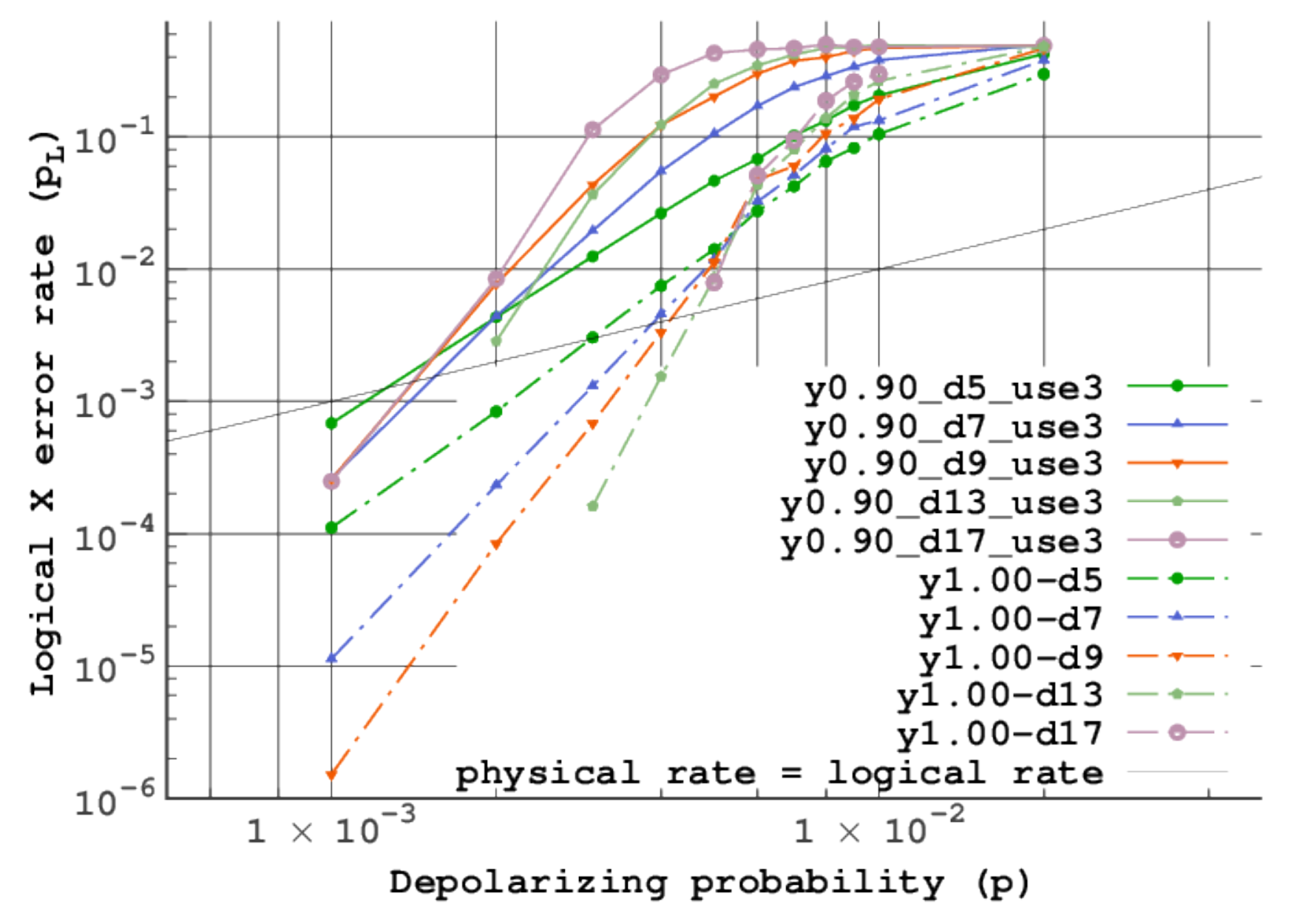}
  \includegraphics[width=7cm]{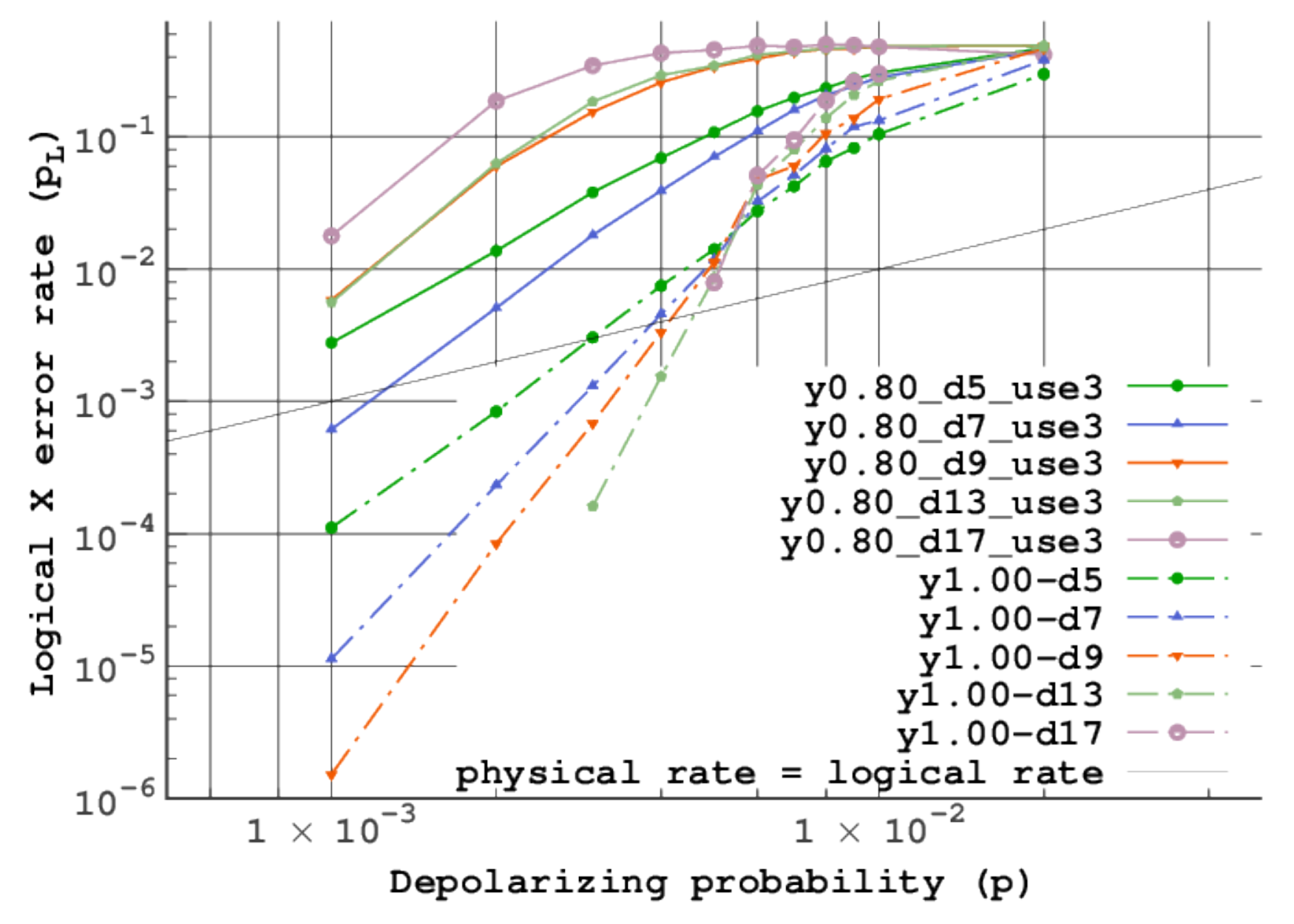}
 \end{center}
\caption{Graphs of randomly defective lattice.
Dashed lines are of the perfect lattices for reference.
The left column is of $y=95\%$,
the middle column is of $y=90\%$
and the right column is of $y=80\%$.
The top row is of all generated lattices,
the middle row is of culling worse 50\%
and the bottom row is of culling worse 90\%.
}
\label{fig:graphs_random}
\end{figure}
\end{landscape}

  Figure \ref{fig:graphs_random} is the main result of this work,
  showing the geometric mean of all encodable lattices, plotting physical error rates versus logical error rates.
  \ref{subsec:scatterplots} shows the scatter plots of raw data.

  The left column in Figure~\ref{fig:graphs_random} show the graphs of $y=95\%$, describing the geometric mean of all encodable lattices,
  of the better 50\%, and of the best 10\% of generated lattices, from the top, respectively.
  Note that those cull percentages are based on the original set of 30 generated lattices,
  not the smaller number of the encodable lattices.
  Some points of longer distance at lower physical error rate are not plotted
  since not enough logical errors are accumulated because of the very low logical error rates.
  
  At $95\%$ functional qubit yield, we see many chips beating break-even at $p=10^{-3}$.
  The threshold is about $0.3\%$, about half of the threshold error rate for a perfect lattice.
  The significant
  penalty in both threshold and residual error rate can be dramatically reduced by culling poorer chips and discarding them.
  At $50\%$ cull at $p=10^{-3}$, the residual error rate for $d=7$
  is about that of $d=5$ with a perfect lattice, and $d=13$ is about that of a perfect $d=7$.  

Naturally, the logical error rates get better as we discard more of the
poorest lattices.  At $p=0.2\%$, $y=95\%$ in
Unculled $y=95\%$ shows that even
distance 17 is just on the break-even line and 90\%-culled $y=95\%$
shows that all 5 distances exceed break-even.  The steepness of the
slope of the curves of culled defective lattices exceeds that of the
curves of lower code distances on the perfect lattice, though it does
not match the perfect lattice of the same distance.  Thus, an
appropriate culling strategy reduces the penalty for a 5\% fault rate
to a manageable level, allowing us to achieve a desired level of error
suppression by using a slightly larger code distance.
  At $p=0.1\%$, by culling $90\%$, the penalty against the perfect lattices changes
  from $12.0\times$ to $1.2\times$ at $d=5$,
  from $39.0\times$ to $6.1\times$ at $d=7$, and
  from $119.9\times$ to $14.2\times$ at $d=9$.
  We do not have data points for $90\%$-discarding at $d=13$ and at $d=17$ since
  not enough logical errors are accumulated on best $10\%$ of their lattices.
  The smaller code distance gets closer to the perfect lattice
  because it has fewer qubits, therefore
  good outliers may be generated with higher probability,
  as shown in Table~\ref{tab:num_faulty}.
  Table~\ref{tab:num_faulty} summarizes the average number of static losses on all the generated lattices,
  on the 50\%-cull lattices and on the 90\%-cull lattices.
  The remaining 3 lattices of 90\%-cull distance 5 have 1, 2 and 2 static losses respectively.
  Table~\ref{tab:num_faulty} also allows us to see the importance of static loss placement
  because the number of static losses does not decrease much
  but all the logical error rates improve.

  The middle column in Figure~\ref{fig:graphs_random} is the graphs of $y=90\%$,
  under the same conditions with those of $y=95\%$.
  The threshold is about $0.15\%$, about a quarter of the threshold for a perfect lattice.
  At $90\%$ cull (in the middle-bottom of Figure~\ref{fig:graphs_random}), 
  at $p=10^{-3}$, the residual error rates for $d=7$, $d=9$ and $d=17$
  are about twice those of $d=5$ with a perfect lattice. $d=13$ would be better than that of perfect $d=5$,
  but it is missing since the logical error rate may be too low to accumulate enough number of logical errors.
  At $p=0.1\%$ of $y=90\%$,
  unculled (middle-top) shows that only distance 13 exceeds the break-even,
  but 90\%-cull (middle-bottom) shows all five distances exceed the break-even.

  The right column in Figure~\ref{fig:graphs_random} is the graphs of $y=80\%$.
  At $y=80\%$, we have already seen that only two-thirds of the chips can even be encoded.
  Our simulations indicate that even those chips for which we could create a circuit are unusable.
  Even at $p=10^{-3}$, there is no evidence of a correctable threshold, and although residual error rates
  do decline as the physical error rate is reduced, only a single data point reaches break-even.
  We conclude that $y=80\%$ is not good enough to build a computer.

Unculled $y=95\%$ shows that distances 13 and 17 are approximately identical at $p=0.2\%$, while other distances show that longer is better.
Unculled distance 13 and 17 for $y=90\%$ do not show that longer is better, though other distances do.
Unculled $y=80\%$ shows that distance 7 exceeds distance 5 at $p=0.1\%$, while other distances do not show an improvement for the longer distance.
Those indicate that the longer code distances cross at lower physical error rate.
We need to consider this fact when deciding the code distance to use.  

\subsection{Metrics for selecting good chips}
\label{subsec:metrics}
  Both to improve our understanding of the root causes of the error rate
  penalty and to provide a simple means of selecting good chips, we evaluated the
  correlation between a set of easy-to-calculate metrics and the simulated residual error rate.
  Tables \ref{tab:linear_correlation} and \ref{tab:exponential_correlation} describe the correlations 
  between eighteen metrics and logical error rates or log(logical error rates), respectively,
  for $p=0.002$ for each combination of yield and code distance.

  The simplest possible metrics, just counting numbers of qubits in various categories, show only modest correlation.

  Steane's $KQ$ metric is the space-time product of a circuit: the number of qubits $Q$ involved, multiplied by the
  circuit depth $K$~\cite{steane02:ft-qec-overhead}.

  The $CDQ$ and $CQ$ is the product of the ``cycle'', which is the average 
  number of steps in a stabilizer measurement
  and the number of data qubits or the total number of qubits 
  including ancillae involved in the stabilizer, respectively.
  The $CDQ$ and $CQ$ reflect the total probabilities of possible physical errors which occur 
  in a measurement of the stabilizer.
  Both Tables \ref{tab:linear_correlation} and \ref{tab:exponential_correlation} 
  indicate that 
  the average of the $CDQ$ and the average of the $CQ$ of $Z$ stabilizers 
  have the strongest and the second strongest correlations with the logical $X$ error rate.
  The average number of qubits in a $Z$ stabilizer and
  the average ``cycle'' of $Z$ stabilizers show the next strongest correlations.
  Those mean that the accumulation of possible errors in a stabilizer may be the factor most 
  strongly correlated to the logical error rate.

  Somewhat to our surprise, both the $KQ$ of the largest stabilizer and the average across the entire lattice 
  do not have good correlations.
  This may be because this form of $KQ$ does not correctly capture the total probabilities of possible physical errors which occur 
  in a measurement of the stabilizer.

  Table \ref{tab:num_faulty} implies that the number of faulty devices is correlated
  with the logical error rate. 
  By culling bad lattices, 
  Table \ref{tab:num_faulty} shows that the average number of faulty devices on a lattice is reduced
  and Figure~\ref{fig:graphs_random}
  shows that the logical error rate gets better.
  However, the average $CDQ$ of $Z$ stabilizers has significantly higher correlation with
  logical $X$ error rate, 0.76, than that of the number of faulty devices, 0.43.
  We calculated the cross-correlation of elements for $y=0.95$ and $d=9$.
  The correlation between the number of faulty devices and the average $CDQ$ of $Z$ stabilizers is 0.79.

  The number of faulty ancilla qubits is the most weakly correlated to the logical error rate.
  This fact indicates that even if the number of faulty ancilla qubit increases,
  the logical error rate does not decline rapidly.
  For a given yield, the placement of faults matters more than the exact number.

\begin{landscape}
   \begin{center}
 \begin{table}[t]
   \begin{center}
     \scalebox{0.7}[0.9]{
  \begin{tabular}[t]{c|c||c|c|c|c||c|c||c|c|c|c||c|c||c|c|c|c||c|c||c|c|c|c||c}
yield& code  &\#stab&\#flty&\#flty&\#flty & $Z$   &\#$Z$& bgst  & ave.  & bgst  & ave.  & dpst  & ave.  & bgst & ave. & bgst & ave. & bgst & ave. & bgst & ave. & bgst & ave. & ave.\\
     & dist. &      & qubit& data & syn.  & redu- &stab& \#qubit&\#qubit&\#data &\#data & depth & depth & $KQ$ & $KQ$ & $KDQ$& $KDQ$& $Z$  & $Z$  & $CQ$ & $CQ$ & $CDQ$& $CDQ$& \#$Z$ \\
     &       &      &      & qubit& qubit & ced   &     & of $Z$& of $Z$& qubit & qubit & of $Z$& of $Z$& of   & of   & of   & of   & cycle& cycle& of   & of   & of   & of   & stab\\
     &       &      &      &      &       & dist. &     & stabs & stabs & of $Z$& of $Z$& stabs & stabs & $Z$  & $Z$  & $Z$  & $Z$  &      &      & $Z$  & $Z$  & $Z$  & $Z$  & msmts\\
     &       &      &      &      &       &       &     &       &       & stabs & stabs &       &       & stabs& stabs& stabs& stabs&      &      & stabs& stabs& stabs& stabs& /step\\
\hline
\hline
0.95&5&-0.64&0.59&0.70&0.08&-0.67&-0.71&0.81&0.83&0.79&0.71&0.26&-0.01&0.24&0.37&0.23&0.16&0.69&0.71&0.74&\bf{0.88}&0.76&0.86&-0.69\\
0.95&7&-0.40&0.45&0.65&-0.13&-0.62&-0.46&0.65&0.44&0.70&0.44&0.05&0.00&0.07&0.11&0.08&0.04&0.78&0.55&0.76&0.78&\bf{0.79}&\bf{0.79}&-0.44\\
0.95&9&-0.50&0.68&0.72&0.33&-0.60&-0.48&0.63&0.81&0.64&0.79&0.42&-0.07&0.46&0.48&0.40&0.24&0.62&0.81&0.67&0.88&0.68&\bf{0.89}&-0.69\\
0.95&13&-0.28&0.48&0.55&0.11&-0.19&-0.27&0.36&0.56&0.37&0.52&-0.02&-0.47&0.02&-0.14&0.03&-0.25&0.30&0.59&0.33&\bf{0.71}&0.34&0.69&-0.40\\
0.95&17&-0.18&0.34&0.27&0.25&-0.10&-0.18&0.20&0.39&0.15&0.40&0.05&-0.32&0.17&-0.06&0.05&-0.16&0.13&0.50&0.11&\bf{0.56}&0.10&0.55&-0.40\\
\hline
0.90&5&-0.19&0.35&0.26&0.29&-0.48&-0.24&\bf{0.62}&0.20&0.55&0.16&-0.23&-0.21&0.06&-0.13&-0.11&-0.16&0.49&0.28&0.69&0.48&\bf{0.62}&0.46&-0.21\\
0.90&7&-0.31&0.37&0.43&0.07&-0.59&-0.33&0.50&0.53&0.49&0.45&0.20&-0.33&0.59&0.23&0.43&-0.01&0.51&0.58&0.71&0.71&0.70&\bf{0.74}&-0.46\\
0.90&9&-0.39&0.40&0.54&0.09&-0.65&-0.42&0.67&0.65&0.63&0.58&-0.29&-0.54&0.01&-0.12&-0.18&-0.32&0.28&0.39&0.43&\bf{0.68}&0.43&0.66&-0.45\\
0.90&13&-0.36&0.47&0.63&-0.01&-0.30&-0.37&0.47&0.77&0.52&\bf{0.81}&-0.26&-0.25&-0.10&0.24&-0.28&0.04&0.46&0.57&0.60&\bf{0.81}&0.62&\bf{0.81}&-0.51\\
0.90&17&-0.55&0.58&0.62&0.10&-0.62&-0.56&0.56&0.76&0.56&0.74&-0.05&-0.25&0.34&0.33&0.10&0.08&0.17&0.65&0.35&0.86&0.37&\bf{0.87}&-0.74\\
\hline
0.80&5&-0.20&0.30&0.22&0.17&-0.43&-0.22&\bf{0.78}&0.59&0.46&0.40&-0.11&-0.13&0.43&0.30&-0.19&0.05&0.64&0.72&\bf{0.78}&0.75&0.63&0.67&-0.49\\
0.80&7&0.22&0.35&0.10&0.40&-0.51&0.13&0.56&0.81&0.44&0.75&0.07&-0.34&0.29&0.58&0.03&0.24&0.56&0.74&0.64&0.79&0.69&\bf{0.83}&-0.48\\
0.80&9&-0.39&0.37&0.31&0.15&-0.47&-0.38&0.63&0.88&0.68&0.68&-0.42&-0.45&0.41&0.58&0.12&-0.01&0.50&0.77&0.58&0.86&0.66&\bf{0.92}&-0.53\\
0.80&13&0.02&0.34&0.18&0.20&-0.43&-0.05&0.65&0.79&0.66&0.80&0.21&0.09&0.45&0.68&0.22&0.52&0.70&0.63&0.65&0.79&0.70&\bf{0.84}&-0.35\\
0.80&17&0.39&0.25&0.19&0.19&-0.24&0.34&0.37&0.60&0.29&0.57&-0.15&-0.41&0.28&0.37&0.04&0.07&0.51&0.55&0.34&0.61&0.33&\bf{0.62}&0.16\\
\hline
\multicolumn{2}{c||}{average}&-0.25&0.42&0.42&0.15&-0.46&-0.28&0.56&0.64&0.53&0.59&-0.02&-0.25&0.25&0.25&0.06&0.04&0.49&0.60&0.56&0.74&0.56&\bf{0.75}&-0.44
  \end{tabular}
  }
  \caption{Linear Correlation between candidate metrics for lattice quality factors and X logical error rates for each combination of yield and code distance.
    Boldface is the strongest correlation in each row.
    ``Dist.'', ``flty'', ``bgst'', ``syn.'', ``stab'' and ``msmt'' stand for distance, faulty, biggest, syndrome, stabilizer and measurement, respectively.
    Averages here are arithmetic means.
    ``Reduced dist.'' is the the minimum distance between corresponding boundaries shortened by merging stabilizers.
    ``\#Z stabs'' is the number of Z stabilizers.
    ``Bgst \#qubit of Z stabs'' and
    ``ave. \#qubit of Z stabs'' is the biggest and the average number of qubits involving data qubits and syndrome qubits in Z stabilizer circuits.
    ``Bgst \#data qubit of Z stabs'' and
    ``ave. \#data qubit of Z stabs'' is the biggest and the average number of data qubits in a Z stabilizer.
    ``Dpst depth of Z stabs'' is the depth of the deepest Z stabilizer circuit.
    ``Ave. depth of Z stabs'' is the average depth of Z stabilizer circuits.
    ``Bgst $KQ$ of Z stabs'' and
    ``Ave. $KQ$ of Z stabs'' is the biggest and average $KQ$ (metric is the space-time product of a circuit: the number of qubits $Q$ involved, multiplied by the circuit depth $K$~\cite{steane02:ft-qec-overhead}) of Z stabilizer circuits.
    ``Bgst $KDQ$ of Z stabs'' and
    ``Ave. $KDQ$ of Z stabs'' is the biggest and average $KDQ$ which is the product of the number of data qubits $DQ$ involved and of the circuit depth $K$ of Z stabilizer circuits.
    ``Cycle'' here indicates every how many steps a stabilizer is measured, including waiting time by scheduling of stabilizers.
    ``Bgst Z cycle'' and
    ``Ave. Z cycle'' is the biggest and the average cycle of Z stabilizers.
    ``CQ'' is the product of the cycle $C$ and the number of qubits $Q$, similar to $KQ$.
    ``Bgst $CQ$'' and
    ``Ave. $CQ$'' is the biggest and the average $CQ$ of stabilizers.
    ``Bgst $CDQ$'' and
    ``Ave. $CDQ$'' is the biggest and average $CDQ$ which is the product of the number of data qubits $DQ$ involved and of the cycle $C$ of Z stabilizer circuits.
    ``Ave. \#Z stabs per step'' is how many stabilizers are measured in a step on average.
  }
  \label{tab:linear_correlation}
  \end{center}
 \end{table}
  \end{center}
 \end{landscape}

\begin{landscape}
   \begin{center}
 \begin{table}[t]
   \begin{center}
     \scalebox{0.7}[0.9]{
  \begin{tabular}[t]{c|c||c|c|c|c||c|c||c|c|c|c||c|c||c|c|c|c||c|c||c|c|c|c||c}
yield& code  &\#stab&\#flty&\#flty&\#flty & $Z$   &\#$Z$& bgst  & ave.  & bgst  & ave.  & dpst  & ave.  & bgst & ave. & bgst & ave. & bgst & ave. & bgst & ave. & bgst & ave. & ave.\\
     & dist. &      & qubit& data & syn.  & redu- & stab&\#qubit&\#qubit&\#data &\#data & depth & depth & $KQ$ & $KQ$ & $KDQ$& $KDQ$& $Z$  & $Z$  & $CQ$ & $CQ$ & $CDQ$& $CDQ$& \#$Z$ \\
     &       &      &      & qubit& qubit & ced   &     & of $Z$& of $Z$& qubit & qubit & of $Z$& of $Z$& of   & of   & of   & of   & cycle& cycle& of   & of   & of   & of   & stab\\
     &       &      &      &      &       & dist. &     & stabs & stabs & of $Z$& of $Z$& stabs & stabs & $Z$  & $Z$  & $Z$  & $Z$  &      &      & $Z$  & $Z$  & $Z$  & $Z$  & msmts\\
     &       &      &      &      &       &       &     &       &       & stabs & stabs &       &       & stabs& stabs& stabs& stabs&      &      & stabs& stabs& stabs& stabs& /step\\
\hline
\hline
0.95&5&-0.63&0.66&0.71&0.18&-0.74&-0.68&0.87&0.76&\bf{0.88}&0.64&0.44&-0.05&0.40&0.33&0.39&0.12&0.72&0.75&0.72&0.82&0.76&0.83&-0.75\\
0.95&7&-0.49&0.62&0.79&-0.05&-0.67&-0.53&0.57&0.59&0.65&0.55&0.27&0.06&0.29&0.26&0.30&0.15&0.79&0.69&0.69&0.85&0.73&\bf{0.87}&-0.54\\
0.95&9&-0.63&0.81&0.83&0.42&-0.63&-0.62&0.70&0.77&0.69&0.72&0.39&-0.15&0.43&0.38&0.36&0.14&0.66&0.83&0.69&0.87&0.70&\bf{0.88}&-0.79\\
0.95&13&-0.31&0.58&0.66&0.15&-0.29&-0.31&0.43&0.71&0.43&0.68&-0.05&-0.30&0.01&0.09&0.03&-0.03&0.31&0.70&0.41&\bf{0.82}&0.40&0.81&-0.52\\
0.95&17&-0.16&0.18&0.17&0.12&-0.08&-0.16&0.25&0.24&0.18&0.29&-0.02&-0.24&0.08&-0.08&-0.02&-0.13&0.15&0.34&0.17&\bf{0.41}&0.16&0.40&-0.28\\
\hline
0.90&5&-0.10&0.41&0.28&0.36&-0.42&-0.15&0.62&0.32&0.53&0.19&-0.14&-0.32&0.13&-0.17&-0.01&-0.24&0.52&0.45&\bf{0.66}&0.60&0.59&0.58&-0.24\\
0.90&7&-0.34&0.45&0.39&0.30&-0.65&-0.35&0.51&0.54&0.46&0.40&0.29&-0.39&0.51&0.16&0.39&-0.11&0.51&0.65&0.64&0.72&0.61&\bf{0.73}&-0.50\\
0.90&9&-0.49&0.51&0.58&0.19&-0.67&-0.52&0.66&0.63&0.63&0.53&-0.22&-0.58&0.11&-0.15&-0.09&-0.36&0.42&0.52&0.46&\bf{0.74}&0.48&0.73&-0.58\\
0.90&13&-0.43&0.69&0.74&0.21&-0.48&-0.46&0.46&0.87&0.49&0.86&-0.14&-0.28&0.03&0.29&-0.15&0.04&0.47&0.77&0.54&0.92&0.55&\bf{0.93}&-0.69\\
0.90&17&-0.50&0.75&0.67&0.30&-0.63&-0.51&0.49&0.76&0.49&0.67&0.05&-0.29&0.40&0.31&0.18&0.03&0.25&0.73&0.35&\bf{0.87}&0.36&0.86&-0.76\\
\hline
0.80&5&-0.20&0.29&0.24&0.14&-0.51&-0.26&0.69&0.65&0.49&0.55&0.01&0.07&0.39&0.48&-0.09&0.28&0.77&\bf{0.81}&0.75&0.79&0.68&0.77&-0.63\\
0.80&7&0.19&0.36&-0.06&0.51&-0.62&0.03&0.65&0.78&0.56&\bf{0.83}&-0.04&-0.31&0.31&0.59&0.01&0.33&0.44&0.67&0.58&0.69&0.60&0.75&-0.62\\
0.80&9&-0.36&0.27&0.24&0.09&-0.54&-0.35&0.59&0.80&0.58&0.60&-0.35&-0.40&0.19&0.46&-0.07&-0.06&0.55&0.77&0.61&0.82&0.64&\bf{0.87}&-0.49\\
0.80&13&0.13&0.20&0.01&0.19&-0.36&0.05&0.59&0.78&0.64&\bf{0.80}&0.27&0.17&0.50&0.73&0.32&0.58&0.56&0.61&0.54&0.71&0.60&0.79&-0.28\\
0.80&17&0.43&0.19&0.14&0.15&-0.13&0.40&0.41&0.59&0.33&0.54&-0.18&-0.40&0.29&0.35&0.01&0.05&0.49&0.50&0.36&\bf{0.61}&0.35&\bf{0.61}&0.25\\
\hline
\multicolumn{2}{c||}{average}&-0.26&0.47&0.43&0.22&-0.50&-0.30&0.57&0.65&0.54&0.59&0.04&-0.23&0.27&0.27&0.10&0.05&0.51&0.65&0.54&0.75&0.55&\bf{0.76}&-0.49
  \end{tabular}
  }
  \caption{Logarithmic Correlation between candidate metrics for lattice quality factors and X logical error rates for each combination of yield and code distance.
    See Table \ref{tab:linear_correlation} for column heading definitions.
  }
  \label{tab:exponential_correlation}
  \end{center}
 \end{table}
  \end{center}
 \end{landscape}

\section{Discussion}
\label{sec:discussion}
We have proposed and analyzed an adaptation of the surface code for static losses,
which are manifested as faulty devices on quantum computation chips occurring during fabrication.
With this fundamental analysis of static loss and its influence,
independent analysis 
has now been conducted for the three major
imperfections of quantum computation for the surface code:
state error, dynamic loss, and static loss.
The ultimate goal of investigating faulty devices is
to support collection of a large pool of sufficiently
fault-tolerant quantum computation chips,
because a realistic large-scale quantum computer must be assembled from many quantum computation chips, coupled by their proximity or via distributed quantum computation
~\cite{van-meter10:dist_arch_ijqi}.
We analyzed our approach against faulty losses by simulation to investigate the relationship
between the logical error rates and lattice characteristics of simulated defective lattices.
Our approach is to merge stabilizers broken by faulty data qubits to a superplaquette.
and to work around faulty ancilla qubits using SWAP gates,
without changing the original role of the qubits.

Our simulation with a single faulty device revealed that faulty qubits at the periphery reduce the logical error rate less than those in the center.
Even a single fault has a large impact on the residual error rate.
%

Our simulation with randomly placed faulty devices showed that
at $95\%$ yield, the impact on net error rate is significant but many of the chips still achieve break-even by $p=10^{-3}$,
and therefore could be used in a real-world setting.
At $90\%$ yield, very few chips achieve break-even.
At $80\%$ yield, almost no chips are usable.
Those facts establish the goals for experimental research to build the surface code quantum computer.

The simulation of randomly placed faulty devices
also showed that discarding bad lattices makes the ensemble better,
showing the trade-off between the cost of culling and the strength of fault-tolerance of an ensemble.
Culling increases our effective error suppression, such that $d_e^{culled} \approx d_e^{unculled} + 2$,
where $d_e$ is the effective code distance,
for lattices with physical distance 9 and 13
at $95\%$ yield.
$90\%$ yield shows error suppression at practical physical error rates, at around $p=0.1\%$,
and culling works for $90\%$ yield.
With a low physical error rate, $90\%$ yield may be sufficient to build a quantum computer.
At $80\%$ yield,
only very weak error suppression is observed
even at $p=0.1\%$,
even when discarding the weakest 90\% of the chips.
We conclude that $80\%$ yield is not suitable for building a quantum computer,
using the surface code without additional architectural support.

The randomly faulty lattice simulation also revealed that
the average of the $CDQ$ and the average of the $CQ$ of $Z$ stabilizers 
show the strongest correlations to simulated residual error rate among
a set of proposed metrics for chip evaluation.
The $CDQ$ and $CQ$ is the product of the ``cycle'', which is the average 
number of steps in the stabilizer measurement
and the number of data qubits/the number of qubits involved in the stabilizer, respectively.
Therefore the accumulated error possibilities in a stabilizer may be the factor most strongly correlated
to the logical error rate.

Faulty data qubits result in merging plaquettes and deepen the stabilizer circuit
hence lengthen the ``cycle''.
Faulty ancilla qubits result in requiring more SWAP gates to walk through data qubits and ancilla qubits
surrounding the faulty ancilla qubits.
However, our data also shows that the number of faulty ancilla qubits has weak correlation to the residual error rate.
This finding will also contribute to efforts to build a large scale quantum computer.

\section*{Acknowledgement}
This work is supported by JSPS KAKENHI Grant Number 25:4103, Kiban B (25280034 \& 16H02812), the JSPS grant for challenging exploratory research and the JST ImPACT project.

\section*{References}
\bibliographystyle{unsrt}
\bibliography{main.bib}

\clearpage
\appendix
\setcounter{section}{0}
\section{Surface code error correction details}
\label{sec:app:ec}
\begin{figure}[t]
 \begin{center}
(a)
  \includegraphics[width=135pt]{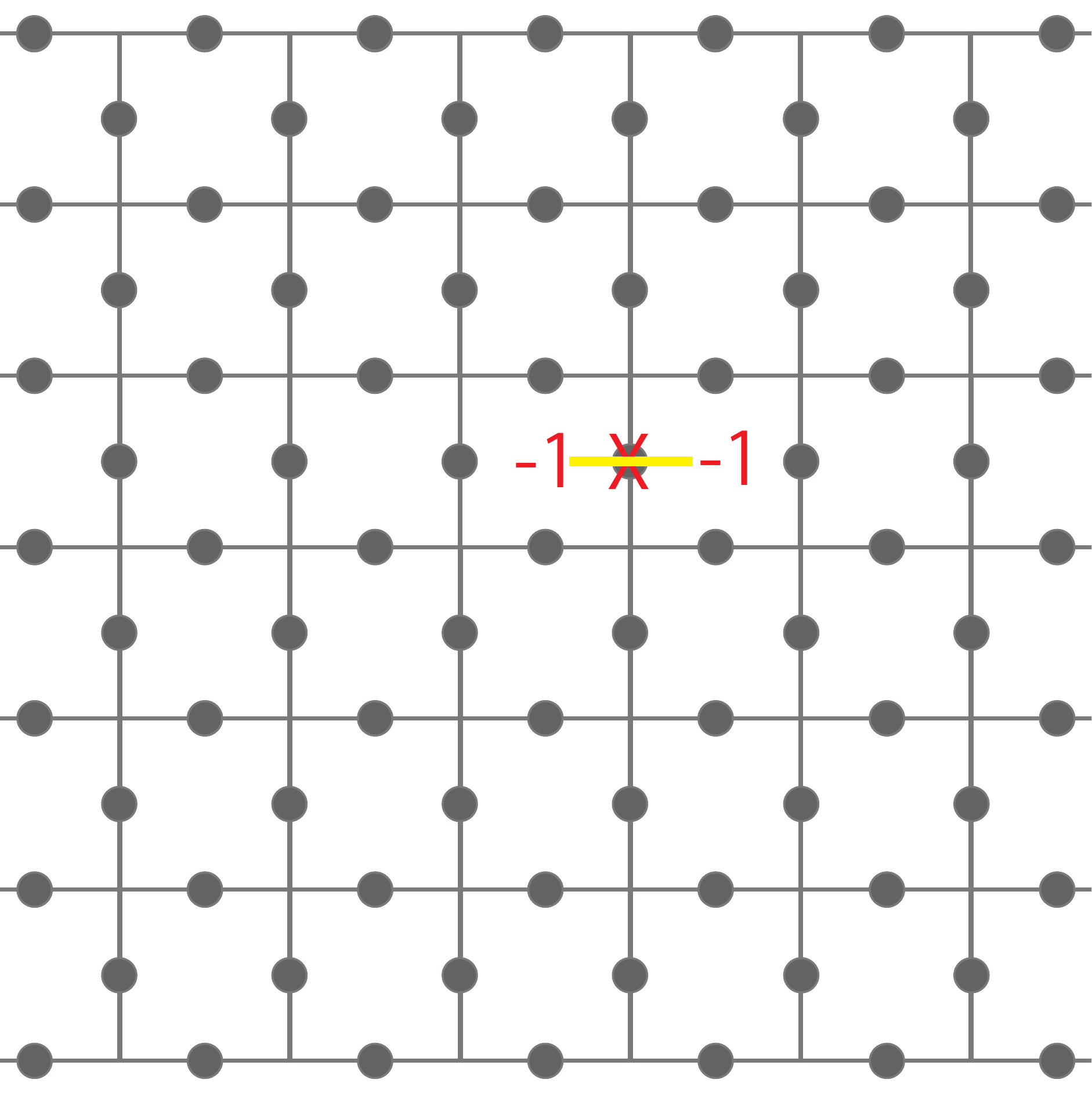}
(b)
  \includegraphics[width=135pt]{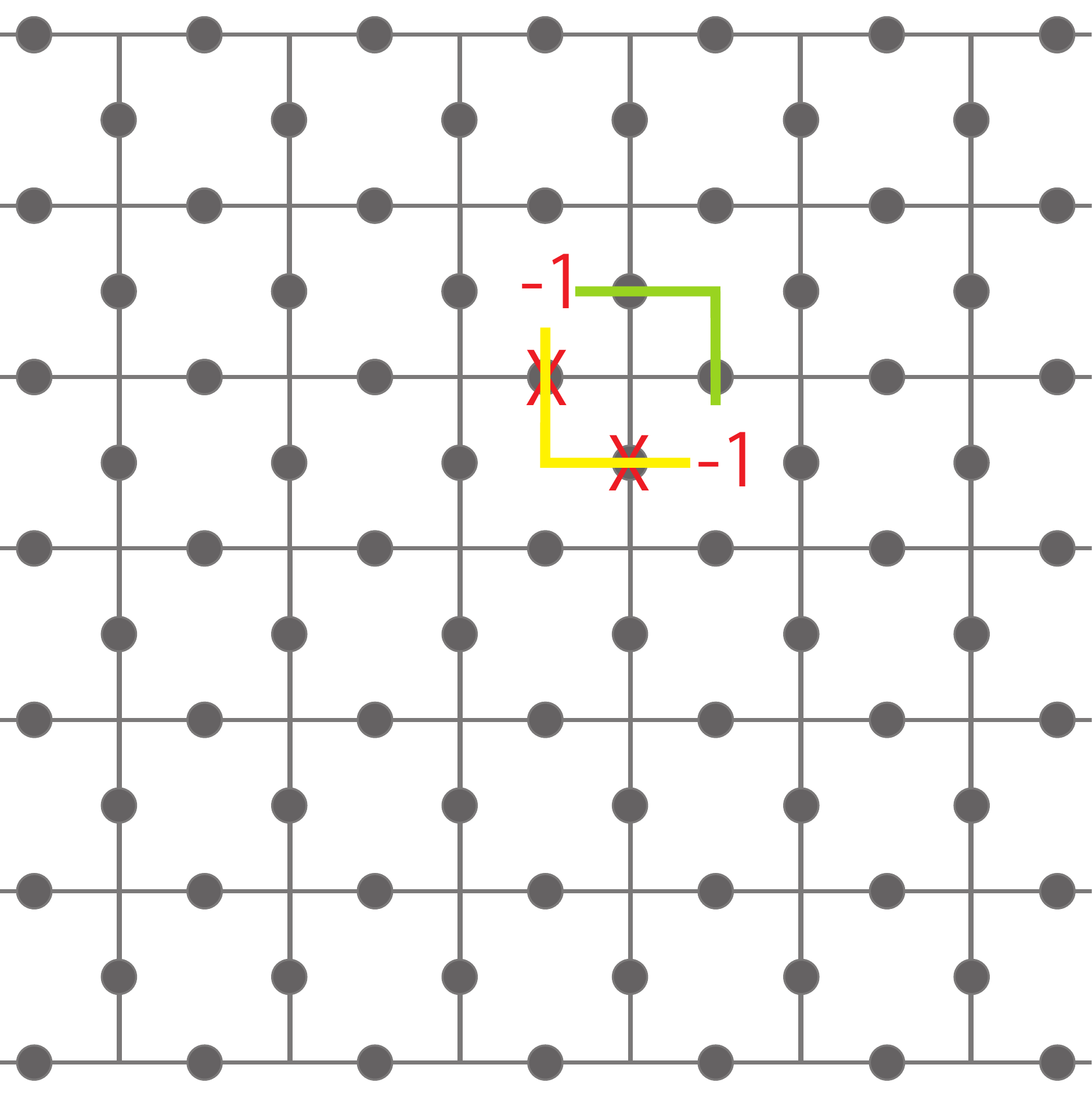}\\
(c)
  \includegraphics[width=135pt]{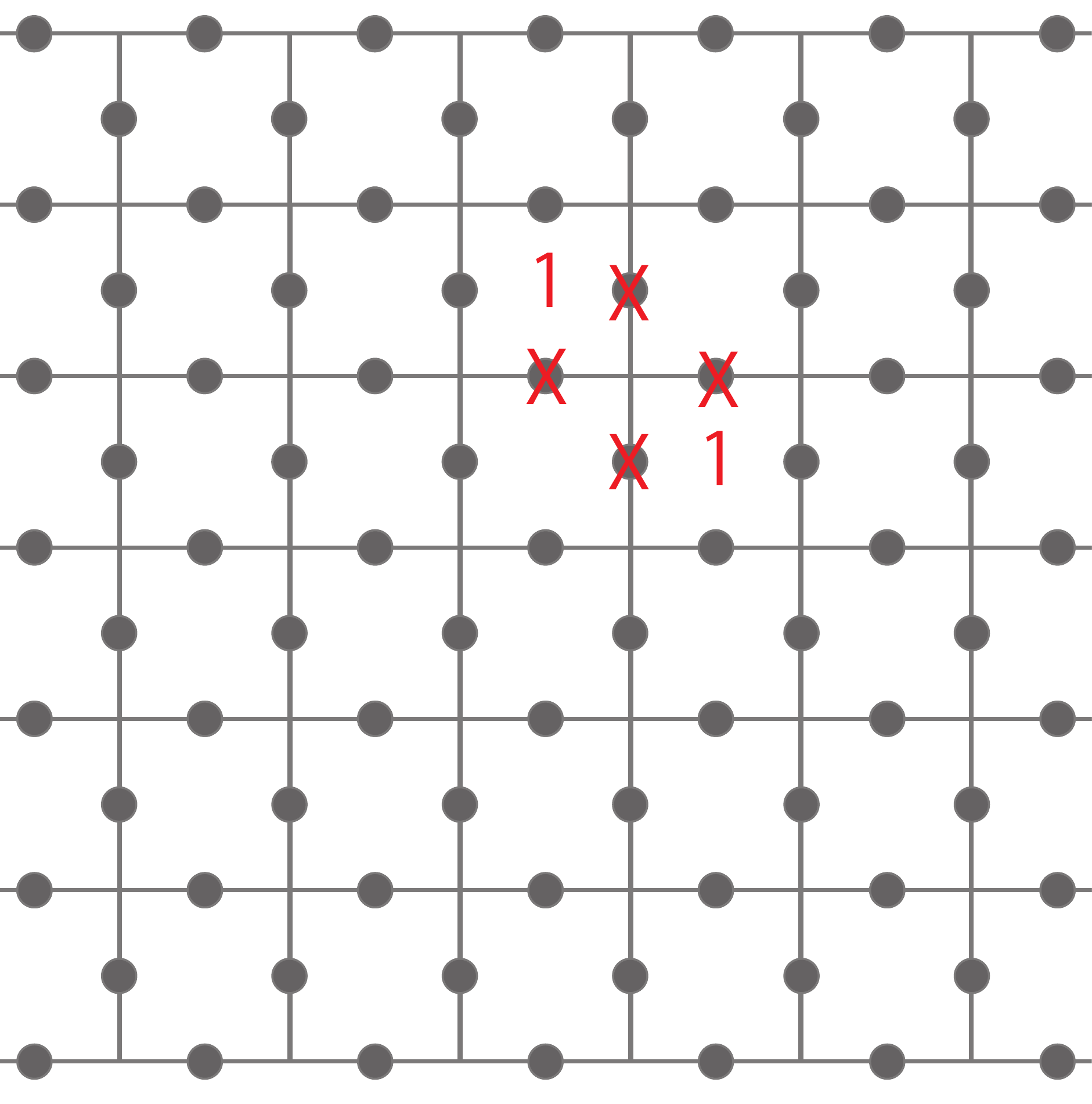}
(d)
  \includegraphics[width=135pt]{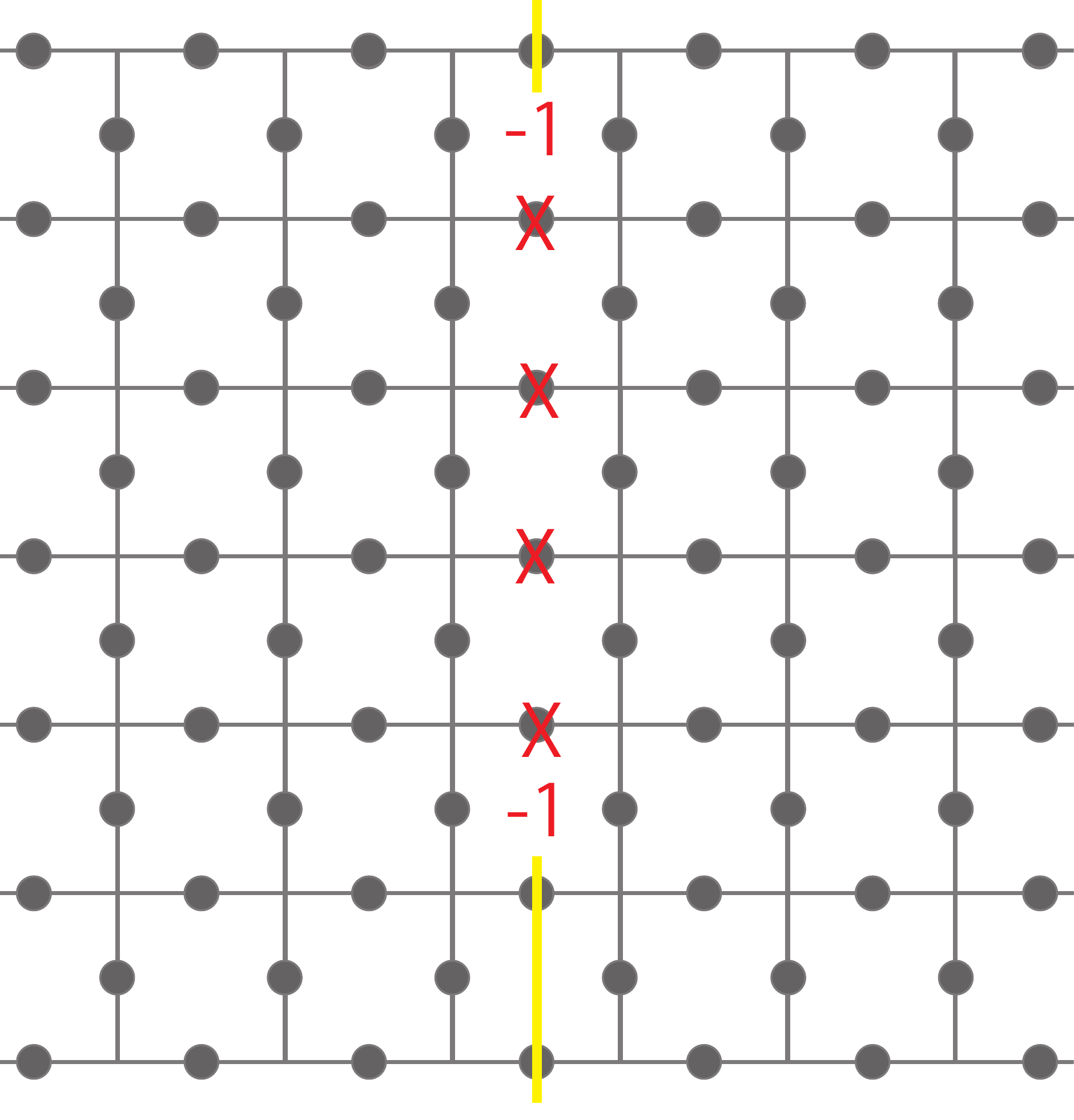}
  \caption{Correctable and uncorrectable errors.
  Data qubits are described with dots and the lines indicate plaquettes.
  Syndrome qubits are omitted for visibility.
  (a) A correctable single error.
  The red 'X' indicates that an X error occurs on the underlying data qubit.
  The corresponding two Z stabilizers which share the data qubit return -1 eigenvalues.
  It is easy to interpret the error chain from the eigenvalues,
  as seen where the yellow line passes through the errored data qubit.
  (b) A correctable case with two errors.
  We can consider two error chains from the eigenvalues of stabilizer measurement, the yellow one and the green one.
  Either of them is valid.
  Obviously the yellow one is valid; executing X gates on qubits underlying the green one generates the trivial error cycle described in (c).
  (c) Topologically trivial error cycle.
  This does not affect the logical state of the surface code; this does not affect the states of data qubits on boundaries.
  (d) Example of a mis-correcting error chain.
  Four X errors occur in the center of the lattice.
  The matching algorithm can pair the two -1 plaquettes, for a distance of 4, or pair each -1 with the neighboring boundary of the X operator, for a total distance of 3.
  Because three errors are more probable than four errors, the matching expects that three errors occur and chooses the yellow error chains.
  After applying X gates on the data qubits under the error chains, a logical X operator is generated, connecting the two boundaries.
  This is a logical error.
}
  \label{fig:sc:errors}
 \end{center}
\end{figure}

In the common case, an isolated X (or Z) error, two neighboring
stabilizers will both show -1 eigenstates, and
the error is easily isolated as shown in Figure \ref{fig:sc:errors} (a).
Because two errors on any plaquette cancel and leave the plaquette in the
+1 eigenstate, a series of errors in a neighborhood likely results
in two -1 plaquettes separated by some distance, surrounded by +1 plaquettes.
If an error chain is connected to the boundary
of the lattice, the termination will be hidden.
So, an error chain running between the two boundaries
will be a logical error.

Applying the same flip operation as the original error
is the obvious means of correcting errors,
because it 
fixes the states of each stabilizer (Figure \ref{fig:sc:errors} (a)).
To achieve this, we have to identify pairs of error terminations by decoding the detected error information.
This problem can be mapped to the graph theory problem known as ``minimum weight perfect matching'',
a common solution for which is the Blossom V algorithm \cite{citeulike:4384001}.

However, many different possible chains can connect two units with -1 eigenvalues, as depicted in Figure \ref{fig:sc:errors} (b).
Fortunately, any chain works equally well.
If the algorithm does not choose the original exact error path, a cycle of errors appears.
Such a trivial error cycle does not affect logical states (Figure \ref{fig:sc:errors} (c)).
Thus, the choice of a chain between -1 units is not a problem.
The important problem in error correction is to pair up the most probable sets of units.
Longer chains of errors occur with lower probability, and the matching algorithm weights such possibilities accordingly.

The distance between the two boundaries for an operator is the code distance of a surface code,
shown in Figure \ref{fig:sc:errors} (d).
The longer the code distance, the higher the tolerance against errors.
In the figure, four errors between the two boundaries for the X operator are fatal,
because the matching algorithm fails to pair them properly.
If the two boundaries were farther apart,
a longer chain of errors would be required to cause the error correction to fail.

In deleting the oldest round of error syndromes, there can be an error syndrome which is temporally matched to a syndrome which is not to be deleted. If this error syndrome is deleted, the left pair will be matched to another syndrome, leading to unintended behaviors. To avoid this behavior, Autotune employs a means that the syndrome to be deleted is retained until its pair is deleted.

\section{Details of the implementation}
\subsection{Irregular whole circuit on account of a fault}
\begin{figure}[t]
 \begin{center}
  \includegraphics[width=300pt]{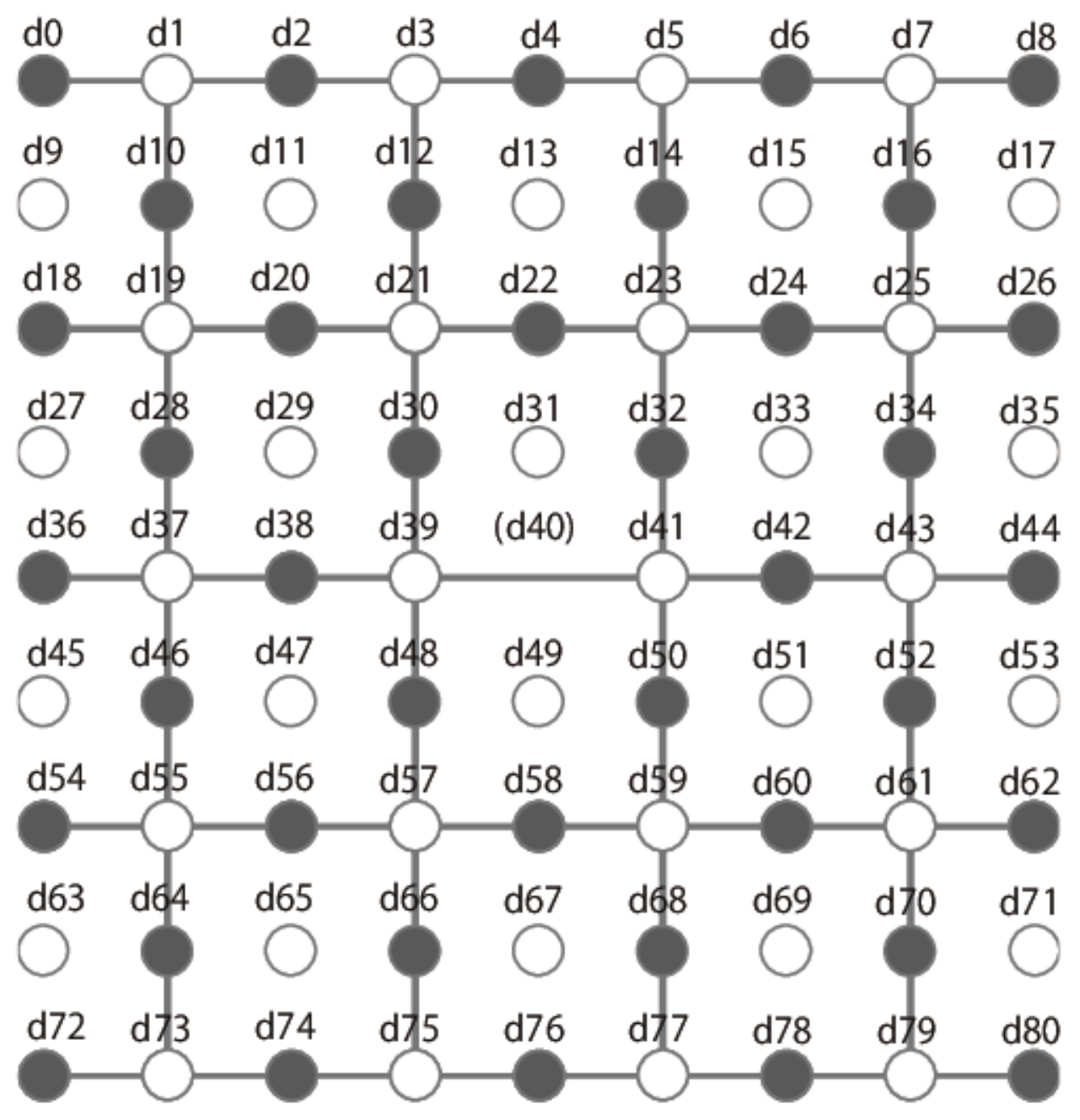}
  \caption{The picture of a lattice corresponding to Figure \ref{fig:sol:whole_circuit_single_faulty_center}.
  Gray dots are data qubits and white dots are syndrome qubits.
  The data qubit labeled (d40) is faulty.}
  \label{fig:sol:single_faulty_center_lattice}
 \end{center}
\end{figure}

\begin{figure}[t]
 \begin{center}
  \includegraphics[height=550pt]{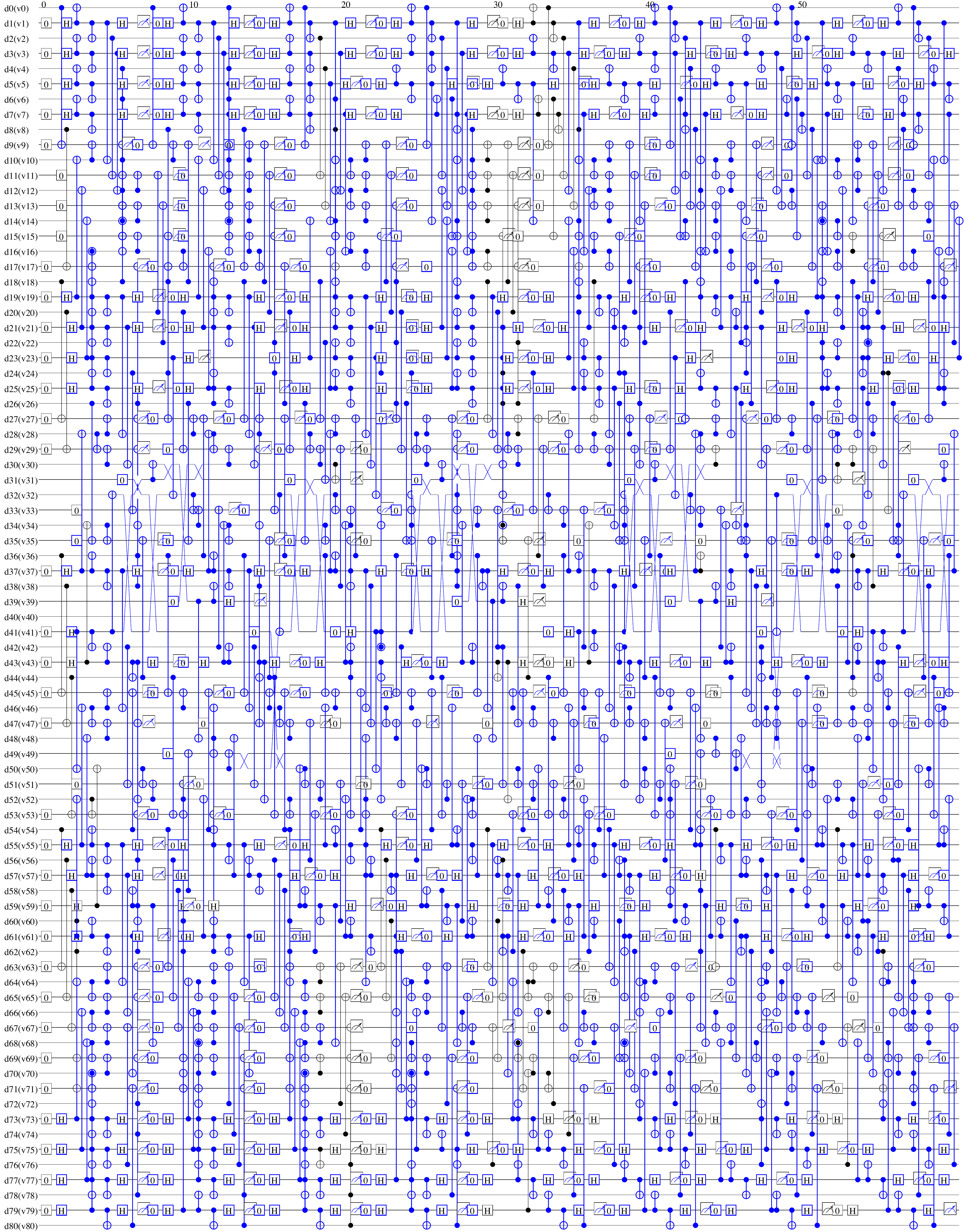}
  \caption{The first sixty steps of a whole circuit for a lattice of d=5 and in which the qubit device d40 in the center is faulty.
  The lattice condition is shown in Figure \ref{fig:sol:single_faulty_center_lattice}.
  We can see swap gates around d40.
  The detail of the irregular stabilizer circuit is shown in Figure \ref{fig:sol:a_2-units_superunit}.}
  \label{fig:sol:whole_circuit_single_faulty_center}
 \end{center}
\end{figure}

Figure~\ref{fig:sol:single_faulty_center_lattice} show an example of
a defective lattice in which the central qubit d40 is faulty.
Figure \ref{fig:sol:whole_circuit_single_faulty_center} shows the first few tens of steps of the whole circuit of the lattice.
We can see that the circuit becomes irregular around the faulty device.

\subsection{Algorithms}
\label{subsec:app:alg}
\IncMargin{1em}

\begin{algorithm*}
 \caption{Stabilizer Circuit Composition}
 \label{alg:sol:stabilizer_circuit}
 \SetKwInput{Input}{Input}
 \SetKwInOut{Output}{Output}
 \SetKwFunction{GatherSyndrome}{GatherSyndrome}
 \SetKwFunction{AddOp}{AddOp}
 \SetKwFunction{IsNotAlreadyGathered}{IsNotAlreadyGathered}
 \SetKwFunction{SolveTravelingSalesman}{SolveTravelingSalesman}
 \SetKwFunction{Combination}{Combination}
 \SetKwFunction{Neighbors}{Neighbors}
 \SetKwFunction{Num}{Num}
 \SetKwFunction{NextInMinPath}{NextInMinPath}
 \Input{$Dat$: Set of data qubits belonging to the stabilizer (typically 4 or 6)}
 \Input{$Anc$: Set of ancilla qubits around the stabilizer (typically 1 or 8)}
 \Input{$G$: Graph of qubits, describing qubits' neighbor relationships}
 \Output{Stabilizer circuit of shallowest depth}
 $MinCost = \_\_INT\_MAX\_\_$; $MinPath = None$\;
 /* Search for the smallest set of ancilla qubits which neighbor all data qubits. */\\
 \For{$n \in (1..\Num(Anc))$}{
   \For{$ancs \in \Combination(Anc,n)$}{
     \If{$d \in \Neighbors(ancs), \forall d \in Dat$}{
       /* Search for the shortest path involving data qubits. */\\
       $cost, path = \SolveTravelingSalesman(ancs)$\;
       \If{$cost < MinCost$}{
         $MinCost = cost$\;
         $MinPath = path$\;
       }
     }
   }
   \If{$MinPath \ne None$}{
     break\;
   }
 }
 /* Add operations along the path found. */\\
 \AddOp(Initializations(MinPath.ancillas))\;
 \For{$q \in MinPath$}{
   \If{$q \in Anc$}{
     \ForEach {$d \in [d | d \in Dat, d \in \Neighbors(q)]$}{
       \If{\IsNotAlreadyGathered($d$)}{
         \AddOp(\GatherSyndrome($d$, $q$))\;
       }
     }
   }
   /* SWAP gate which moves the ancilla qubit variable which holds error syndrome to the next hop in MinPath. */\\
   \If{$q.next \ne Null$}{
     \AddOp(SWAP($q$, $q.next$))\;
   }
   /* SWAP gate which returns the data qubit variable which was swapped to the previous hop to the original positions. */\\
   \If{$q \in Dat$}{
     \AddOp(SWAP($q.prev$, $q$))\;
   }
 }
 /* Measure the ancilla qubit which holds error syndrome. */\\
 \AddOp(Measurement($q$))\;

\end{algorithm*}

Algorithm~\ref{alg:sol:stabilizer_circuit}, discussed in Section \ref{subsec:stab}, composes our stabilizer circuits for the individual superunits.
\IncMargin{1em}

\begin{algorithm*}
\caption{Scheduling algorithm}
\label{alg:sol:scheduling_algorithm}
\SetKwInput{Input}{Input}
\SetKwInOut{Output}{Output}
\SetKwFunction{Sort}{Sort}
\SetKwFunction{Head}{Head}
\SetKwFunction{AfterHead}{AfterHead}
 \SetKwFunction{Schedule}{Schedule}
 \SetKwFunction{Any}{Any}
\SetKwFunction{NotOverCurrentStep}{NotOver\_CurrentStep}
\Input{$SC$: The set of stabilizer circuits}
\Input{$MaxStep$: The number of time steps to output}
 \Output{$WholeCircuit$}
/* Sort stabilizers in order of depth, longest first. If they tie, stabilizers on top-left of the lattice have priority. */\\
$sortedSC = \Sort(SC)$\;
$deepest = \Head(sortedSC)$\;
$afterHead = \AfterHead(sortedSC)$\;

$WholeCircuit = NULL$\;
$wholeCeil = 0$\;
\While{$wholeCeil \leq MaxStep$}{ 
 /* wholeCeil is the step when the deepest stabilizer last scheduled finishes. */\\
 $deepest.ceil = wholeCeil = \Schedule(deepest,\textrm{ }WholeCircuit$)\; 
 \ForEach{$s \in afterHead$}{
  /* schedule every stabilizer once */\\
 $s.ceil = \Schedule(s)$\; 
 }
 /* loop until every s.Ceil $>$ wholeCeil */\;
\While{$\Any(s.ceil \leq wholeCeil| s \in afterHead)$}{ 
  \ForEach{$s \in afterHead$}{
   \If{$s.ceil \leq wholeCeil$}{
    $s.ceil = \Schedule(s,\textrm{ }WholeCircuit)$\; 
   }
  }
 }
}

\end{algorithm*}

Algorithm~\ref{alg:sol:scheduling_algorithm}, discussed in Section \ref{subsec:whole}, schedules the stabilizer circuits into a whole circuit for the chip.

\subsection{Solving conflicts in scheduling}
\label{sec:appendix:conflict}
\begin{figure}[t]
 \begin{center}
  (a)\includegraphics[width=7cm]{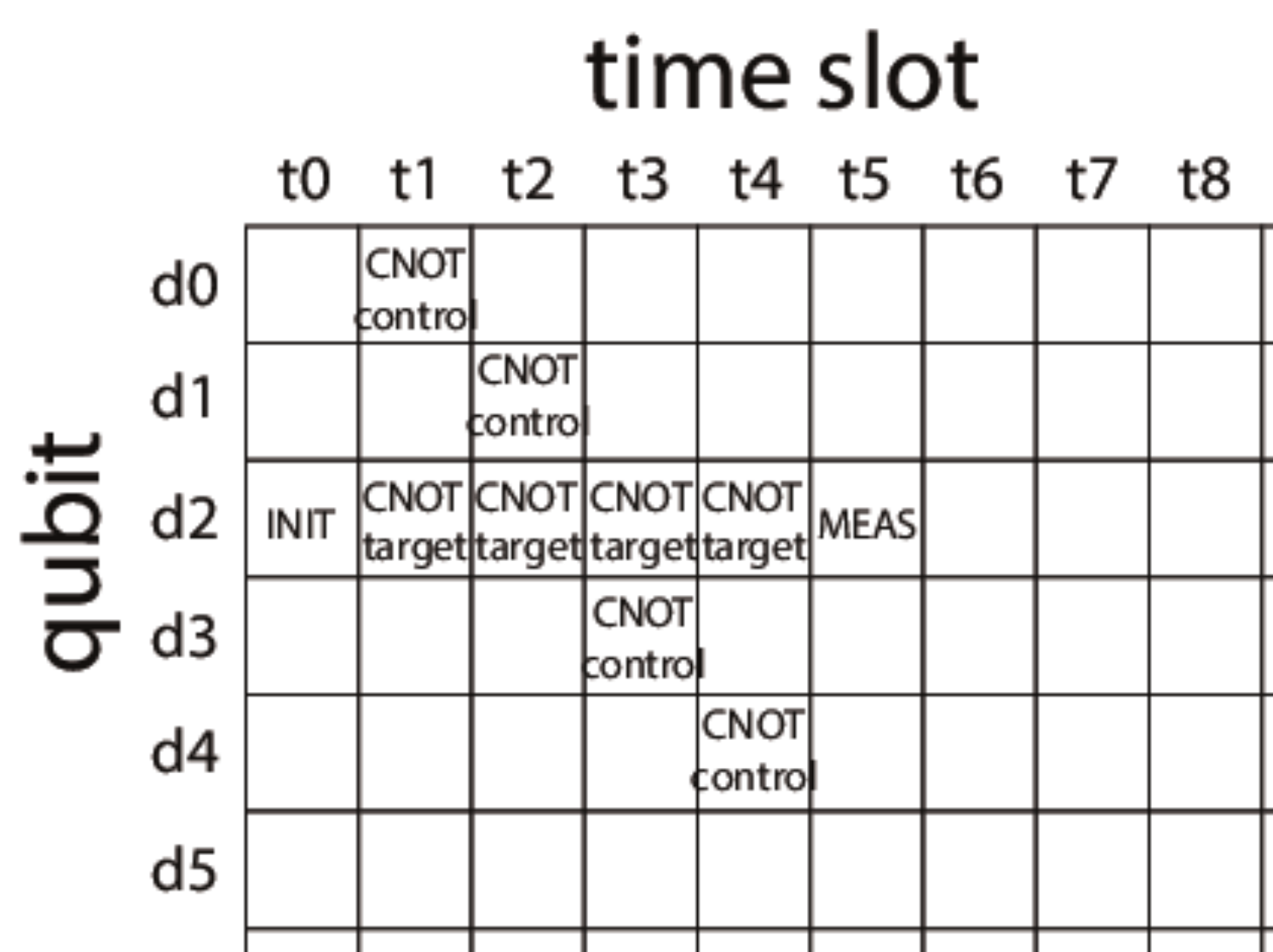}
  (b)\includegraphics[width=7cm]{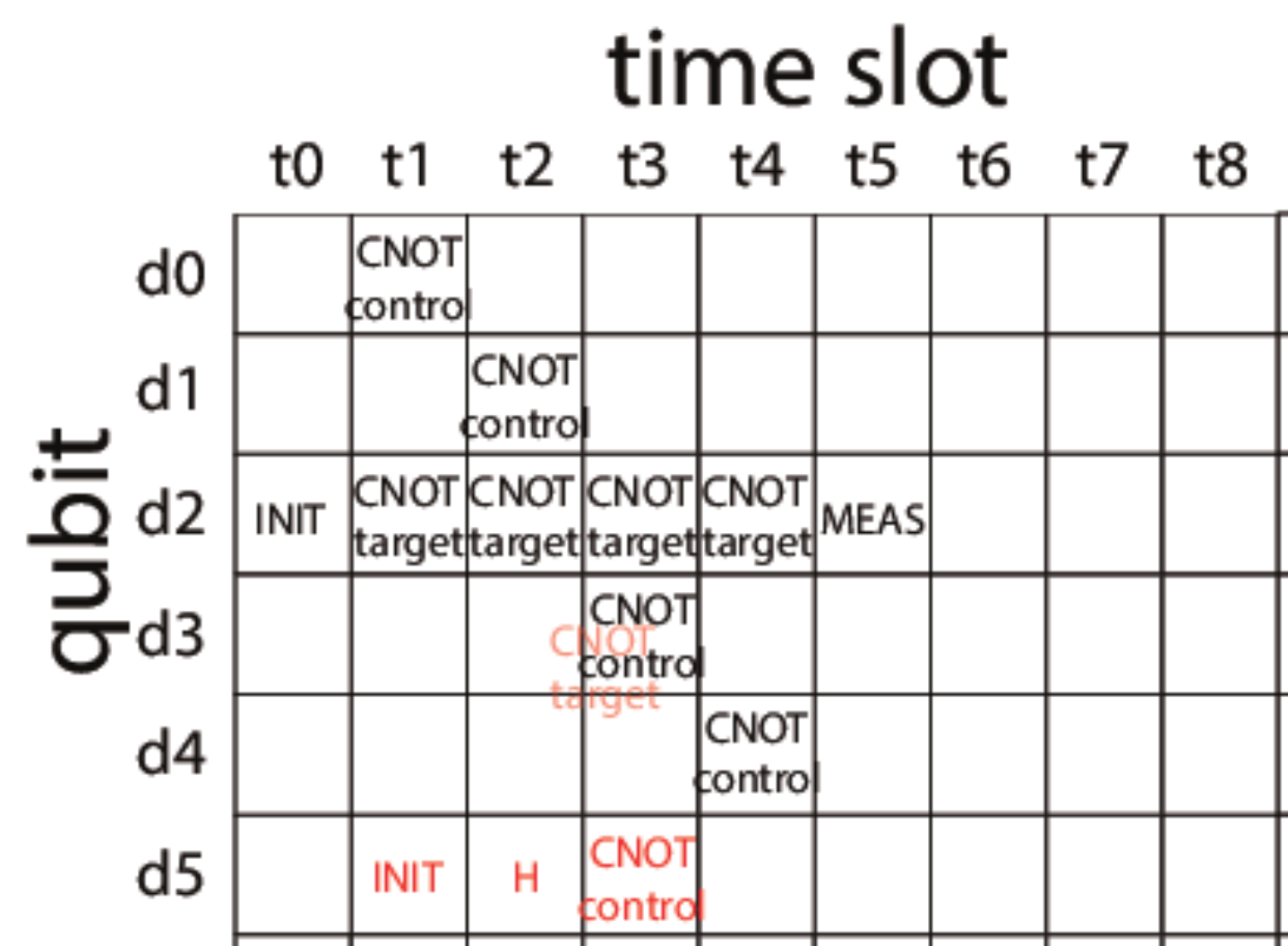}\\
  (c)\includegraphics[width=7cm]{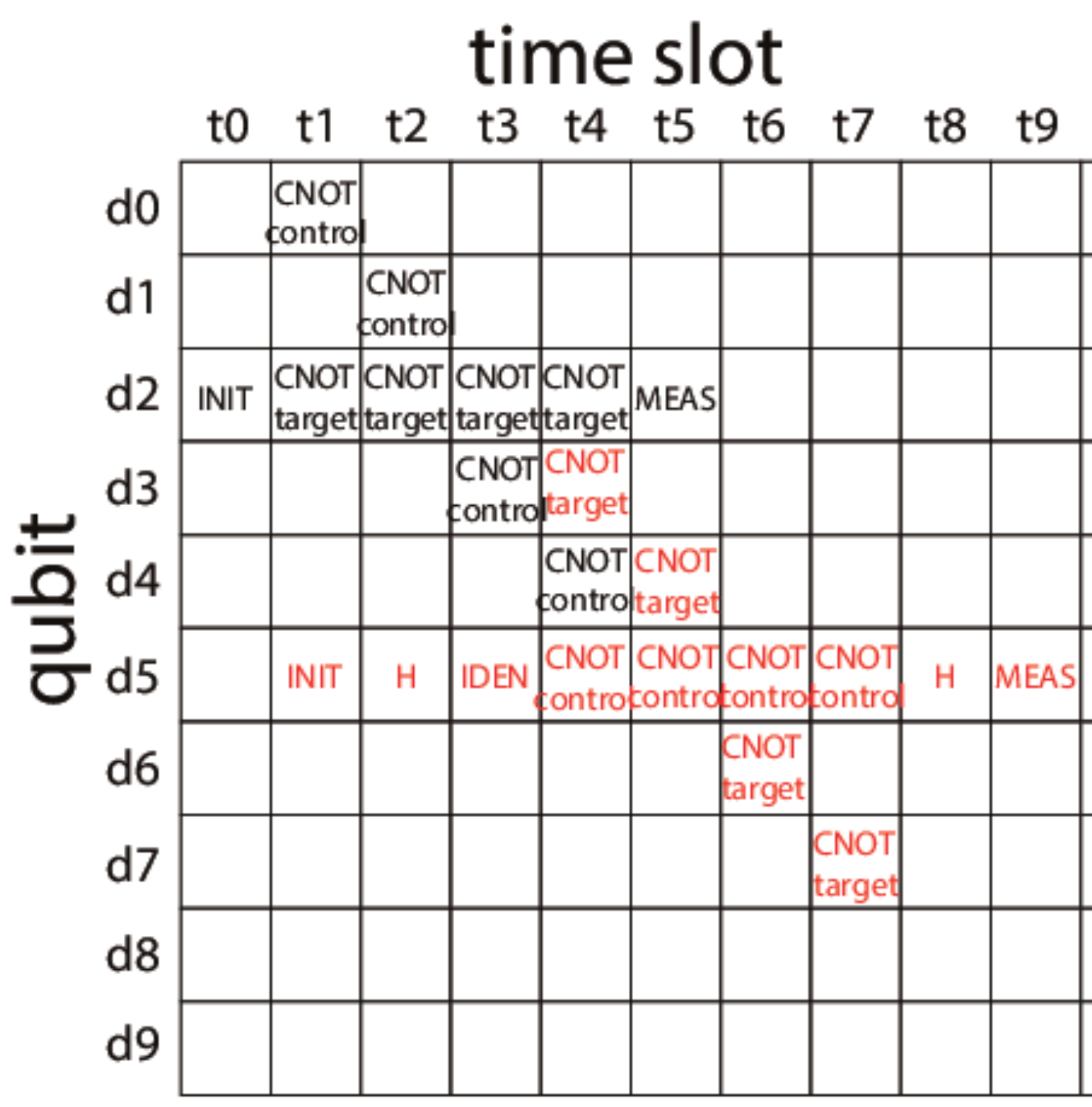}
  (d)\includegraphics[width=7cm]{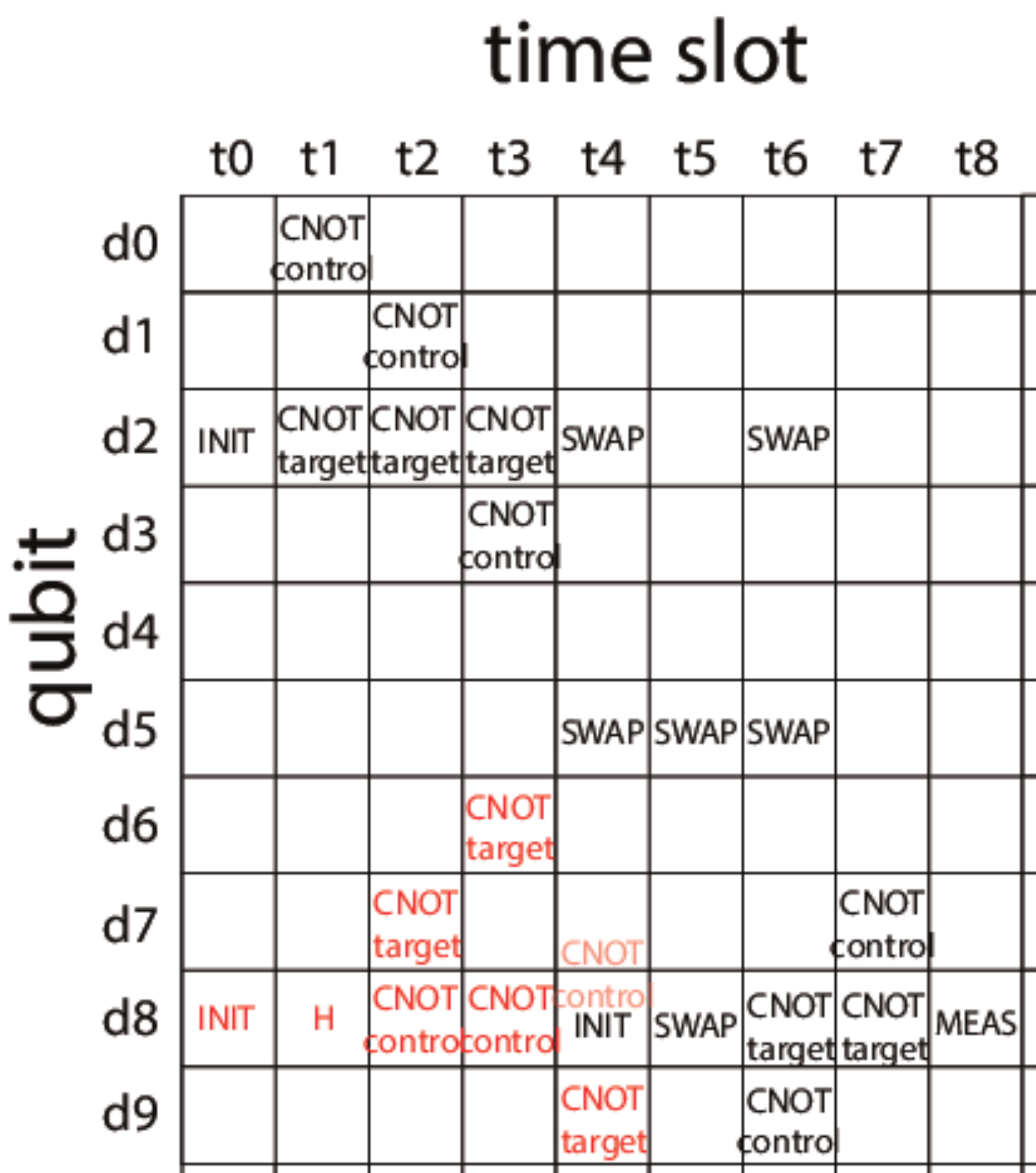}\\
  \caption{
  (a) Scheduling chart for the six 
  gates comprising a Z stabilizer circuit.
  Each box is called a slot and NULL slots are vacant.
  (b) Example of resource allocation conflict occurring between two stabilizers, with the higher priority one in black and the
  lower priority one in red.
  d3-t3 is allocated to the black stabilizer.
  The red stabilizer cannot lock the slot d3-t3, as indicated by the light-colored CNOT target.
  (c) Solution to the contention for a data qubit. The red stabilizer waits for the slot of d3-t3 to become unlocked.
  (d) The slot of d8-t4 is allocated to the black stabilizer. The black SWAP gate already locks the slot of d8-t4 and the red CNOT cannot lock it.
  }
  \label{fig:sol:slots1}
 \end{center}
\end{figure}

\begin{figure}[t]
 \begin{center}
  (e)\includegraphics[width=14cm]{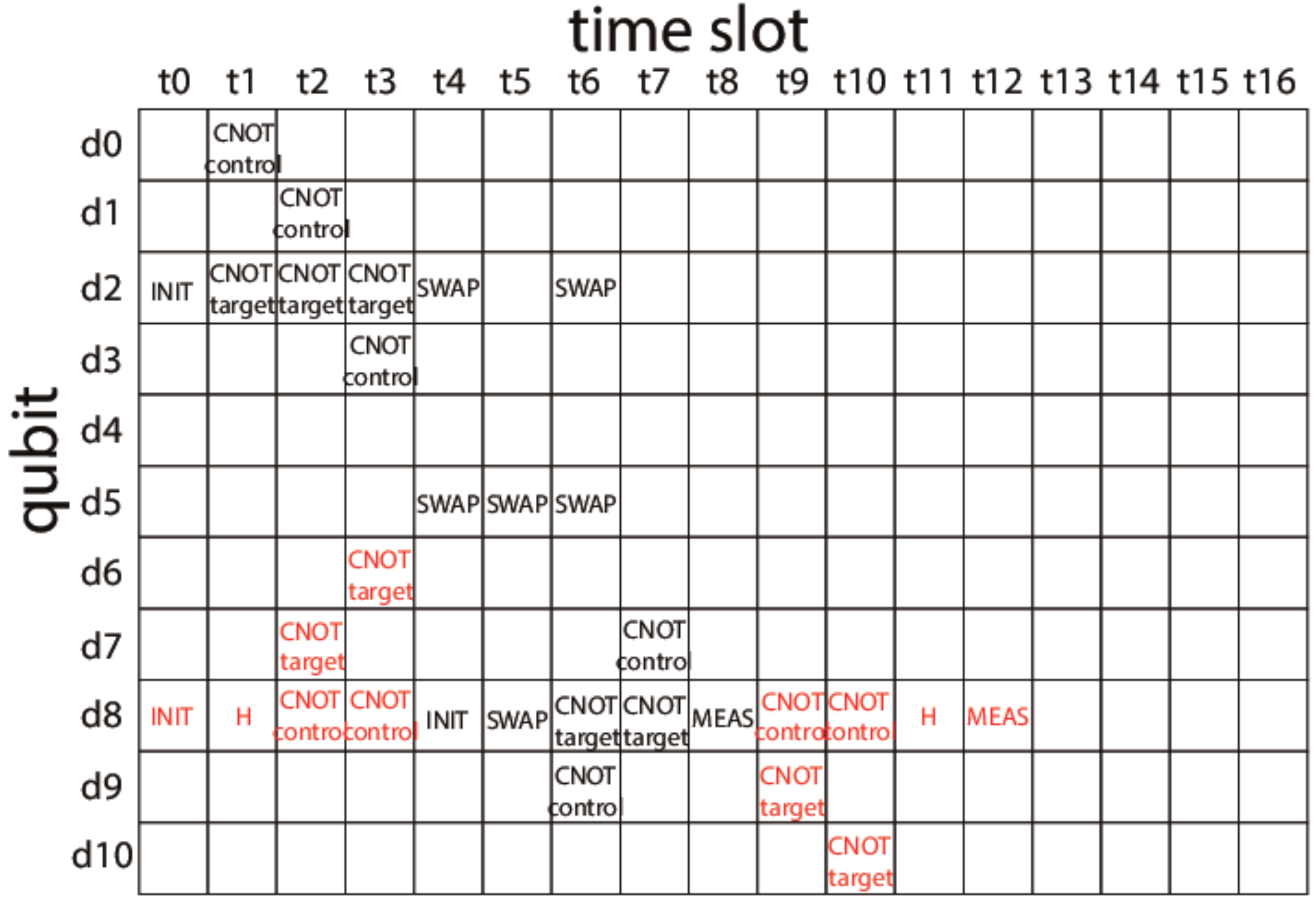}
  (f)\includegraphics[width=14cm]{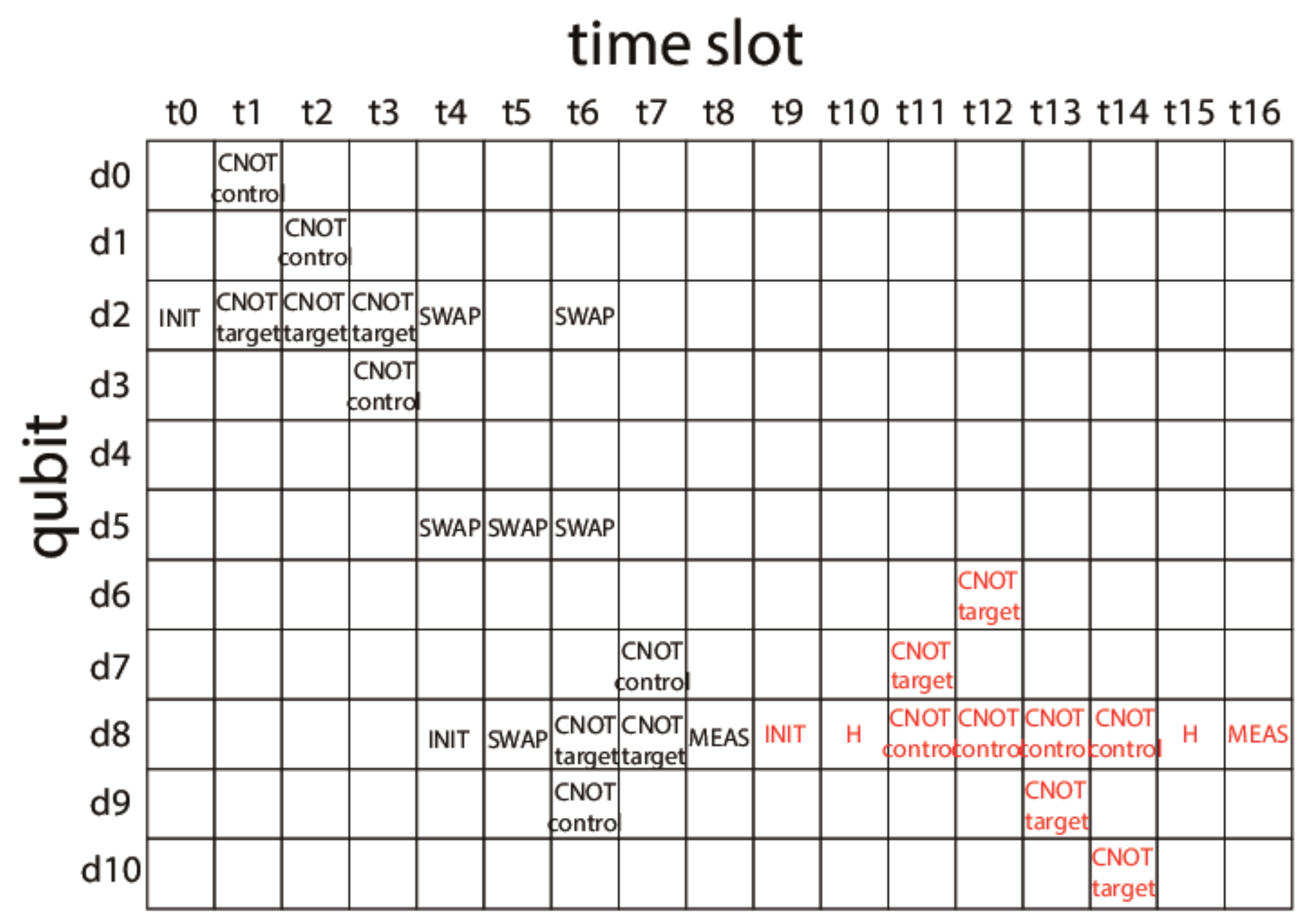}
  \caption{
  (e) Attempting to solve the competition for a syndrome qubit by waiting. The red stabilizer is split and the former half of its error syndromes are deleted by the initialization gate in d8-t4. This is invalid.
  (f) Solution to the competition for a syndrome qubit. The red gates are all rescheduled after the black stabilizer.
  }
  \label{fig:sol:slots2}
 \end{center}
\end{figure}

Figures~\ref{fig:sol:slots1} and~\ref{fig:sol:slots2} illustrate various conflicts that occur during scheduling and our solutions.
Our scheduling is implemented to allocate ``slots'' to gates of stabilizer circuits, as shown in Figure \ref{fig:sol:slots1} (a).
Each qubit on each step has a slot and
only one gate can operate in a slot.
When a gate is set in a slot, the slot gets locked.
When a swap gate is set in the slot of a data qubit, the data qubit is locked
until the data qubit variable returns to the original data qubit device.
If conflicts occur, we add identity gates
as shown in Figure \ref{fig:sol:slots1} (b) and Figure \ref{fig:sol:slots1} (c).
This method does not work for conflict on syndrome qubits.
This is because a data qubit may be unlocked after a single time step but a syndrome qubit
may not be unlocked for several steps, as shown in Figure \ref{fig:sol:slots1} (d).
Any sequence of gates on syndrome qubits in a stabilizer starts with an initialization gate,
which removes the error syndromes which have already been gathered by the syndrome qubit,
as illustrated in d8-t4 in Figure \ref{fig:sol:slots1} (d).
If the stabilizer currently being scheduled (red gates) waits for the syndrome qubit to be unlocked,
the initialization gate deletes the syndrome qubit variable with some error syndromes of
the stabilizer as shown at d8-t4 in Figure \ref{fig:sol:slots2} (e).
To avoid this problem, the currently scheduled stabilizer is completely rescheduled
after the previously scheduled stabilizer finishes, as shown in Figure \ref{fig:sol:slots2} (f).

\subsection{Change of the matching nest}
\label{sec:appendix:matching}
\begin{figure}[t]
 \begin{center}
  \includegraphics[width=15cm]{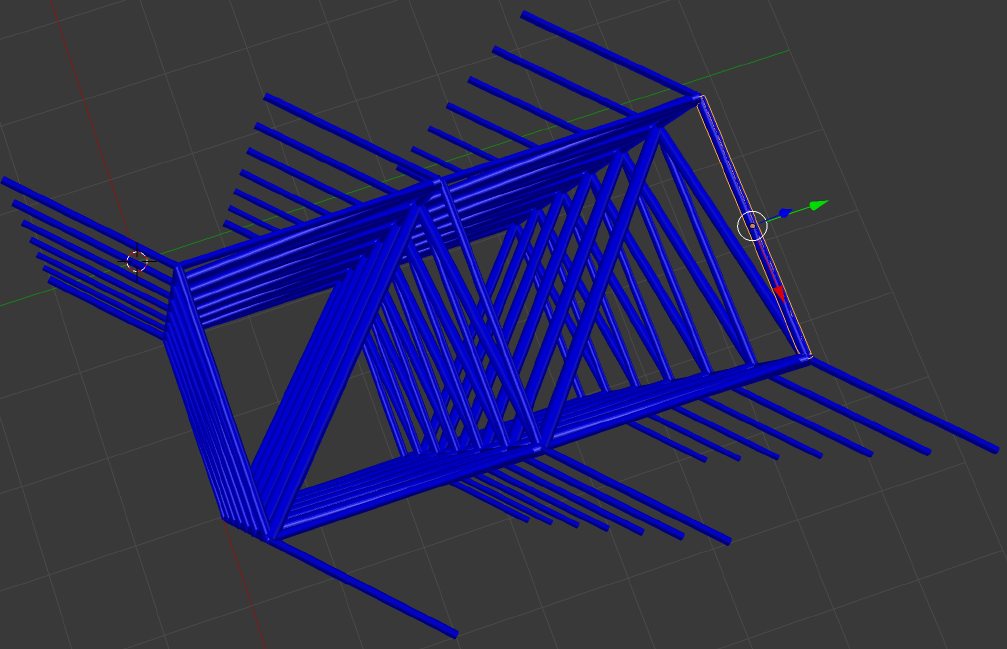}
  \caption{The matching nest for the distance 3 surface code.
  A stabilizer measurement corresponds to a vertex and a qubit error corresponds to a edge.
  The ends of the nest correspond to the boundaries of the lattice, hence they do not have measurement values.
  If an error occurs on a data qubit, the stabilizers the data qubit is stabilized
  by will get a different measurement result than the prior round.
  Then the corresponding vertices create and hold nodes for the minimum weight perfect matching.
  Lines for the minimum weight perfect matching are created by Dijkstra's algorithm
  on this nest, searching from a node for other nodes~\cite{Dijkstra1959}.
  The weight of a line is defined by the weights of edges that the search goes through to create the line,
  which corresponds to the possibility of the errors which result in the line.
  }
  \label{fig:sc:normal_nest}
 \end{center}
\end{figure}

A \textit{nest} is used to prepare a network for minimum weight perfect matching.
Figure \ref{fig:sc:normal_nest} depicts a nest of a perfect lattice
of the surface code, output by the Autotune Software created by Fowler et al.~\cite{Fowler:2012autotune}.
Each vertex of the nest corresponds to a stabilizer value
and each edge corresponds to a possible error.
Edges which do not have two vertices are at the boundaries of the lattice.
As time advances, the nest expands along the Z axis, creating new vertices and edges when measuring stabilizers.
A stabilizer which measures a different eigenvalue from the last stabilizer measurement creates and holds a node on the corresponding vertex.
Because an ancilla error (or measurement error) will happen only once, three cycles with an error on the middle cycle would produce the eigenvalue sequence +1, -1, +1. The two transitions will be recorded in the nest as two nodes. A data qubit error results in errors on two neighboring stabilizers, so that an error after the first measurement would give the sequence +1, -1, -1 in two separate places (or only one if the qubit is on a boundary). In this case, the two transitions result in the creation of two horizontal neighbor nodes in the nest. The matching algorithm will match the two vertical nodes of a stabilizer error or the two horizontal nodes of a qubit error. 
Lines for the matching between nodes are created with Dijkstra's algorithm on the nest~\cite{Dijkstra1959}.
The weight of a line is given by the sum of weights of the edges which compose the line.
Minimum weight perfect matching based on those weights selects the most probable physical errors,
therefore it works as error correction.

\begin{figure}[t]
 \begin{center}
  \includegraphics[width=15cm]{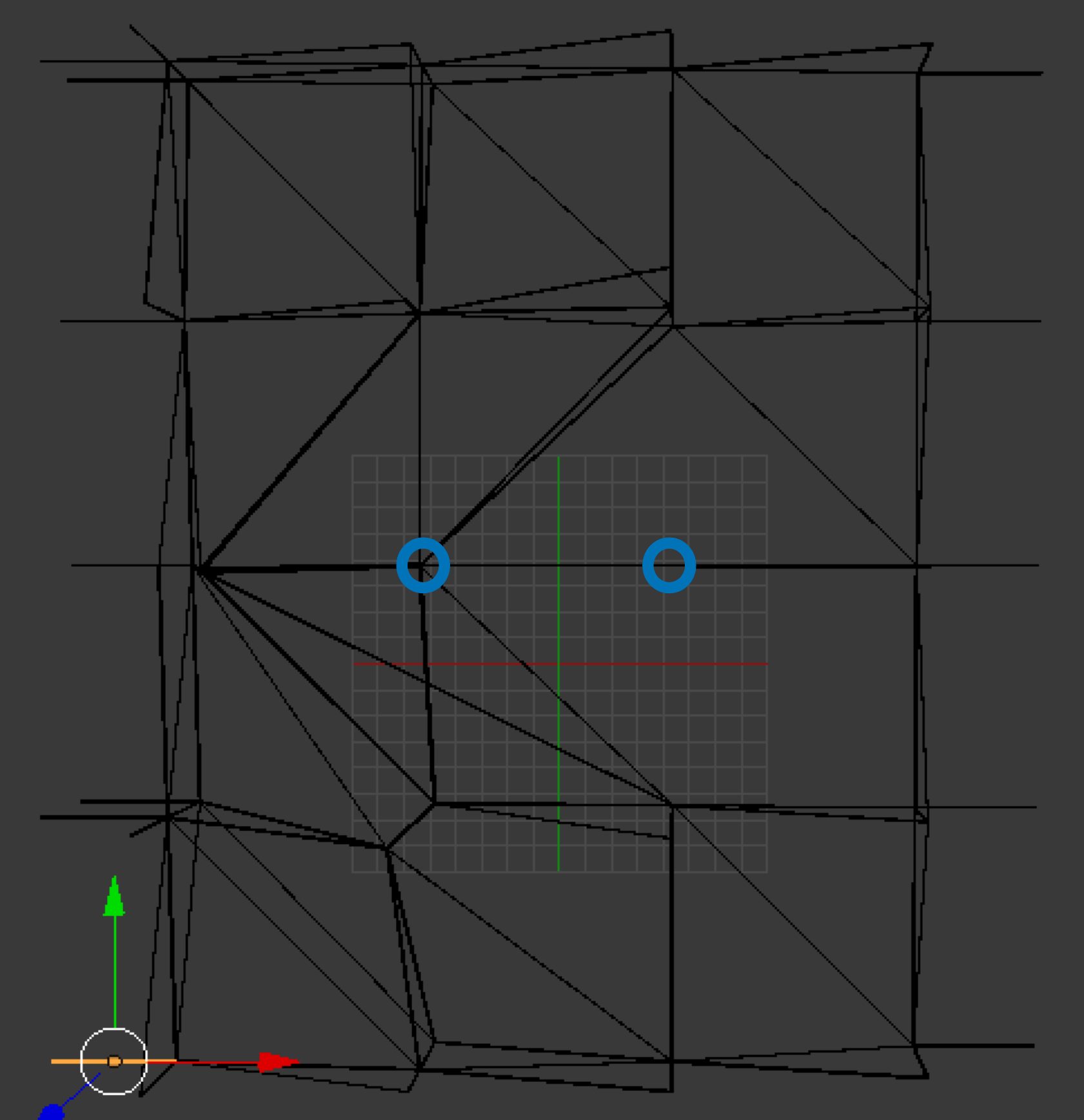}
  \caption{A top view of a visualization of an asynchronous nest. Each vertex describes a syndrome measurement and each edge connects two syndrome measurements which might be changed by the same error. The two blue circles indicate the positions where two unit stabilizers originally existed and now are merged into a superunit stabilizer placed in the left blue circle.}
  \label{fig:sim:nest-topview}
 \end{center} 
\end{figure}

\begin{figure}[t]
 \begin{center}
  \includegraphics[width=15cm]{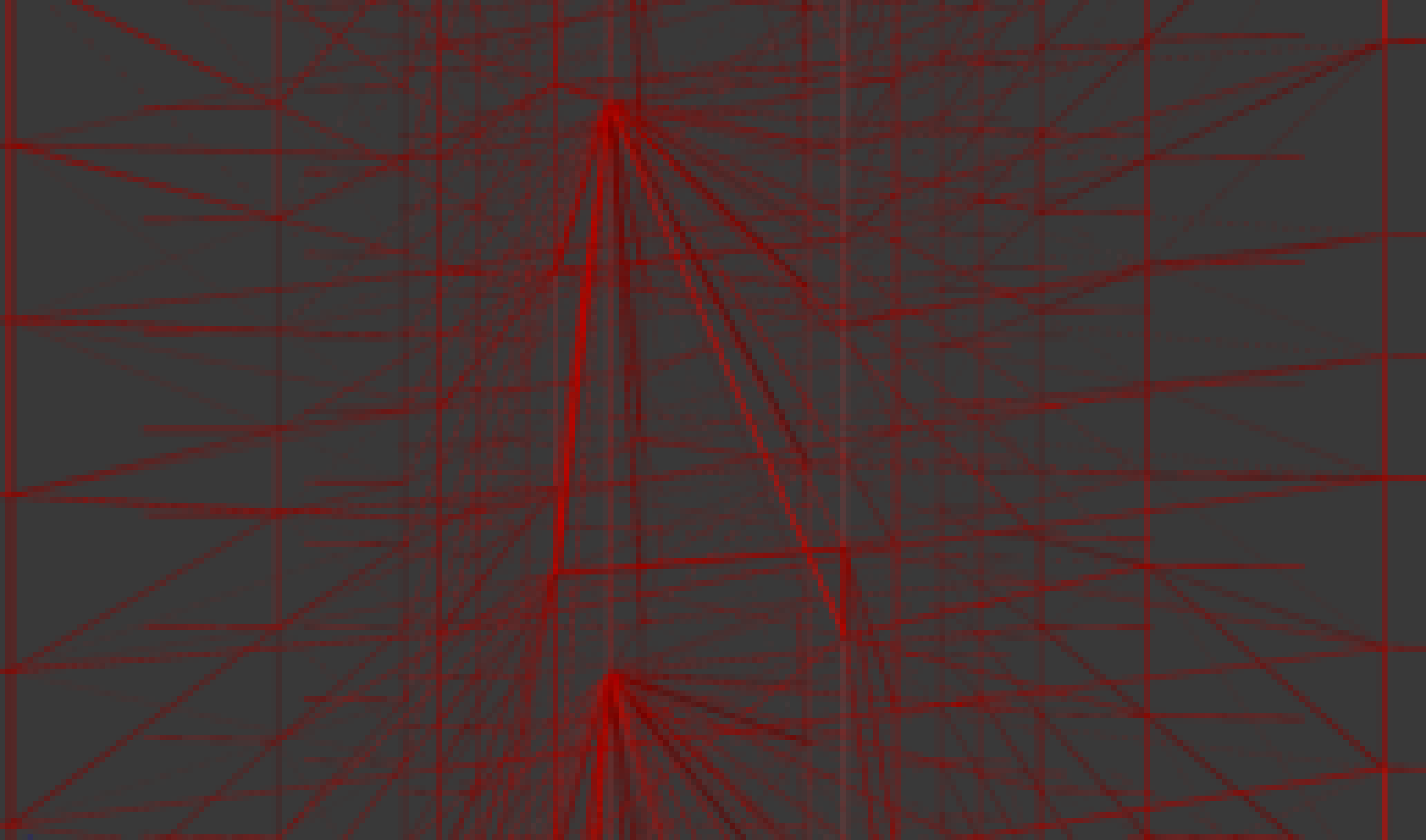}
  \caption{A front-view of a visualization of an asynchronous nest. The diameter of each edge
  is proportional to the probability of observing detection events at the endpoints of the edge.
  }
  \label{fig:sim:nest-frontview}
 \end{center}
\end{figure}

Irregular stabilizer circuits degrade the parallelism of stabilizer measurements of the whole circuit so that the surface code on a defective lattice has irregular error matching nests, as shown in Figures \ref{fig:sim:nest-topview} and \ref{fig:sim:nest-frontview}.
These figures are output by Autotune during simulation of our surface code on a defective lattice~\cite{Fowler:2012autotune}.
We can see the irregularity of the nest.
A superunit stabilizer is measured in a longer cycle than normal stabilizers and
the vertex of a superunit stabilizer has many edges, some of which are thick.
This thickness is proportional to the error probability.
The network on which Blossom V runs is generated on this adapted nest to
find the best possible solution.

\subsection{System architecture}
\begin{figure}[t]
 \begin{center}
 \includegraphics[width=400pt]{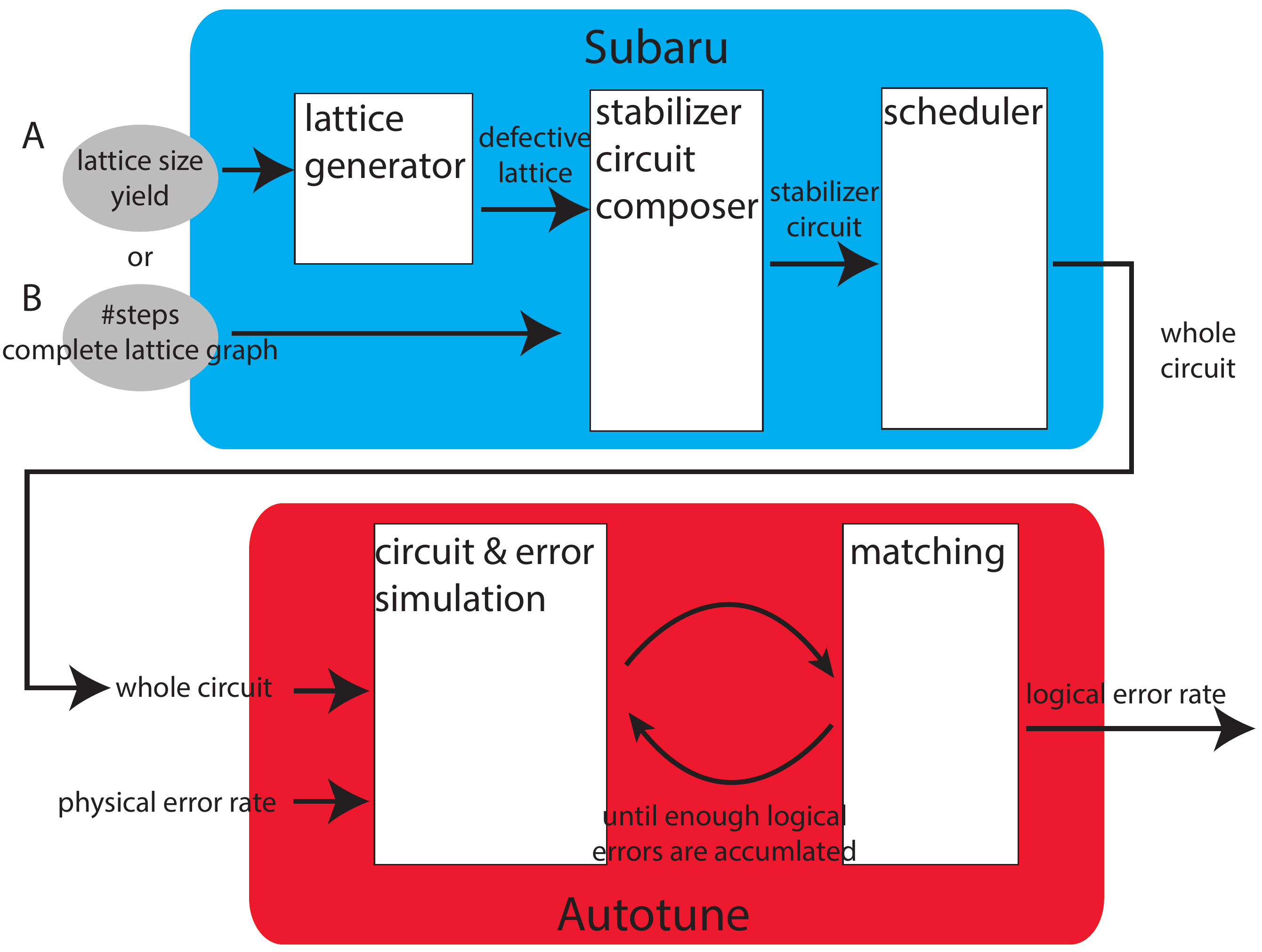}
  \caption{Simulation system design.
  Subaru can have one of two mutually exclusive inputs -- a pair of yield and a lattice size (labeled ``A'') or a full description of a lattice (labeled ``B'').
  With the former input, it randomly generates a defective lattice, then builds circuits to suit.
  With the latter input, it builds circuits for the provided defective lattice. It outputs a whole circuit of the requested number of steps.
  The circuit and physical error rate are input into Autotune, and Autotune outputs the logical error rate corresponding to the inputs.}
  \label{fig:sim:system_design}
 \end{center}
\end{figure}

Figure \ref{fig:sim:system_design} shows the major software components for compiling a circuit for the surface code and simulating its behavior on a defective lattice.
Subaru produces a whole circuit for a defective lattice and Autotune simulates the operation of the surface code along the circuit \cite{Fowler:2012autotune}.
Subaru performs the tasks described in section \ref{sec:solution}.
Subaru can have alternative inputs -- yield or a lattice.
The yield is the probability of fabricating qubits which work properly.
Instead of a yield, a complete lattice can be input into Subaru.
This enables us to investigate particular conditions using hand-constructed lattices.

\section{Supplemental graphs}
This appendix shows supplemental graphs to visualize the effect of culling
and raw data.
\subsection{Graphs to compare culled pools}
\label{subsec:overlapped_graphs}
\begin{landscape}
\begin{figure}[t]
 \begin{center}
  \includegraphics[width=7cm]{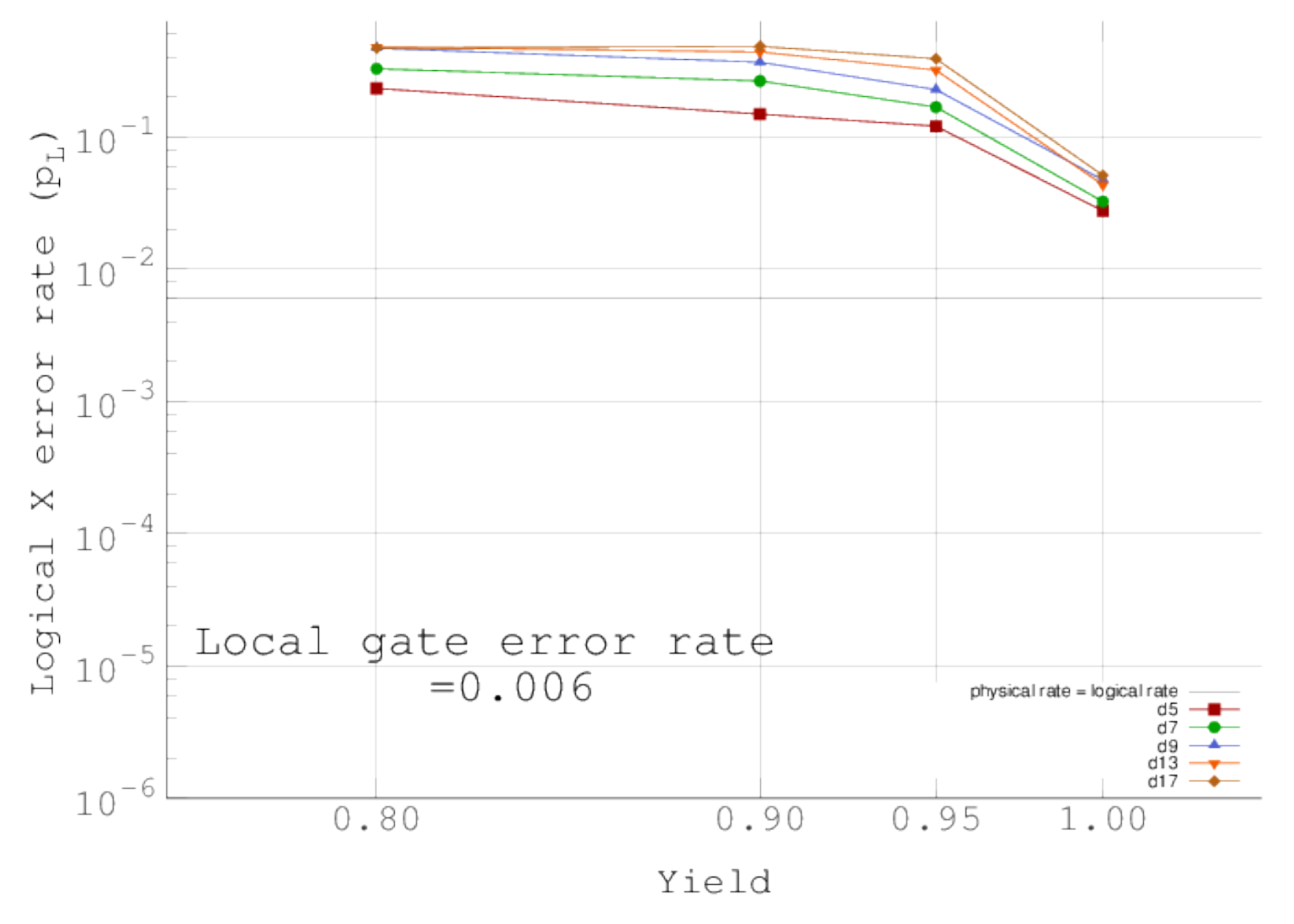}
  \includegraphics[width=7cm]{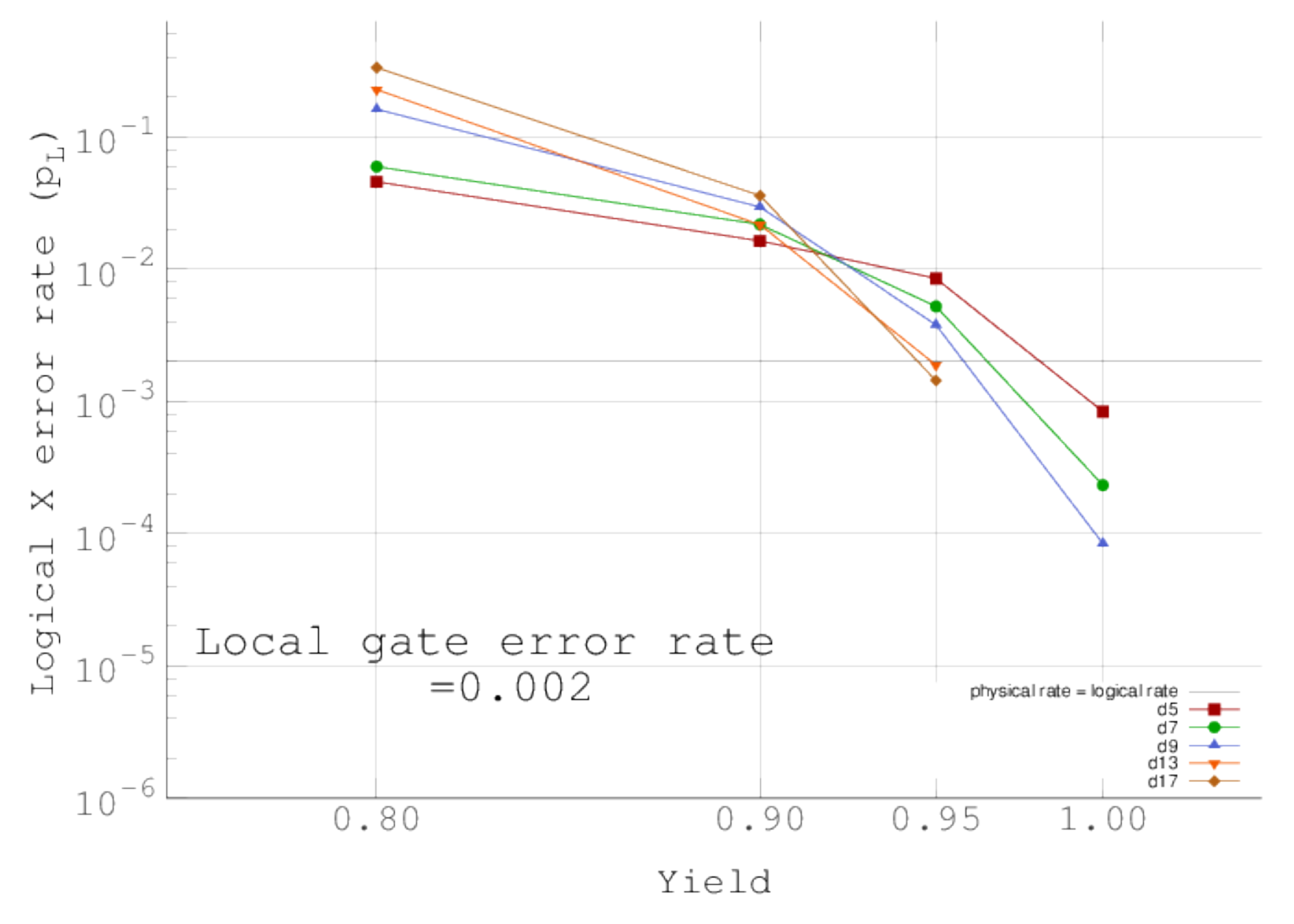}
  \includegraphics[width=7cm]{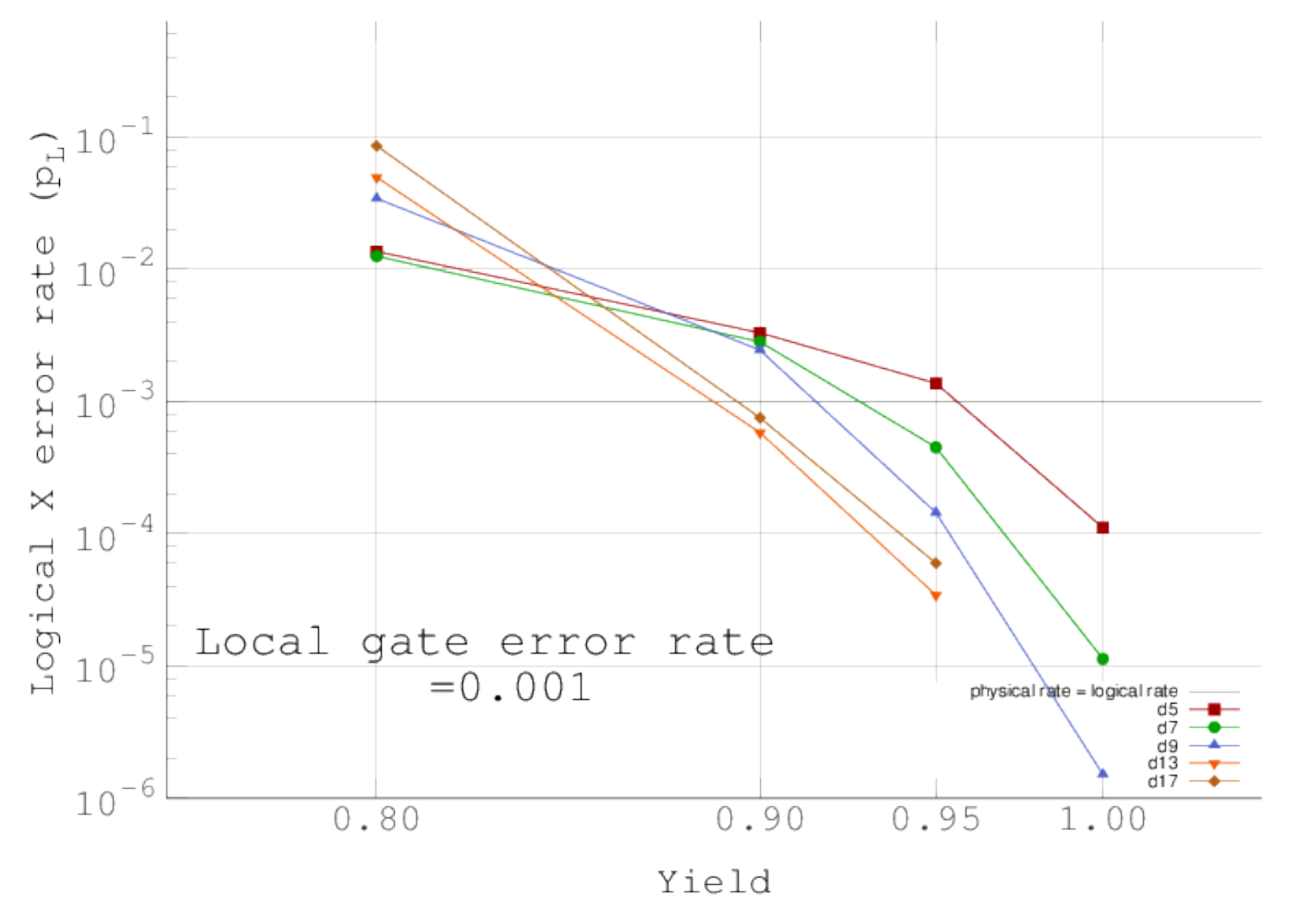}\\
  \includegraphics[width=7cm]{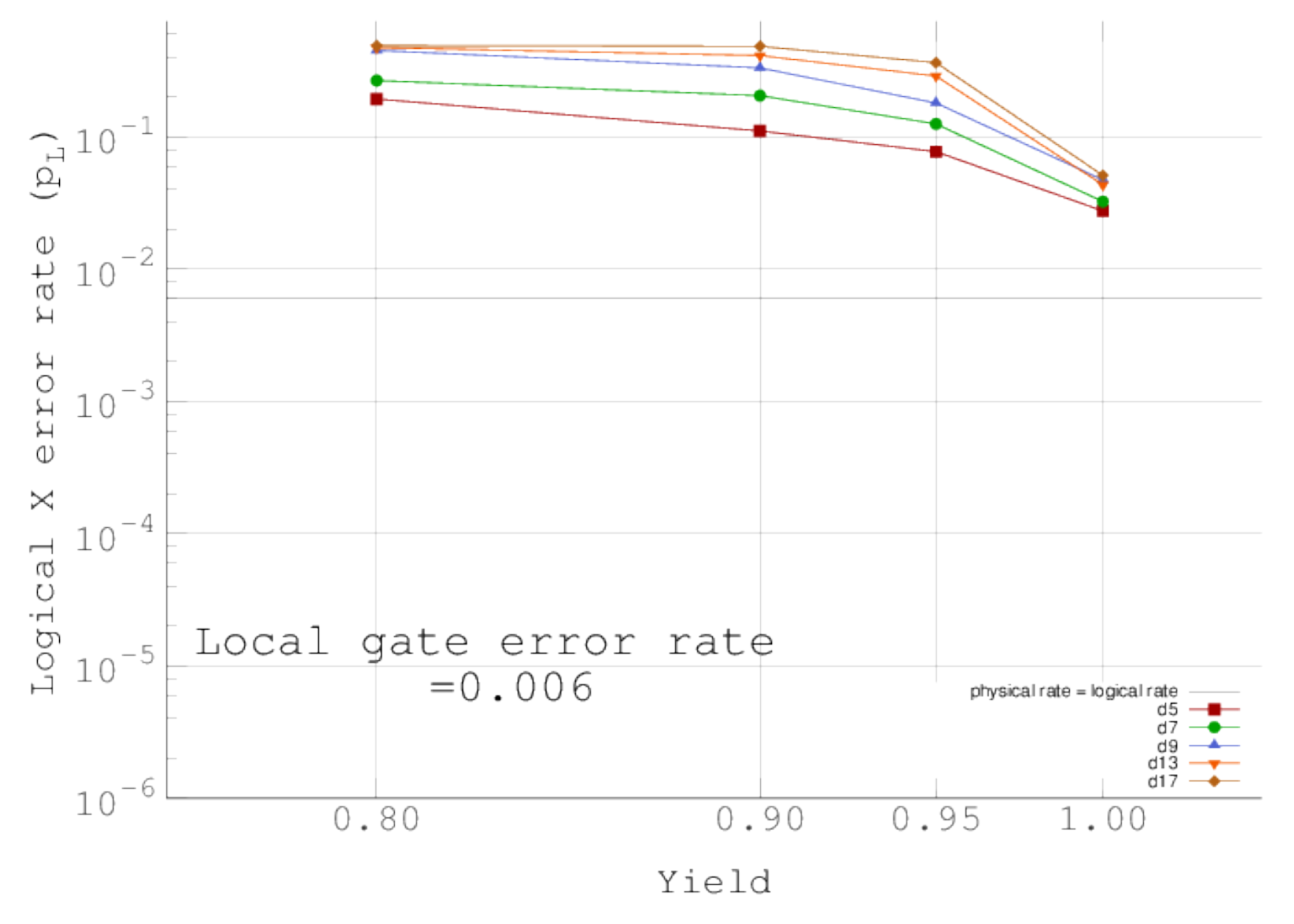}
  \includegraphics[width=7cm]{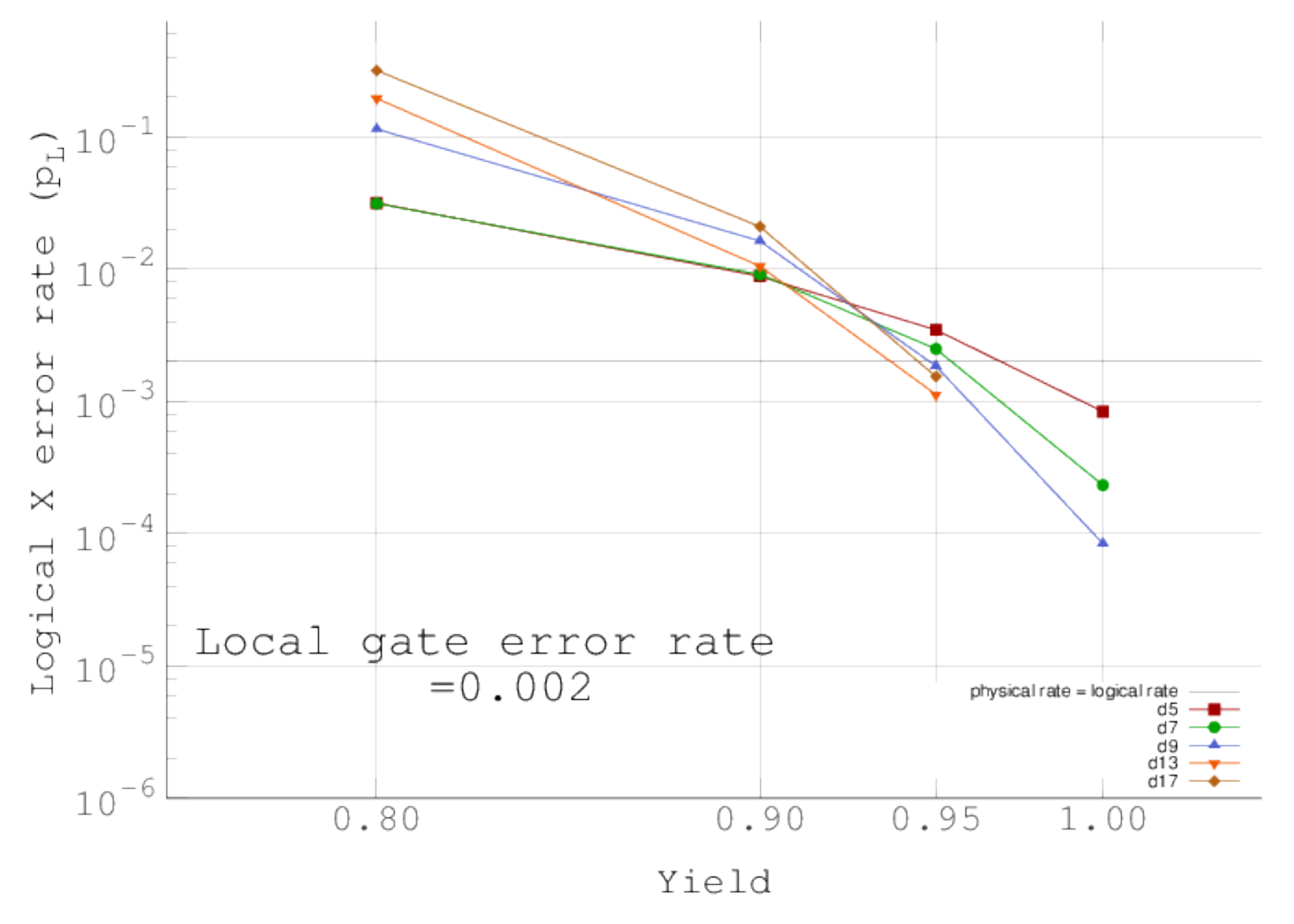}
  \includegraphics[width=7cm]{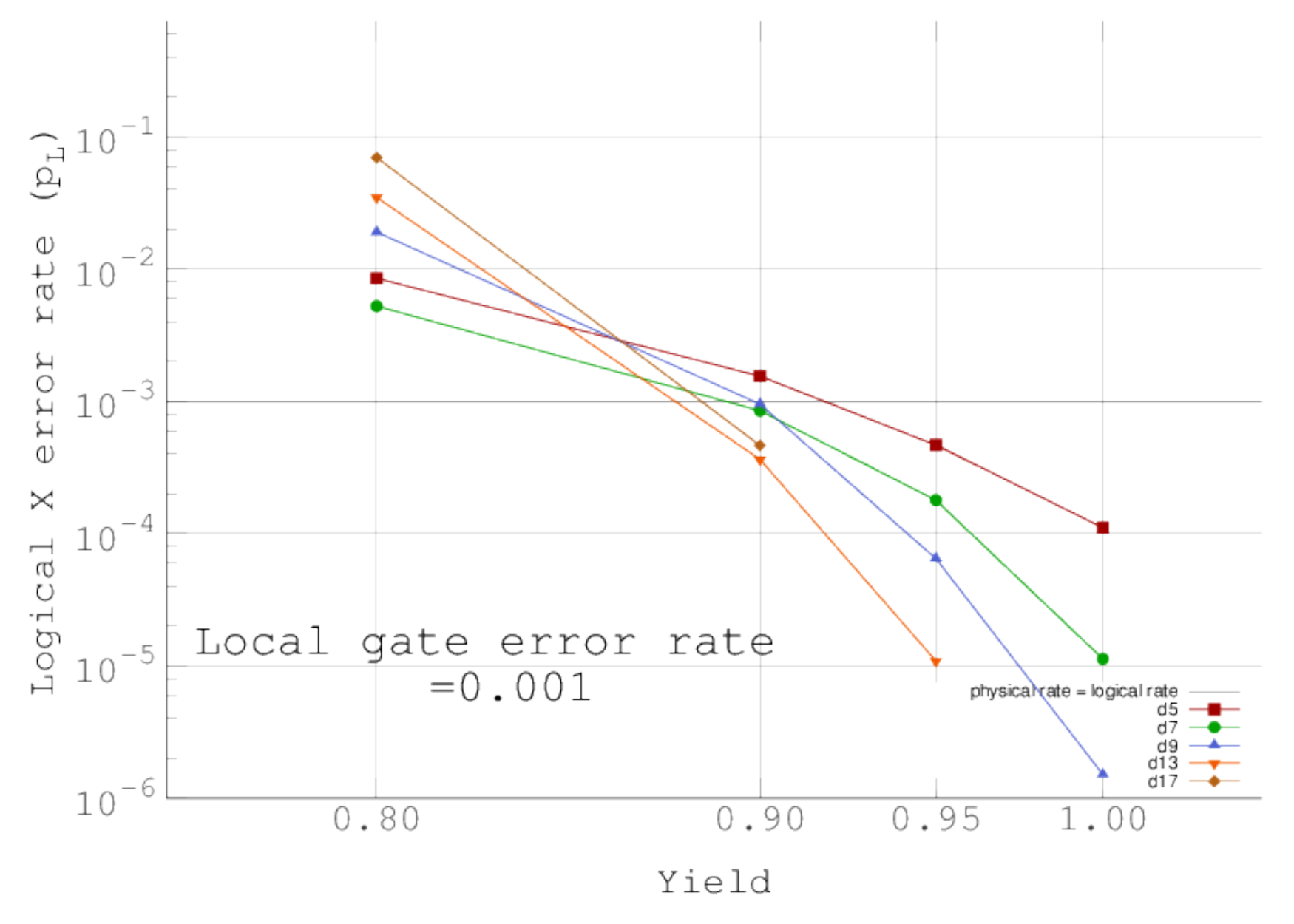}\\
  \includegraphics[width=7cm]{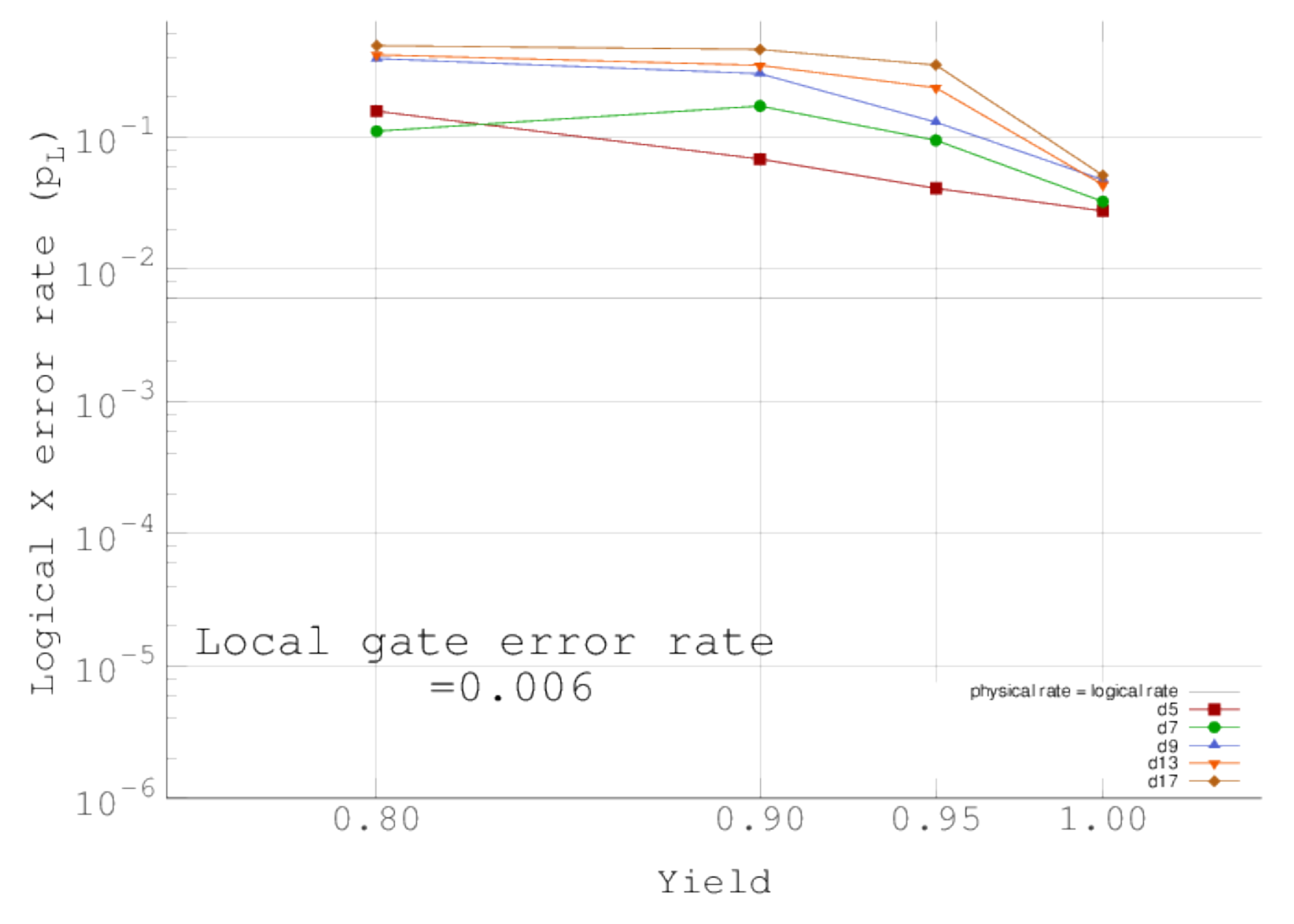}
  \includegraphics[width=7cm]{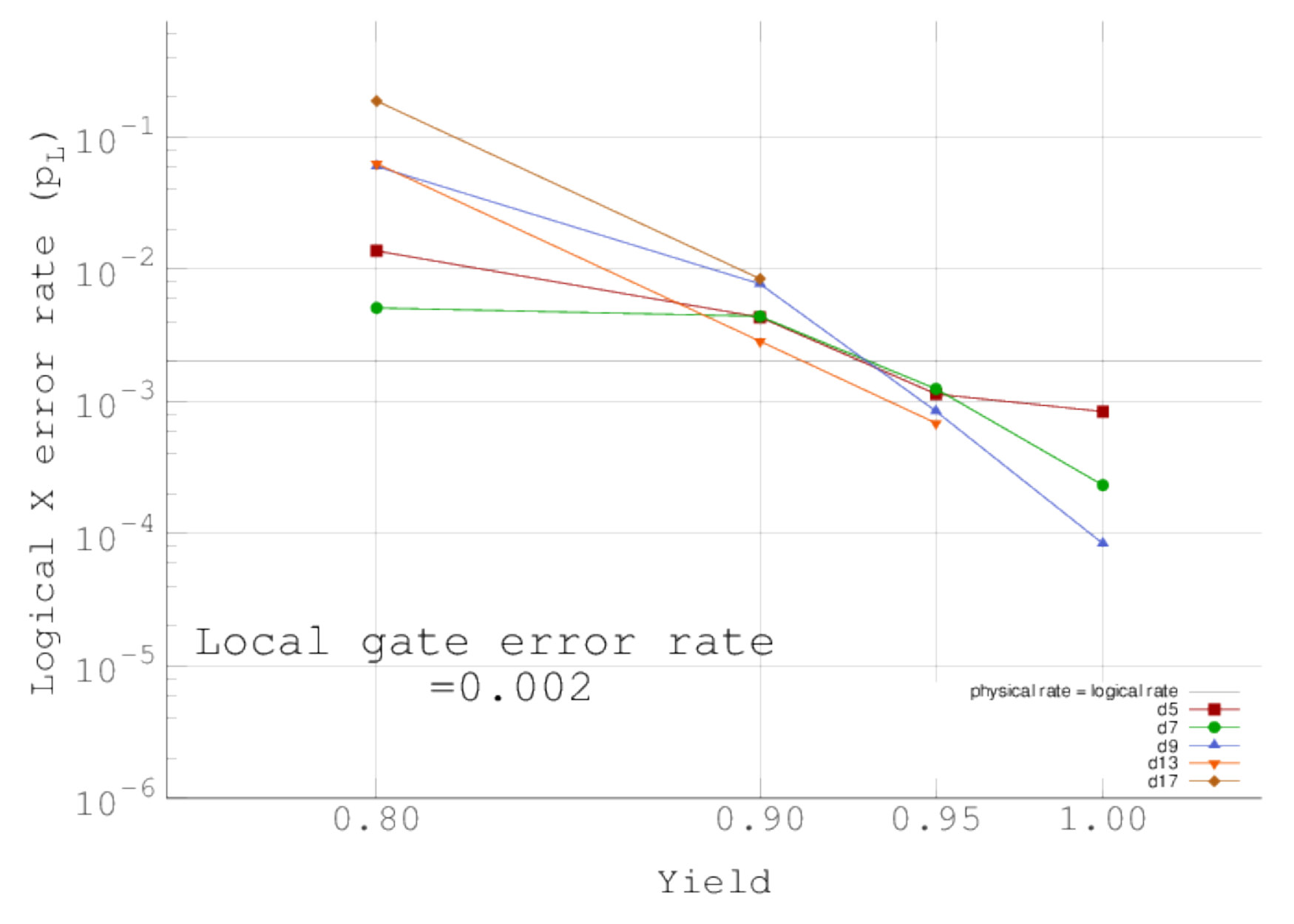}
  \includegraphics[width=7cm]{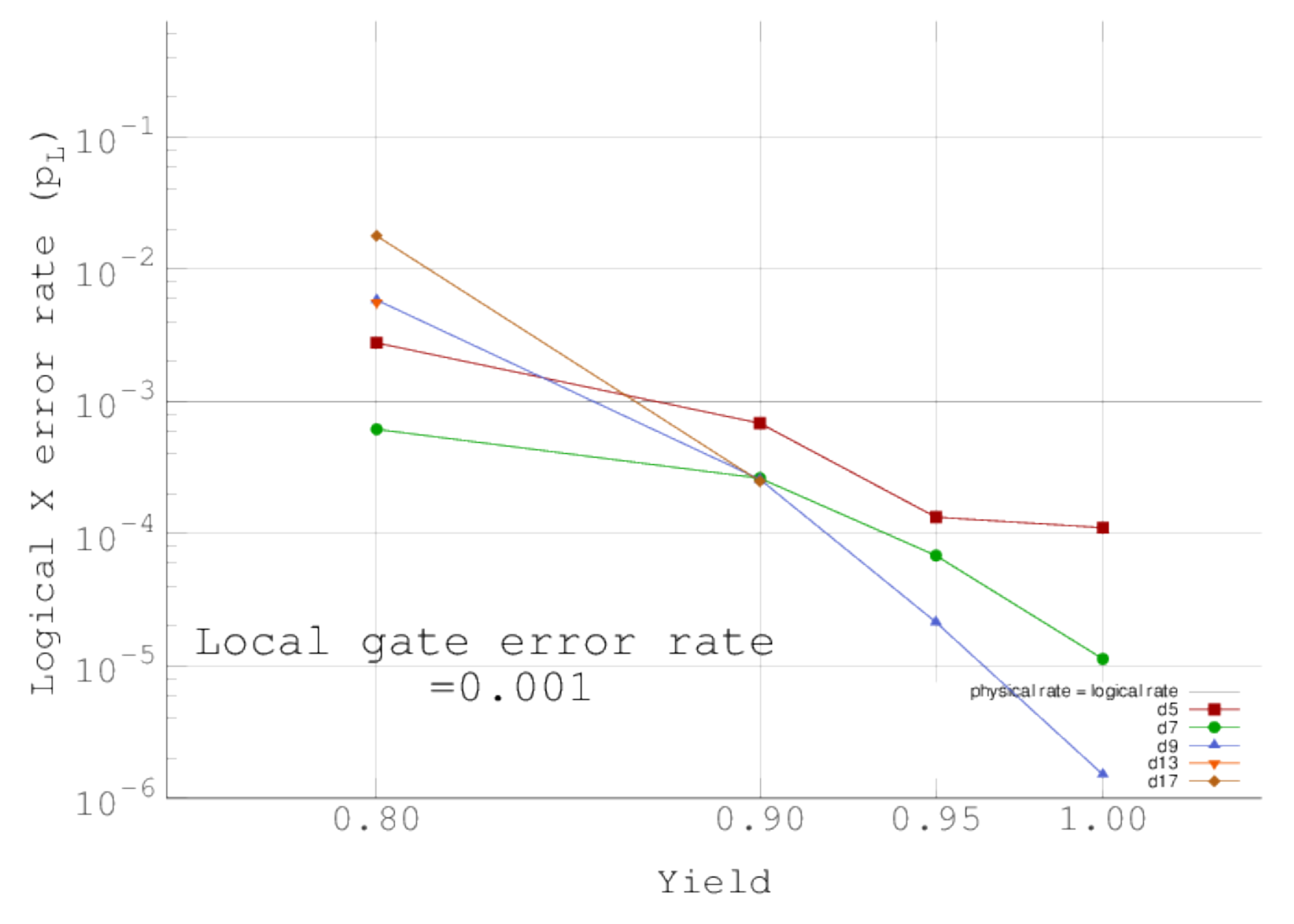}
  \caption{shows the graphs between yields and logical error rates at specific physical error rates.
The top row is of no-culled graphs,
the middle row is of 50\%-culled graphs
and the bottom row is of 90\%-culled graphs.
The left column is for $p=0.6\%$, 
the middle column is for $p=0.2\%$ 
and the right column is for $p=0.1\%$.
}
  \label{fig:cull-comp}
 \end{center}
\end{figure}
\end{landscape}

Figure \ref{fig:cull-comp}
shows the graphs between yields and logical error rates at specific physical error rates.

\subsection{Scatterplots of randomly defective lattices}
\label{subsec:scatterplots}
\begin{figure}[t]
 \begin{center}
  \includegraphics[width=18cm]{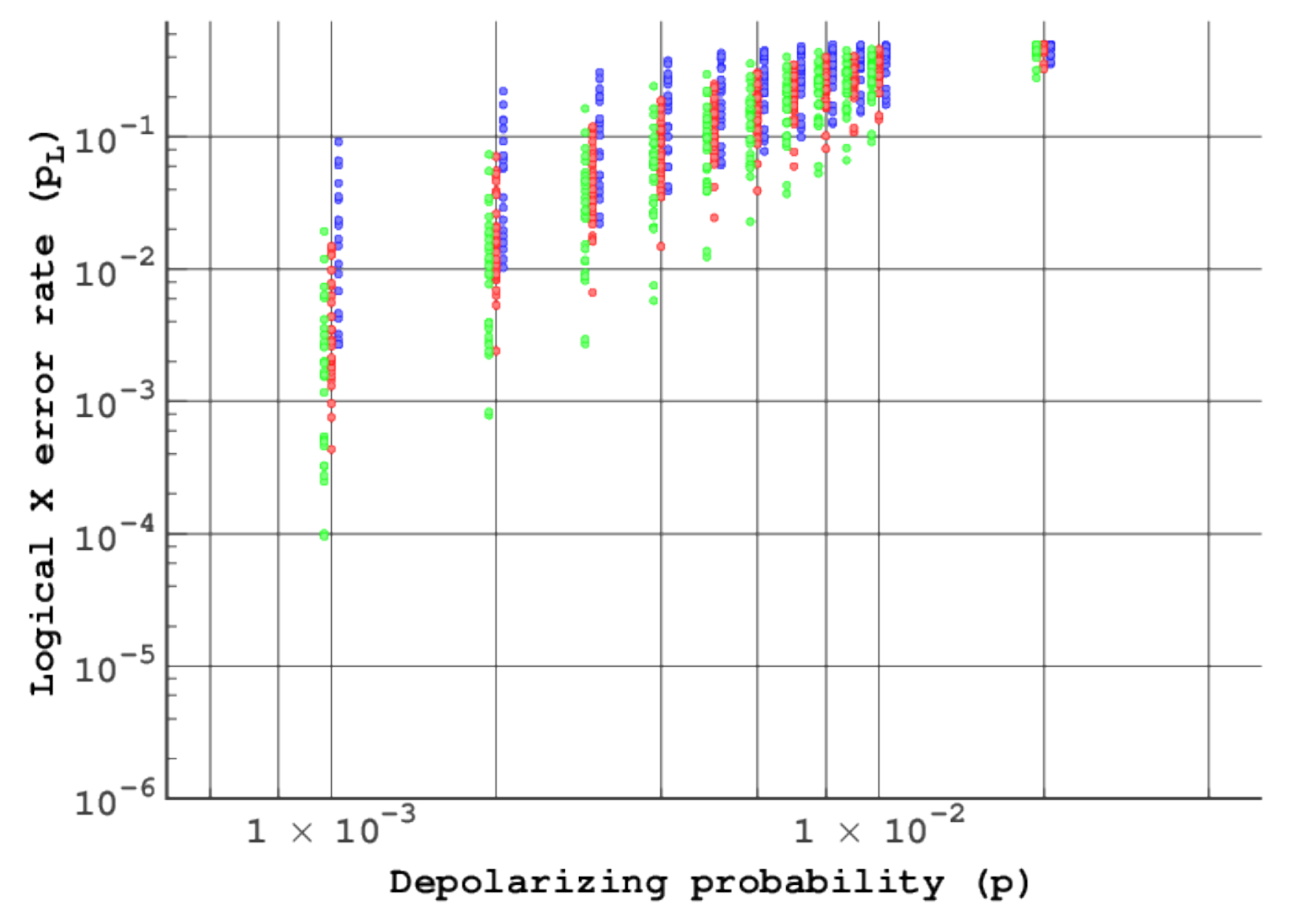}
  \caption{Scatterplot of $d=5$ with one dot per chip. 
  Green dots are of $y=95\%$, red dots are of $y=90\%$ and blue dots are of $y=80\%$.
  Blue and green data are offset from the vertical line for visibility.
  }
  \label{fig:scatter_d5}
 \end{center}
\end{figure}

\begin{figure}[t]
 \begin{center}
  \includegraphics[width=18cm]{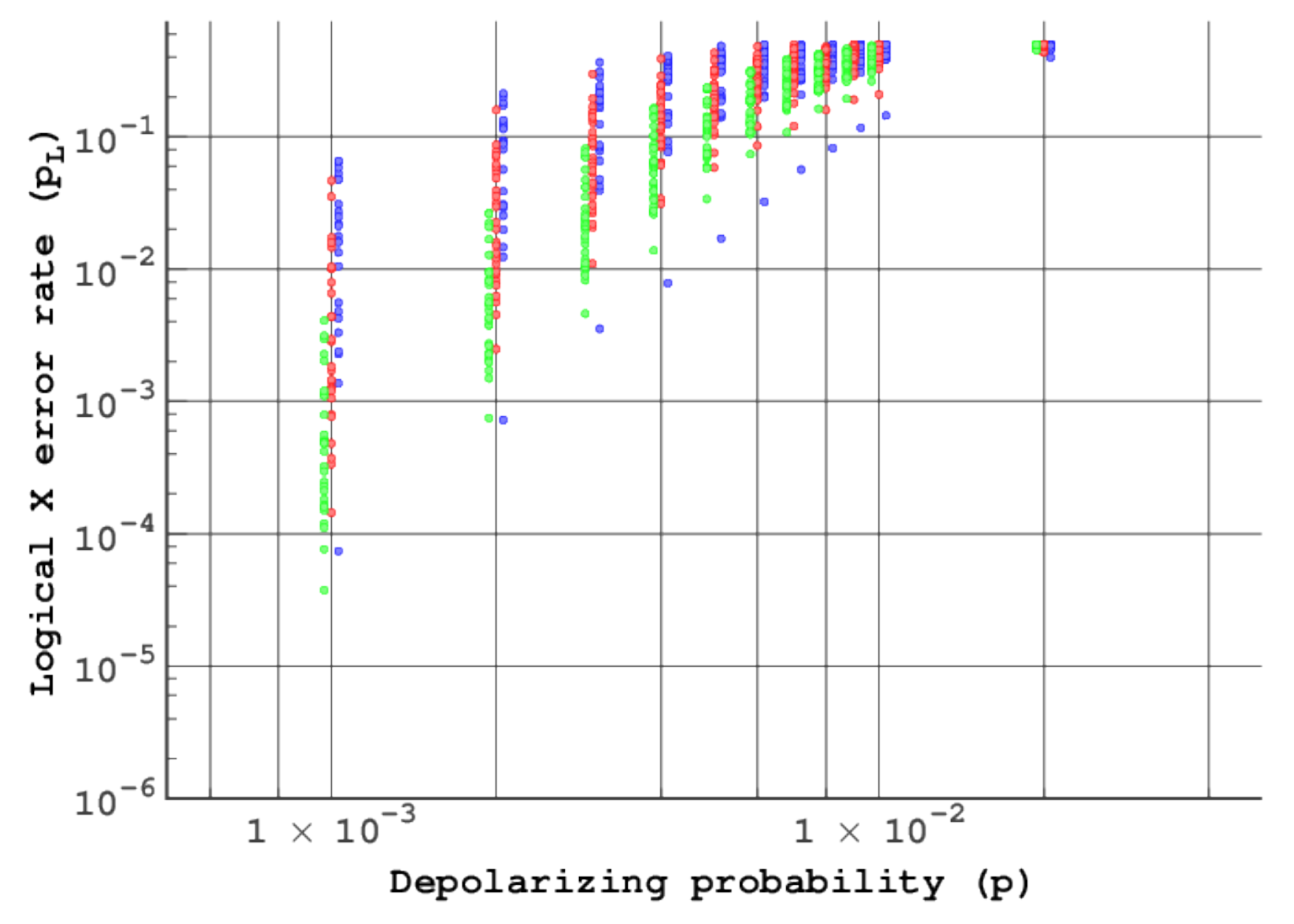}
  \caption{Scatterplot of $d=7$ with one dot per chip. 
  Green dots are of $y=95\%$, red dots are of $y=90\%$ and blue dots are of $y=80\%$.
  Blue and green data are offset from the vertical line for visibility.
  }
  \label{fig:scatter_d7}
 \end{center}
\end{figure}

\begin{figure}[t]
 \begin{center}
  \includegraphics[width=18cm]{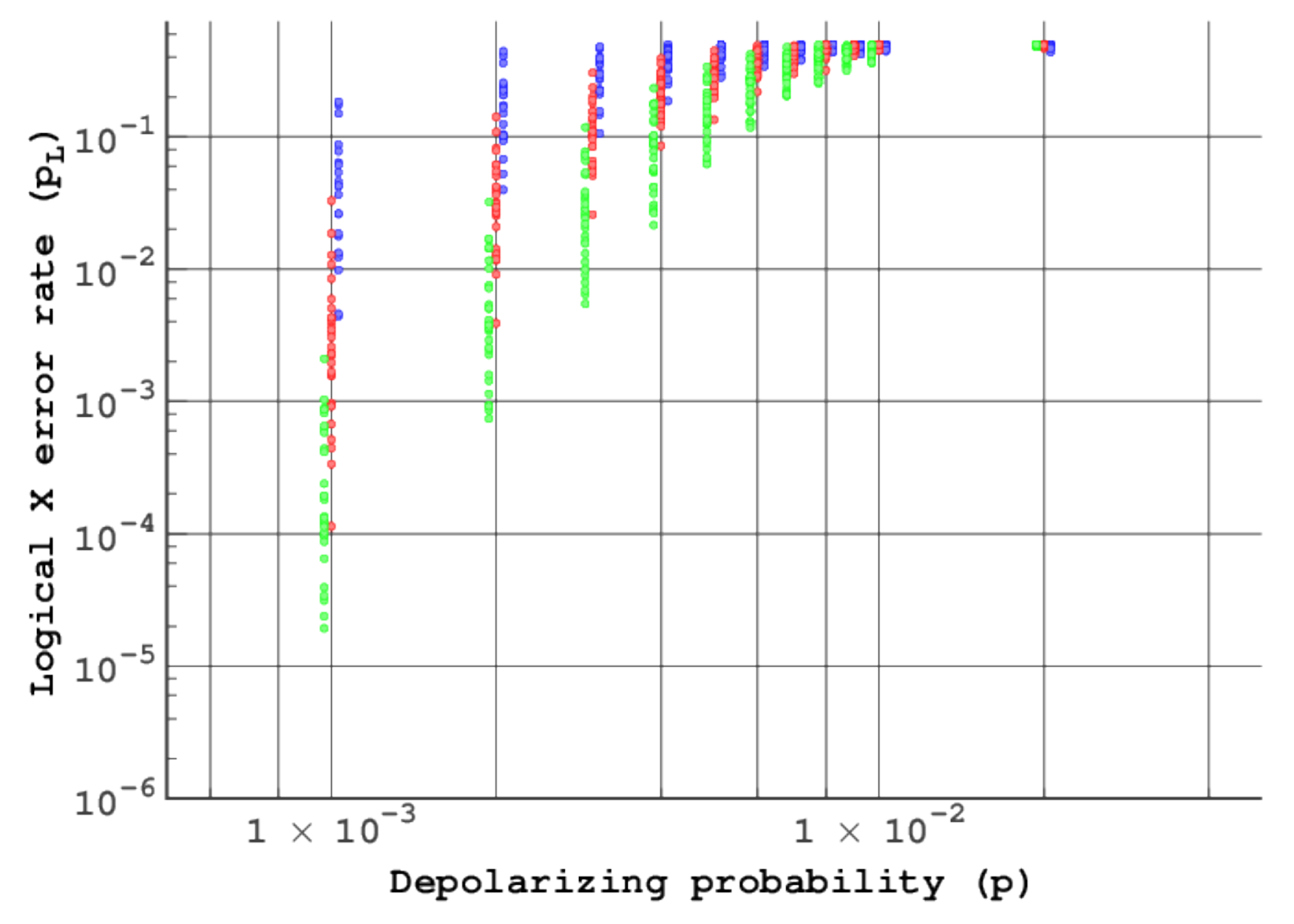}
  \caption{Scatterplot of $d=9$ with one dot per chip. 
  Green dots are of $y=95\%$, red dots are of $y=90\%$ and blue dots are of $y=80\%$.
  Blue and green data are offset from the vertical line for visibility.
  }
  \label{fig:scatter_d9}
 \end{center}
\end{figure}

\begin{figure}[t]
 \begin{center}
  \includegraphics[width=18cm]{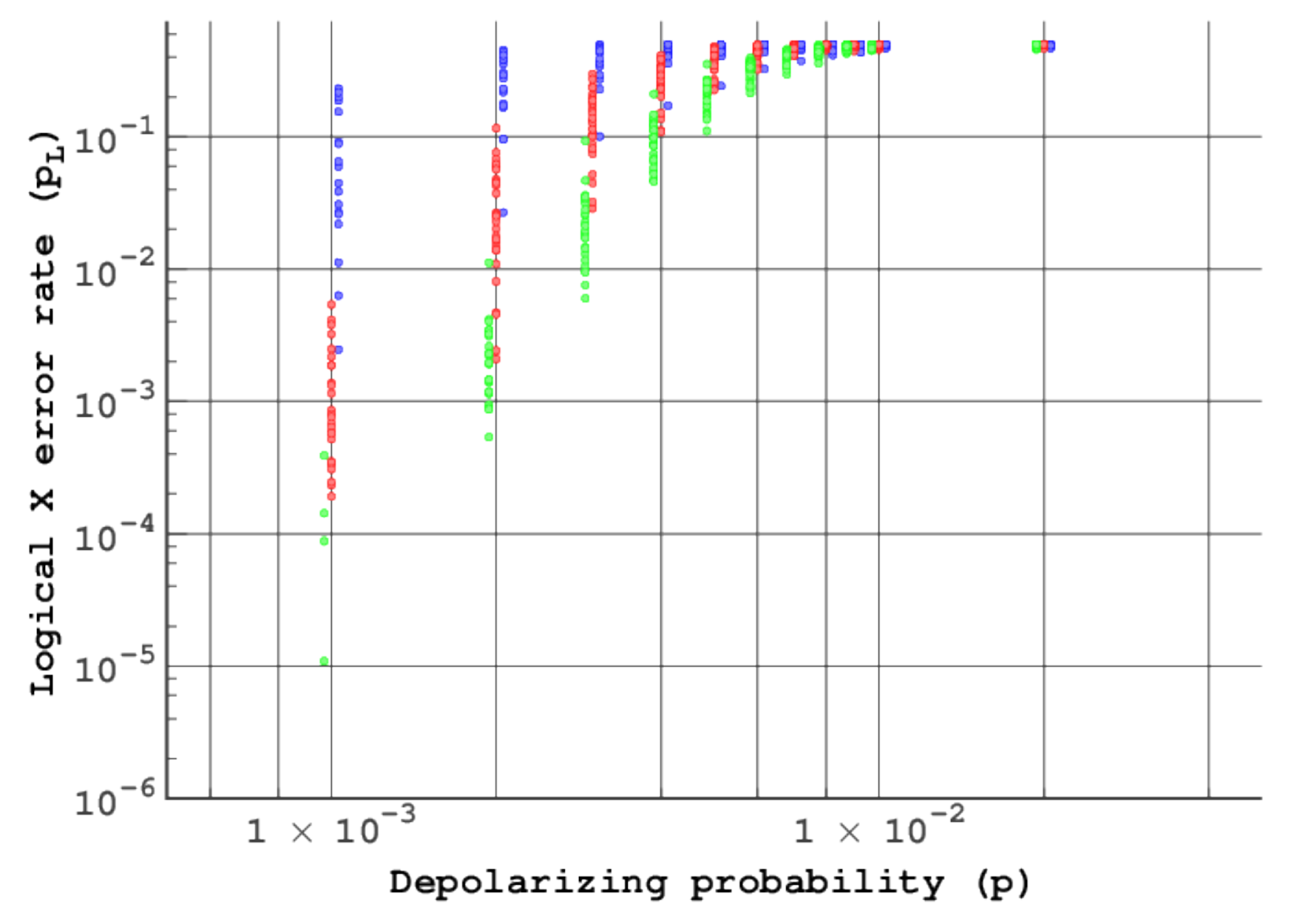}
  \caption{Scatterplot of $d=13$ with one dot per chip. 
  Green dots are of $y=95\%$, red dots are of $y=90\%$ and blue dots are of $y=80\%$.
  Blue and green data are offset from the vertical line for visibility.
  }
  \label{fig:scatter_d13}
 \end{center}
\end{figure}

\begin{figure}[t]
 \begin{center}
  \includegraphics[width=18cm]{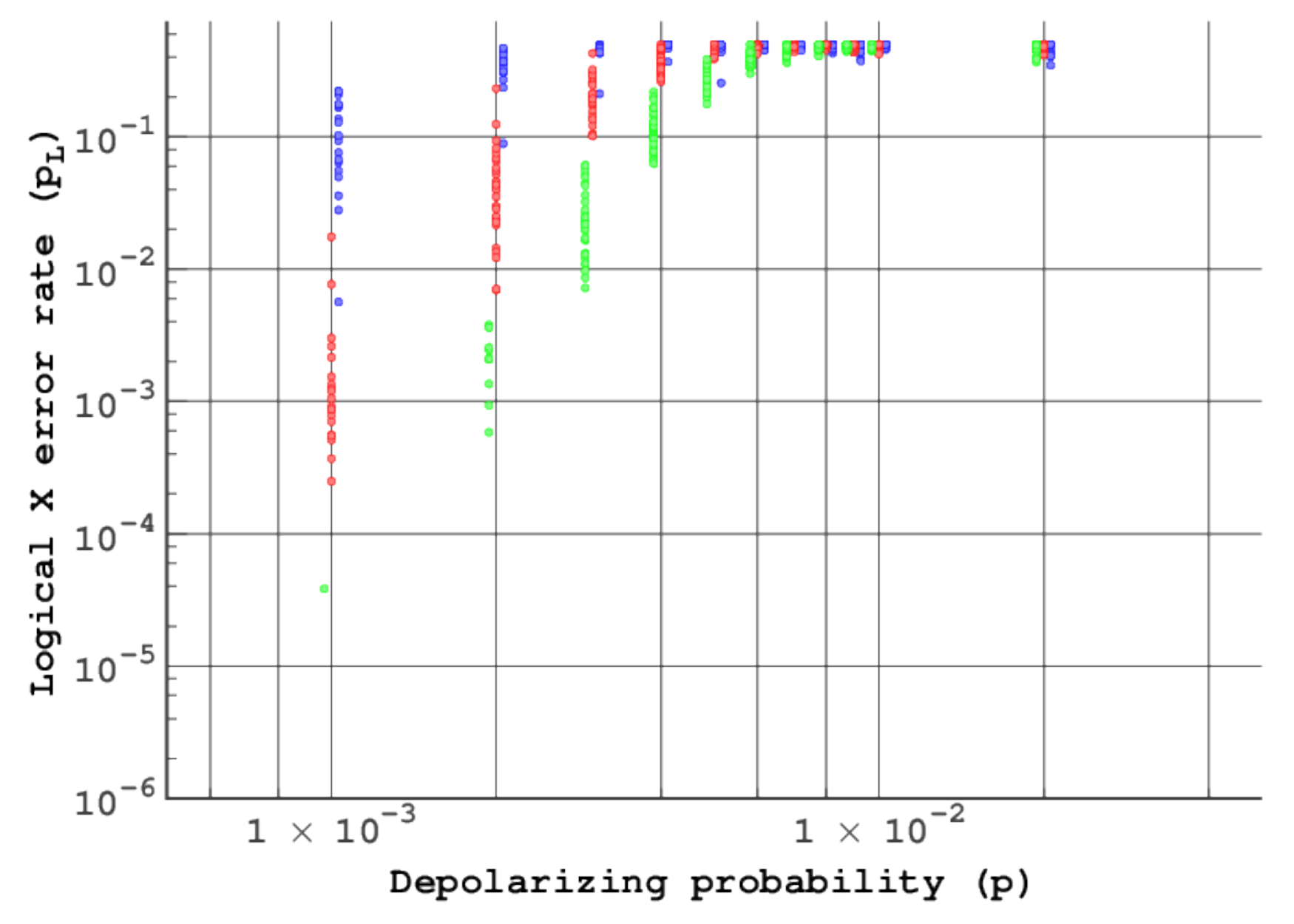}
  \caption{Scatterplot of $d=17$ with one dot per chip. 
  Green dots are of $y=95\%$, red dots are of $y=90\%$ and blue dots are of $y=80\%$.
  Blue and green data are offset from the vertical line for visibility.
  }
  \label{fig:scatter_d17}
 \end{center}
\end{figure}

Figures \ref{fig:scatter_d5}, \ref{fig:scatter_d7}, \ref{fig:scatter_d9}, \ref{fig:scatter_d13} and \ref{fig:scatter_d17}
show the scatter plots of raw data of randomly defective lattices.
  Figure \ref{fig:scatter_d7} shows an outlier chip.
  Actually the lattice on the chip has $40 \times$ worse logical $Z$ error rate as logical $X$ error rate.
  The lattice by chance has faulty devices 
  which deform the left and the right boundaries to be close
  with preserving the top and the bottom boundaries apart.
  Because of the largely deformed shape of the lattice,
  the usable area of the lattice is narrow and there are only few faulty devices on the usable area
  which increase logical error rates.
  Hence the chip exhibits stronger tolerance against logical $X$ error than others.


\end{document}